\documentclass[prx,preprintnumbers,amsmath,amssymb,nofootinbib,superscriptaddress]{revtex4-2}
\usepackage{graphicx,color}
\usepackage{bm}
\usepackage{epsfig}
\usepackage{hyperref}
\usepackage{float}
\usepackage{enumerate}  
\definecolor{amethyst}{rgb}{0.6, 0.4, 0.8}
\usepackage{amsmath}
\usepackage{stmaryrd}
\usepackage{soul}
\usepackage{bbold}
\usepackage{footmisc}
\usepackage{comment}
\usepackage{tikz}

\usetikzlibrary{calc, positioning, decorations.pathreplacing}

\newcommand{\be}{\begin{equation}}
\newcommand{\ee}{\end{equation}}
\newcommand{\beq}{\begin{equation}}
\newcommand{\eeq}{\end{equation}}
\newcommand{\bea}{\begin{eqnarray}}
\newcommand{\eea}{\end{eqnarray}}

\hypersetup{
    colorlinks=true,       
    linkcolor=red,          
    citecolor=blue,        
    filecolor=magenta,      
    urlcolor=cyan           
}

\begin{document}

\title{Exact Stationary State of a $d$-dimensional Run-and-Tumble Particle in a Harmonic Potential}

\begin{abstract}
We derive the exact nonequilibrium steady state of a run-and-tumble particle (RTP) in $d$ dimensions confined in an isotropic harmonic trap $V(\mathbf r)=\mu r^{2}/2$, with $r=\|\mathbf r\|$. Rotational invariance reduces the problem to the stationary single-coordinate marginal $p_X(x)$, from which the radial distribution $p_R(r)$ and the full joint stationary density follow by explicit integral transforms. We first focus on a generalized trapped RTP in one dimension, where post-tumble velocities are drawn from an arbitrary distribution $W(v)$. Using a Kesten-type recursion, we represent its stationary position in terms of a stick-breaking (or Dirichlet) process, yielding closed-form expressions for its distribution and its moments. Specializing $W(v)$ to the projected velocity law of an isotropic RTP, we reconstruct $p_R(r)$ and the full joint distribution of all the coordinates in $d=1,2,3$. In $d=1$ and $d=2$, the radial law simplifies to a beta distribution, while in $d=3$, {we derive closed-form expressions for $p_R(r)$ and the stationary joint distribution $P(x,y,z)$, which differ from a beta distribution. In all cases,} we characterize a persistence-controlled shape transition at the turning surface $r=v_0/\mu$, where $v_0$ is the self-propulsion speed. We further include thermal noise characterized by a diffusion coefficient $D>0$, showing that the stationary law is a Gaussian convolution of the $D=0$ result, which regularizes turning-point singularities and controls the crossover between persistence- and diffusion-dominated regimes as $D \to 0$ and $D \to \infty$ respectively. All analytical predictions are systematically validated against numerical simulations.
\end{abstract}


\author{Mathis Gu\'eneau}
\affiliation{Max-Planck-Institut f\"ur Physik komplexer Systeme, N\"othnitzer Straße 38, 01187 Dresden, Germany}
\author{Satya N. Majumdar}
\affiliation{LPTMS, CNRS, Univ.  Paris-Sud,  Universit\'e Paris-Saclay,  91405 Orsay,  France}
\author{Gr\'egory Schehr}
\affiliation{Sorbonne Universit\'e, Laboratoire de Physique Th\'eorique et Hautes Energies, CNRS UMR 7589, 4 Place Jussieu, 75252 Paris Cedex 05, France}

{
\let\clearpage\relax
\maketitle
\pretocmd{\tableofcontents}{\hypersetup{linkcolor=black}}{}{}
\apptocmd{\tableofcontents}{\hypersetup{linkcolor=red}}{}{}
\tableofcontents
}

\section{Introduction}

Confinement is the rule rather than the exception for active motion~\cite{soft, Ramaswamy2017, Marchetti2017, Schweitzer, ActiveMatterGKS, ActiveMatterRoadMap, flocking1, flocking2, separation1, separation2, separation3, separation4}: motile cells and synthetic swimmers navigate channels, pores, droplets, and traps, where geometric constraints compete with persistence and reshape steady-state statistics~\cite{BechingerRev,Denissenko,PorousRTP,Spagnolie,ReviewMicroMotors,spiralvortex,Berke2008}. 
Trapping provides a particularly clean route to probe nonequilibrium steady states: even a single active particle violates detailed balance, and its stationary distribution can become strongly non-Boltzmann even without any interactions~\cite{Berke2008, Cates2009, Potosky2012, Solon2015, Angelani2019, DKMSS19, sebastianActiveNoises, Bressloff, naftaliSS}. 
While minimal models such as the active Brownian particle (ABP)~\cite{ABP2012, ChristinaFranosch, ABP2018, ABP2019}, the active Ornstein–Uhlenbeck particle (AOUP)~\cite{AOUP, Wijland21}, and the run-and-tumble particle (RTP)~\cite{Tailleur_RTP, RTP_free, PRWWeiss} have been extensively explored, exact stationary distributions in smooth traps remain scarce beyond one dimension~\cite{CaraglioFranosch, Malakar2019, Frydel1, Frydel2, Frydel3}. 
In this work, we derive the exact stationary state of a $d$-dimensional run-and-tumble particle in an isotropic harmonic potential, closing the gap of an exact solution in three dimensions ($d=3$). Our results yield explicit and testable predictions for the full stationary statistics, directly comparable to experiments.\\

Run-and-tumble dynamics is a minimal and widely used model of bacterial motility, notably for E. coli~\cite{Berg2004, Naturecoli}. In its simplest implementation, an RTP performs straight runs interrupted by isotropic reorientation events and already reproduces key microscopic features of bacterial motion~\cite{PRLcoli, PREcoli, ISFRTP, Angelani2013}. Its motion is intrinsically out of equilibrium, affecting not only steady states but also time-dependent observables, including relaxation and first-passage properties~\cite{Singh2019, generalRTP, RTPsurvivalMori, MoriCondensation, Benjamin1, MFPTlong, MFPT_1D_RTP, ExitProbaShort, FrydelSplitting, SurvivalRPTlinear, GrangeYuan, Kafri2, TVB12, RBV16, EPJEAngelani, Angelani2014, Targetsearch, BressloffStickyBoundaries}. A harmonic trap is a natural and versatile minimal model of confinement as it describes the vicinity of a stable minimum to leading order. Experimentally, optical and acoustic tweezers are often approximated by a quadratic potential near the trapping point and provide a controlled way to manipulate individual bacteria~\cite{Optical, OpticalBacteria,OpticalBacteria2,OpticalBacteria3, AcousticBacteria}. Exact stationary distributions for harmonically trapped run-and-tumble particles are known in one dimension~\cite{DKMSS19, resettingNoise, RTP3state, Rosalba} and in a few specific two-dimensional settings~\cite{SLMS2022,Frydel1, Frydel2, Frydel3}, whereas closed-form results in higher dimensions remain, to our knowledge, unavailable -- see e.g. \cite{Frydel3}. {While one may try to obtain the stationary state from a Fokker–Planck description, for RTPs this route leads to a nonlocal integro-differential equation that does not appear analytically tractable. Here, we instead solve the problem via a different route, based on a Kesten recursion~\cite{resettingNoise, Kesten73} and its Dirichlet-process representation (via a stick-breaking construction)~\cite{Ferguson, CifarelliRegazzini, Sethuraman}, which provides the stationary distribution in closed form for any dimension.}\\

For bounded-speed dynamics such as run-and-tumble motion, confinement generically creates turning points (or, in higher dimensions, a turning surface) where the force generated from the potential exactly balances self-propulsion and the deterministic velocity vanishes~\cite{DKMSS19, LeDMS, MFPT_1D_RTP, MFPTlong,Hah25, HPVO}. Near these locations, the stationary distribution undergoes a transition as the ratio between the tumbling rate and the trap relaxation rate is varied. In one dimension, this interplay is known to yield a persistence-controlled \emph{shape transition}~\cite{DKMSS19}: the stationary density crosses over from a regular, bell-shaped profile at low persistence to an active regime in which the probability mass accumulates at the turning points, producing integrable edge singularities. 
The structure of this transition in higher dimensions -- most notably, the associated radial statistics that directly quantify accumulation near the turning surface -- remains comparatively less explicit. In experimental settings, motile microorganisms are additionally subject to thermal fluctuations~\cite{PRLcoli, PREcoli}, which are expected to regularize turning-point singularities, and govern the crossover between persistence-dominated and diffusion-dominated stationary regimes, motivating an explicit treatment of the combined active and thermal noises~\cite{sebastianActiveNoises}.\\

The remainder of the paper is organized as follows. In Sec.~\ref{ModelSection}, we introduce the harmonically trapped RTP in $d$ dimensions and summarize our main results, emphasizing how isotropy allows to reduce the full stationary problem to the analysis of the one-coordinate marginal. In Sec.~\ref{GeneralizedRTPSection}, we derive the stationary state of a generalized one-dimensional run-and-tumble process in a harmonic trap, where post-tumble velocities are redrawn from an arbitrary distribution. This yields an explicit stationary law for a single component of the $d$-dimensional RTP. In Sec.~\ref{2dSSSection}, we specialize these results to standard RTPs in $d=1,2,3$ and {compute the stationary radial distribution and the full joint distribution of {the coordinates of the} position}, including a detailed characterization of the shape transition in each case. In Sec.~\ref{DonSection}, we include thermal diffusion ($D>0$), and show that the relative strength of active noise and thermal diffusion regularizes the turning-point singularity and controls the crossover between persistence-dominated and diffusion-dominated regimes in $d=1,2,3$. Finally, in Sec.~\ref{NstateSection}, we discuss a discrete $N$-state RTP as a further application. We conclude in Sec.~\ref{ConclusionSection} with a discussion of the main implications of our exact results and a brief outlook on extensions and open questions.

\section{Model and Main Results}\label{ModelSection}


We consider a run-and-tumble particle in $d$ dimensions ($d>1$), confined in a harmonic potential. Its position is
\begin{equation}
    \textbf{r}(t) = \bigl(x_1(t),x_2(t),\cdots,x_d(t)\bigr)\in \mathbb{R}^d
\end{equation}
and it evolves according to
\begin{equation}\label{ddimRTPLange}
    \dot{\textbf{r}}(t) \;=\; -\mu\, \textbf{r}(t) \;+\; v_0\, \textbf{n}(t)\, ,
\end{equation}
where $\mu>0$ is the trap strength, $v_0>0$ the constant self-propulsion speed, and
\begin{equation}
    \textbf{n}(t)=\bigl(n_1(t),\cdots,n_d(t)\bigr), \qquad \|\textbf{n}(t)\|=1,
\end{equation}
is the orientation vector. Between tumbles, $\mathbf n(t)$ remains constant. Tumbling events occur at a fixed rate $\gamma$ such that the run durations are exponentially distributed
\begin{equation}
    p(\tau) =  \gamma e^{-\gamma\tau}\, .
\end{equation}
It is convenient to introduce the dimensionless ratio of timescales
\begin{equation}\label{DefAlpha}
    \alpha = \frac{\gamma}{\mu}\, ,
\end{equation}
which compares the mean run time $1/\gamma$ to the trap relaxation time $1/\mu$. This ratio controls the crossover from persistence-dominated dynamics ($\alpha \ll 1$) to a passive-like behavior ($\alpha \gg 1$). At each tumbling event, the orientation is reset and drawn \emph{isotropically}, i.e., uniformly over the unit sphere $S^{d-1}\subset\mathbb R^d$.
Since both the trap and the reorientations are isotropic, the stationary state is rotationally invariant. As a result,
each Cartesian component $x_i(t)$ evolves as a one-dimensional run-and-tumble process driven by the projected velocity
$v(t)=v_0 n_i(t)$, whose stationary statistics are the same for all $i$. We therefore drop the index $i$ and focus on a
single component, whose Langevin equation reads
\begin{equation}\label{generalizedRTP}
    \dot{x}(t) \;=\; -\mu\, x(t) \;+\; v(t)\,.
\end{equation}
During the $n^{\text{th}}$ run, the velocity component $v(t) = {\sf v}_n$ is constant and drawn from the distribution 
\begin{equation}\label{WprojEq}
    W_{\rm proj}(v) \;=\; \frac{1}{v_0}\,\frac{\Gamma\!\bigl(\tfrac{d}{2}\bigr)}{\sqrt{\pi}\,\Gamma\!\bigl(\tfrac{d-1}{2}\bigr)}\,
    \left(1-\frac{v^2}{v_0^2}\right)^{\frac{d-3}{2}}, \qquad -v_0\le v\le v_0 \, ,
\end{equation}
where the index ``proj'' indicates that this is the distribution of a single coordinate projection of the vector $v_0\, \textbf{n}(t)$. In Fig.~\ref{fig-GeneralizedRTP}, we show a realization of such a process while in Appendix~\ref{WprojApp}, we derive the expression of $W_{\rm proj}(v)$. When $d=2$, the distribution of a single velocity component reduces to the arcsine law, while for $d=3$ it becomes uniform:
\begin{equation}
W_{\text{proj}}(v) =
\begin{cases}
\dfrac{1}{\pi\sqrt{v_0^2 - v^2}}, & d=2\ , \\[2ex]
\dfrac{1}{2 v_0}, & d=3\, ,
\end{cases}
\qquad -v_0 \le v \le v_0 \, .
\end{equation}

Our main goal is to determine the stationary distribution $p_R(r)$ of the particle's radial coordinate. Let
$\textbf{x} = (x_1, \cdots, x_d)$ denote a random vector distributed according to the stationary law of $\textbf{r}(t)$. We define
\begin{equation}
    R \overset{d}{=} \|\textbf{x} \| = \sqrt{x_1^2 + \cdots + x_d^2}\, ,
\end{equation}
Owing to rotational invariance, the distribution of the radius $R$ is completely characterized by the stationary distribution $p_X(x)$ of any single coordinate, say $X \overset{d}{=} x_i$. The two are related through an integral transform: for $d\geq 2$,  
\begin{equation}\label{pXtopR}
    p_X(x) \;=\; \frac{\Gamma\!\bigl(\tfrac{d}{2}\bigr)}{\sqrt{\pi}\,\Gamma\!\bigl(\tfrac{d-1}{2}\bigr)}
    \int_{|x|}^{v_0/\mu} dr \, \frac{p_R(r)}{r}\left(1-\frac{x^2}{r^2}\right)^{\frac{d-3}{2}} \, .
\end{equation}
{The relation in Eq.~(\ref{pXtopR}), which is derived in Appendix~\ref{radial_single_App}, expresses the coordinate distribution as a projection of the radial law and was demonstrated explicitly in~\cite{MoriCondensation} for the unconfined RTP.}
As shown in Appendix~\ref{momentsXRapp}, it is also possible to relate the moments of $|X|$ to those of $R$ for any $d>1$. In the special cases $d=2$ and $d=3$, the inverse relation can be obtained explicitly -- see also Appendix~\ref{radial_single_App}. One finds
\begin{equation}\label{pXtopRd2d3}
p_R(r) =
\begin{cases}
 -2r\int_r^{\frac{v_0}{\mu}} dx\, \frac{p_X'(x)}{\sqrt{x^2-r^2}}, & d=2\, , \\[2ex]
-2r\, p'_X(r), & d=3\, .
\end{cases}
\end{equation}
Finally, thanks to rotational invariance, the joint density of all coordinates can be expressed in terms of the radial distribution -- see Appendix~\ref{jointToRadial}. Specifically,
\begin{equation}
    P(x_1, \cdots, x_d) = f_R\!\left(r =\sqrt{x_1^2 + \cdots + x_d^2}\right) = \frac{p_R(r)}{\Omega_{d-1}r^{d-1}}\, , \qquad \Omega_{d-1} = \frac{2\pi^{d/2}}{\Gamma(d/2)}\, ,
    \label{generalJointPDF}
\end{equation}
where $\Omega_{d-1}$ is the surface area of the unit sphere in $\mathbb{R}^d$.\\

\begin{figure}[t]
    \centering
    \includegraphics[width=0.4\linewidth]{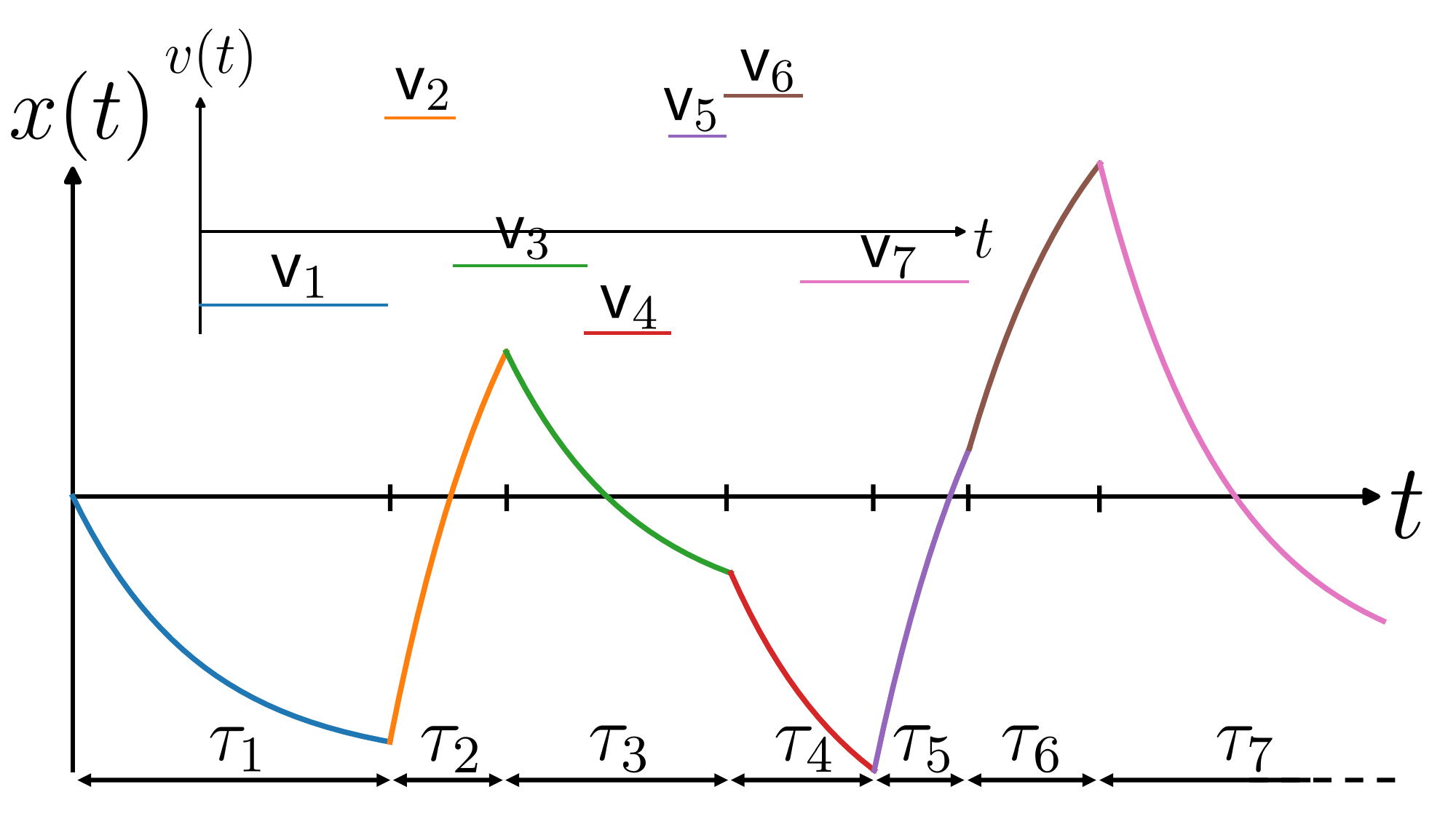}
    \hspace{2cm}
    \includegraphics[width=0.4\linewidth]{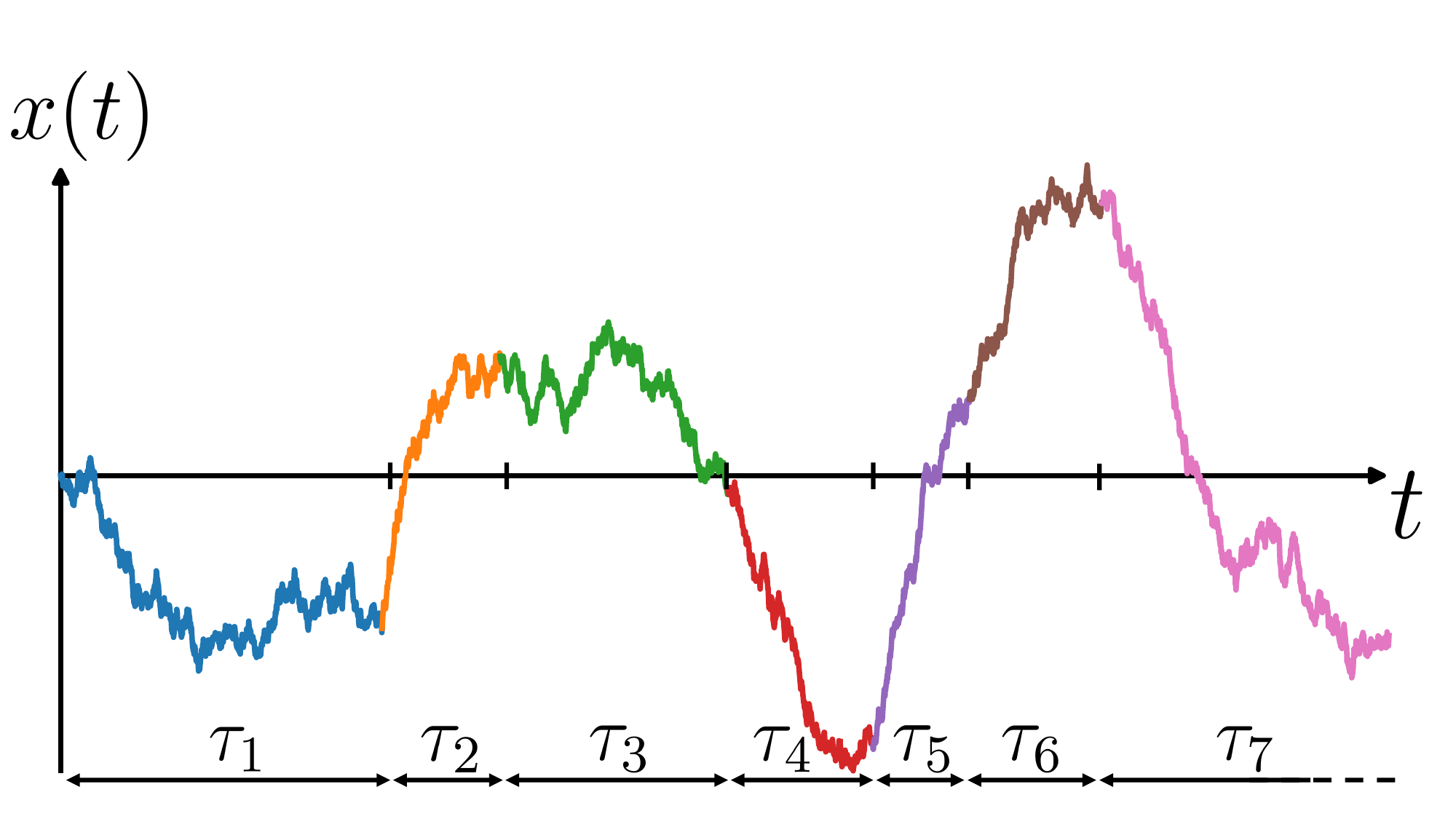}
    \caption{Illustration of a trajectory of the generalized RTP process governed by the Langevin equation~(\ref{generalizedRTP}).
\textbf{Left:} A sample trajectory for $D=0$, where the inter-tumbling velocities ${\sf v}_i$ are drawn from an arbitrary distribution $W(v)$.
\textbf{Right:} The corresponding trajectory in the presence of white noise ($D>0$). In this case, the stationary distribution is given by the convolution of the $D=0$ distribution with the Gaussian distribution $\mathcal{N}(0,D/\mu)$ -- see Section~\ref{DonSection}.}
    \label{fig-GeneralizedRTP}
\end{figure}

If the RTP is additionally subjected to thermal noise, Eq.~(\ref{ddimRTPLange}) acquires an additive Gaussian white-noise term, and the equation of motion becomes
\begin{equation}\label{ddimRTPLangeThermal}
    \dot{\textbf{r}}(t) \;=\; -\mu\, \textbf{r}(t) \;+\; v_0\, \textbf{n}(t) + \sqrt{2D}\, \boldsymbol{\eta}(t)\, ,
\end{equation}
where $\boldsymbol{\eta}(t)=(\eta_1(t),\ldots,\eta_d(t))$ satisfies
\begin{equation}\label{ddimWN}
\langle \eta_i(t) \rangle = 0 \, , \qquad \langle \eta_i(t)\eta_j(t') \rangle = \delta_{ij}\delta(t-t') \, .
\end{equation}
For $D>0$, the stationary position probability distribution function (PDF), denoted by $p_Z(z)$, can be expressed as the convolution of the stationary PDF in the noiseless case $D=0$ (i.e., the Langevin dynamics~(\ref{generalizedRTP})), denoted by $p_X(x)$, with a Gaussian distribution of mean $0$ and variance $D/\mu$. Equivalently,
\begin{equation} \label{pZ}
p_Z(z) = \sqrt{\frac{\mu}{2\pi D}} \int_{-\frac{v_{0}}{\mu}}^{\frac{v_{0}}{\mu}} e^{-\frac{\mu (z-x)^2}{2D}}\, p_X(x)\, dx \, ,
\end{equation}
where $p_X(x)$ is supported on $[-v_0/\mu, v_0/\mu]$. This simplification follows from the linearity of Eq.~\eqref{generalizedRTP} together with the independence of the two noise sources -- see Section~\ref{DonSection}. For $D>0$, Eqs.~\eqref{pXtopR}--\eqref{generalJointPDF} remain valid upon the replacement~$p_X \to p_Z$, and the upper integration limit in Eqs.~(\ref{pXtopR}) and (\ref{pXtopRd2d3}) is extended to $+\infty$ since $p_Z(z)$ is supported on $(-\infty,+\infty)$.\\

The above relations show that the stationary statistics of the $d$-dimensional RTP are entirely encoded in the one-coordinate marginal $p_X(x)$: once $p_X(x)$ is known, one can reconstruct the radial law $p_R(r)$, all moments, and the full joint distribution of all the coordinates. In the next section, we introduce a method to compute the stationary distribution of the Langevin dynamics~\eqref{generalizedRTP} based on a stick-breaking (Dirichlet-process) representation. This construction yields an explicit characterization of $p_X(x)$ and thereby provides access to the complete stationary properties of the RTP. We now summarize the main results of the paper.

\subsection{Main Results}

\noindent\textbf{Stationary State of a Generalized RTP.} We refer to the dynamics in Eq.~(\ref{generalizedRTP}) as a generalized RTP, when at each tumble the velocity is redrawn from an arbitrary distribution $W(v)$ supported on $[v_{\min},v_{\max}]$ -- see~\cite{RTPsurvivalMori}. In this case, we find that the stationary distribution of the dynamics  admits an exact representation as a mean functional of a Dirichlet process~\cite{CifarelliRegazzini} (equivalently, a stick-breaking construction). This yields an explicit closed-form expression for the stationary density
\begin{equation}
p_X(x) = \dfrac{1}{\pi} 
\int_{v_{\min}/\mu}^x \! dt \, (x-t)^{\alpha-1}\, \frac{d\phi_\alpha(t)}{dt},
\qquad x \in \left[\tfrac{v_{\min}}{\mu}, \tfrac{v_{\max}}{\mu}\right], \qquad \forall \alpha > 0\, ,
\label{MRSSX}
\end{equation}
where $\alpha=\gamma/\mu$ is given in Eq.~(\ref{DefAlpha}) and
\begin{eqnarray}\label{phi_defMR}
\phi_{\alpha}(t) = \sin\!\left( \pi \alpha \int_{v_{\min}}^{\mu t}dv\, W(v) \right)
\exp\!\left( -\alpha \int_{v_{\min}}^{v_{\max}} dv\, 
\log\!\left| t - \tfrac{v}{\mu} \right|\, W(v)\right)\, . 
\end{eqnarray}
In addition, all moments $\langle X^n \rangle$ can be expressed in closed form in terms of Bell polynomials~\cite{BellComtet, BellPoly} and the moments of $W(v)$ denoted by $\langle v^n\rangle$
\begin{eqnarray}
\langle X^{n} \rangle = \frac{1}{\mu^n} \frac{\Gamma(\alpha)}{\Gamma(\alpha+n)} \;
B_n\left(1!\, \alpha\, \frac{\langle v \rangle}{1}, \ldots,n!\, \alpha\, \frac{\langle v^{n} \rangle}{n} \right)\, .
\label{exactmomentsMR}
\end{eqnarray}
Specializing this general result to $W_{\text{proj}}(v)$ given in Eq.~(\ref{WprojEq}) immediately produces the exact stationary statistics of a single coordinate of a $d$-dimensional RTP in a harmonic trap for any $d\geq2$.\\

\noindent\textbf{Stationary State in $d=2$.}
Remarkably, just like in $d=1$~\cite{DKMSS19} (see Section~\ref{SS1dRTPnoD}), the stationary radial distribution simplifies to a beta distribution. In $d=2$, this expression was previously inferred from its moments (see Refs.~\cite{Frydel1,Frydel2})
\begin{equation}\label{MR2}
p_R(r)  = 2\alpha \frac{\mu^2 r}{v_0^2}
\left[ 1 - \left( \frac{\mu r}{v_0} \right)^2 \right]^{\alpha - 1}\, ,
\qquad 0 \le r \le \frac{v_0}{\mu}\, .
\end{equation}
Its edge behavior exhibits a transition at $\alpha=1$, diverging at the boundary $r_0 = v_0/\mu$ for $\alpha<1$ (active/persistent regime) and vanishing for $\alpha>1$ (passive regime). In the presence of thermal noise, we show that the radial distribution is given by
\begin{eqnarray}\label{pR2dDneq0}
    p_R\!\left(\tilde r = \frac{\mu r}{v_0}\right) =  \sqrt{\frac{2}{D\mu}} \frac{v_0^2}{ \pi D} \frac{\Gamma\!\left(1+\alpha\right)}{ \Gamma\!\left(\frac{1}{2}+\alpha\right)} \, \tilde r \int_{\tilde r}^{+\infty}\frac{d\tilde z}{\sqrt{\tilde z^2-\tilde r^2}}\int_{-1}^{1}d\tilde x\,  (\tilde z - \tilde x)(1-\tilde x^2)^{\alpha-\frac{1}{2}}\, e^{- \frac{v_0^2}{2 D \mu}(\tilde x- \tilde z)^2}\, .
\end{eqnarray}
In Appendix~\ref{2dDto0App}, we show that, in the limit $D\to0$, Eq.~(\ref{pR2dDneq0}) reduces to Eq.~(\ref{MR2}).\\

\noindent\textbf{Stationary State in $d=3$.}
By contrast, a closed-form expression for the stationary radial distribution in $d=3$ does not appear to be available in the existing literature~\cite{Frydel3}. We compute it and find that it is given by 
\begin{equation}
    p_R(r) = - \frac{\alpha\,e^{\alpha}}{2\pi}\,\frac{\mu^2}{v_0^2}\, r\left[\pi A(r)^{\alpha-1} + \int_{0}^{A(r)} du\,  (A(r)-u)^{\alpha-1}f'(u)\right]\, , \qquad 0 \leq r \leq \frac{v_0}{\mu} \, ,
\end{equation}
where
\begin{eqnarray}
A(r) = \tfrac{1}{2}\!\left(\tfrac{\mu r}{v_0} + 1\right)\, , \qquad 
f(u)=
e^{-\alpha\,[\,u\ln u + (1-u)\ln(1-u)\,]}\,
\left[\pi\cos(\pi\alpha u)
-\ln\!\left(\frac{u}{1-u}\right)\,\sin(\pi\alpha u)\right]\, ,
\end{eqnarray}
while $p_R(r) =0$ outside $[0, v_0/\mu]$.
In $d=3$ also, we find that the PDF $p_R(r)$ exhibits a shape transition at
$\alpha=1$, as in $d=2$. In particular, at this transition when $\alpha =1$, the expression simplifies to 
\begin{align}\label{pX_a1MR}
p_R(r) = \frac{e\,\mu^{2}\,r}{\pi v_0^{2}}
\left(1-\frac{\mu r}{v_0}\right)^{-\frac12\left(1-\frac{\mu r}{v_0}\right)}
\left(1+\frac{\mu r}{v_0}\right)^{-\frac12\left(1+\frac{\mu r}{v_0}\right)}
\Biggl[
2\,\operatorname{arctanh}\!\left(\frac{\mu r}{v_0}\right)\!
\cos\!\left(\frac{\mu\pi r}{2v_0}\right)
+\pi\sin\!\left(\frac{\mu\pi r}{2v_0}\right)
\Biggr]\, .
\end{align}
We also compute the exact radial distribution in the presence of additional thermal noise in Section~\ref{HighDimDon3}.\\

\noindent\textbf{Thermal noise ($D>0$) and finite-$D$ shape transitions.}
When an additional Gaussian white noise of diffusivity $D>0$ is present, the equation of motion is given in Eq.~(\ref{ddimRTPLangeThermal}). This equation being linear, the stationary law is obtained by a Gaussian convolution of the $D=0$ steady state (see Eq.~(\ref{pZ})). This representation makes explicit how temperature (i) rounds off the turning-point singularities, (ii) generates a Gaussian tail beyond the edge of the support $r_0=v_0/\mu$, and (iii) controls the crossover between persistence-dominated and diffusion-dominated regimes. It is convenient to quantify the relative strength of thermal fluctuations to active noise through the dimensionless parameter
\begin{equation}
\theta=\frac{2\mu D}{v_0^2} =\frac{2}{\alpha}\frac{D}{D_{\rm eff}}\, ,
\label{theta_MR}
\end{equation}
where $D_{\rm eff}=v_0^2/\gamma$ is the effective diffusion coefficient of a free RTP.
As $\theta$ decreases, the stationary distribution exhibits a qualitative change: in $d=1$ and $d=2$ the stationary density smoothly crosses over from a unimodal profile peaked at the origin to a bimodal profile with two maxima near the turning points. In $d=3$, the corresponding transition becomes richer, with a coexistence region in which an inner and an outer maximum compete. In some cases, the position of the global maximum jumps discontinuously from $r=0$ to a shell near $r=r_0$ (see Fig.~\ref{figcriticald3}).\\

\noindent\textbf{$N$-state Model.} As a direct consequence of Eq.~(\ref{MRSSX}), one can compute the stationary state of a $N$-state model
\begin{equation}
    W(v) \;=\; \sum_{i=1}^N p_i \, \delta(v - v_i)\,,
    \qquad 0<p_i<1, \qquad \sum_{i=1}^N p_i = 1 \, .
    \label{NWstateMR}
\end{equation}
We find that $p_X(x)$ is the distribution of a linear combination of a Dirichlet random vector. Its expression is given by
\begin{equation}
    p_X(x) = \frac{\Gamma(\alpha)}{\prod_{i=1}^N \Gamma\left(\alpha\, p_i\right)}\, \int_0^1dz_1\cdots\int_0^1dz_N  \, \delta\left(x-\frac{1}{\mu}\sum_{i = 1}^N  v_i z_i\right)\, 
\left( \prod_{i=1}^N z_i^{\alpha p_i - 1} \right)\,
\delta\!\left( 1 - \sum_{i=1}^N z_i \right)\, .
\end{equation}
In particular, it turns out that the density is piecewise continuous on the intervals $x \in (v_k/\mu,v_{k+1}/\mu)$ where $v_k<v_{k+1}$ for all $k = 1,\cdots,N-1$ -- see Eq.~(\ref{CRRepNState}) and Fig.~\ref{3statefig}. {If $N=2$, $v_1 = -v_2=v_0$ and $p_1=p_2=1/2$, this corresponds to the standard one-dimensional RTP~\cite{Tailleur_RTP, DKMSS19, Klyatskin, LHKI, TC2009}}.\\


\section{Generalized Run-and-Tumble Dynamics in one Dimension}\label{GeneralizedRTPSection}

To calculate the steady state distribution of a $d$-dimensional RTP, we first consider a generalized one-dimensional run-and-tumble particle evolving in a harmonic trap, starting from the initial condition $x(0)=0$. The dynamics is governed by Eq.~(\ref{generalizedRTP}), where $v(t)$ is piecewise constant and changes only at the tumbling times (see the left panel of Fig.~\ref{fig-GeneralizedRTP} for an example trajectory). At each tumbling event, which occurs with a fixed rate $\gamma$, the velocity $v$ is redrawn from a prescribed distribution $W(v)$ supported on $[v_{\min}, v_{\max}]$~\cite{RTPsurvivalMori}. Importantly, in this section, we focus on a completely general $W(v)$ which may be either discrete or continuous. In particular, when the distribution $W(v)$ is chosen as in Eq.~(\ref{WprojEq}), the dynamics in Eq.~(\ref{generalizedRTP}) {corresponds} to the one of a single component of the $d$-dimensional RTP introduced in the previous section.

\subsection{Kesten Recursion Relation}\label{KestenSubsect}

One way to describe the dynamics given in Eq.~(\ref{generalizedRTP}) is to consider the deterministic motion between successive tumbling times $\{t_1, t_2, \ldots, t_n\}$ \cite{resettingNoise}. Introducing the variable $x_n$ to denote the position of the particle immediately after the $n$-th tumbling event, we can derive a recursive relation for $x_n$. Indeed, integrating Eq.~(\ref{generalizedRTP}) over one run, i.e., between $t_{n-1}$ and $t_n$, and defining $\tau_n = t_n - t_{n-1}$, we obtain
\begin{eqnarray}
x_n = x_{n-1}\, e^{-\mu \tau_n} + \frac{{\sf v}_n}{\mu}\left(1 - e^{-\mu \tau_n}\right)\, , \qquad \qquad x_0=0\, ,
\label{Kest_rel}
\end{eqnarray}
where we recall that $p(\tau_n) = \gamma\, e^{-\gamma \tau_n}$ and ${\sf v}_n$ is drawn from the PDF $W(v)$. This recursion has the characteristic form of a Kesten relation~\cite{Kesten73, resettingNoise},
\begin{eqnarray}
x_n = U_n x_{n-1} + V_n\, , \qquad\qquad U_n = e^{-\mu \tau_n}\, , \qquad V_n = \frac{{\sf v}_n}{\mu}\left(1 - U_n\right)\, ,
\label{KestenRecursion}
\end{eqnarray}
where $U_n$ and $V_n$ are random variables. Given that $\tau_n$ follows an exponential distribution with fixed rate $\gamma$, it is straightforward to show that $U_n \sim \mathrm{Beta}(\alpha, 1)$, i.e.
\begin{eqnarray}\label{UPDF}
    P(U) = \alpha\, U^{\alpha -1}\, , \qquad 0\leq U\leq1\, , \qquad\alpha = \frac{\gamma}{\mu}\, .
\end{eqnarray}
The parameter $\alpha$ represents the ratio between the two characteristic timescales of the system: the average relaxation time of the particle within the harmonic trap, $1/\mu$, and the mean time between two tumbling events, $1/\gamma$. This important parameter effectively controls the level of activity of the particle: the smaller $\alpha$ is, the more persistent its motion becomes.\\

The stationary properties of such Kesten relations are usually studied by integrating over all paths connecting $x_{n-1}$ to $x_n$ under the constraint of Eq.~(\ref{KestenRecursion}). One indeed obtains~\cite{Kesten73, resettingNoise, thesis, GodrecheLuck}
\bea \label{kesten.1bis} 
p_X(x,n)= \int dU \int dV \int dx'\,  P(U,V)\, p(x', n-1)\,  \delta(x- U x'-V)\, ,
\eea
where $p_X(x,n)$ is the PDF of the random variable $x_n$ and {$P(U,V)$ is the joint PDF of the random variables $U$ and $V$}.
Taking the limit $n \to +\infty$ in Eq.~(\ref{kesten.1bis}), and assuming $p_X(x) = \lim_{n\to +\infty}p_X(x,n)$ exists, leads to an integral equation satisfied by the stationary distribution
\begin{equation}
p_X(x)= \int dU \int dV \int dx'\, P(U,V) \, 
p_X(x')\,  \delta(x-U\, x'-V)\, .
\label{kesten.2}
\end{equation}
{Since $\tau_n$'s (hence $U_n$'s -- see Eq.~(\ref{KestenRecursion})) are independent of the post-tumble velocity ${\sf v}_n$ drawn from $W(v)$, the joint law factorizes as $P(U,V) = P(U)\,P(V | U)$. For fixed $U$, the change of variables $v = \mu V/(1-U)$ yields 
\begin{equation}
P(V | U) = \frac{\mu}{1-U}\, W\!\left(\frac{\mu V}{1-U}\right)\, ,
\end{equation}
with support $V \in \big[(1-U)v_{\min}/\mu,\,(1-U)v_{\max}/\mu\big]$.
Combining this result with Eq.~(\ref{UPDF}) yields the joint law of $U_n$ and $V_n$ defined in Eq.~(\ref{KestenRecursion}), namely
\begin{eqnarray}
    P(U,V)=P(U)P(V|U) =\alpha\, U^{\alpha-1}\, \frac{\mu}{1-U}\, W\left(\frac{\mu V}{1-U}\right)\, .
\end{eqnarray}}
Therefore, the steady state distribution of the generalized RTP satisfies the following integral equation
\begin{equation}
p_X(x)= \int dU \int dV \int  dx'\, \alpha\, U^{\alpha-1}\, \frac{\mu}{1-U}\, W\left(\frac{\mu V}{1-U}\right)\, 
p_X(x')\,  \delta(x-U\, x'-V)\, .
\label{kesten.RTPG1}
\end{equation}
Although this integral equation is very complicated, it can be solved explicitly for the moment generating function (MGF) of $X$, which is defined as
\begin{equation}
    \tilde{p}_X(q) = \langle e^{qx} \rangle = \int_{-\infty}^{+\infty}dx\,  e^{qx} p_X(x)\, .
\label{MGF_BLT_Main}
\end{equation}
We show in Appendix \ref{KestenApp} that the solution for $\tilde{p}_X(q)$ reads
\begin{eqnarray}
\tilde{p}_X(q)
= \Gamma\!\left( \alpha\right)
\int_{\Gamma_B} \frac{ds}{2\pi i}\, 
\exp\!\left(s -\alpha \int_{v_{\min}}^{v_{\max}} dv\, \log\!\left( s - \frac{v}{\mu}q \right) W(v) \right)\, ,
\label{MGF_Main}
\end{eqnarray}
where $\Gamma_B$ represents the Bromwich contour in the complex $s$ plane. Obtaining explicitly the stationary PDF in real space is highly challenging as it involves two nontrivial tasks: (i) performing the Bromwich integral in (\ref{MGF_Main}) for a fixed $q$ and (ii) inverting the transform in (\ref{MGF_BLT_Main}). {Instead, we adopt an alternative approach by explicitly solving the Kesten recursion~(\ref{KestenRecursion}). This reveals a connection to stick-breaking processes and allows us to obtain the full stationary state $p_X(x)$ for an arbitrary distribution $W(v)$.}

\subsection{Connection to the Stick-Breaking Representation of a Dirichlet Process}\label{KestenSectionExplicit}


\begin{figure}[t]
    \centering
    \begin{tikzpicture}[
        execute at begin picture={
            \definecolor{prxblue}{RGB}{70, 100, 140} 
            \definecolor{prxgray}{RGB}{230, 230, 230}
        },
        xscale=13, 
        yscale=1.0, 
        font=\small,
        box/.style={
            draw=black!80, 
            line width=0.6pt, 
            minimum height=0.55cm,
            inner sep=2pt,
            align=center
        },
        fillY/.style={fill=prxblue!25},
        fillRem/.style={fill=prxgray},
        label/.style={anchor=east, font=\bfseries\small, text width=2cm, align=right, xshift=-0.2cm}
    ]

    \useasboundingbox (0, -4.5) rectangle (1, 0.5);

    \def\uone{0.20}   
    \def\utwo{0.24}   
    \def\hh{0.275}

    \node[label] at (0, 0) {Step 0:};
    \draw[box, fillRem] (0, -\hh) rectangle (1, \hh) node[midway] {$1$};

    \def\yTwo{-1.1}
    \node[label] at (0, \yTwo) {Step 1:};
    \draw[box, fillY] (0, \yTwo-\hh) rectangle (\uone, \yTwo+\hh)
        node[midway] {$Y_1=\overline{U}_1$};
    \draw[box, fillRem] (\uone, \yTwo-\hh) rectangle (1, \yTwo+\hh)
        node[midway] {$1-\overline{U}_1$};

    \def\yThree{-2.2}
    \node[label] at (0, \yThree) {Step 2:};
    \draw[box, fillY] (0, \yThree-\hh) rectangle (\uone, \yThree+\hh) node[midway] {$Y_1$};
    \draw[box, fillY] (\uone, \yThree-\hh) rectangle (\uone+\utwo, \yThree+\hh)
        node[midway] {$Y_2=\overline{U}_2(1-\overline{U}_1)$};
    \draw[box, fillRem] (\uone+\utwo, \yThree-\hh) rectangle (1, \yThree+\hh)
        node[midway] {$(1-\overline{U}_2)(1-\overline{U}_1)$};

    \def\yFour{-3.8}
    \node at (0.5, -3.0) {$\vdots$};
    \node[label] at (0, \yFour) {Step $\boldsymbol{n}$:};

    \draw[box, fillY] (0, \yFour-\hh) rectangle (\uone, \yFour+\hh) node[midway] {$Y_1$};
    \draw[box, fillY] (\uone, \yFour-\hh) rectangle (\uone+\utwo, \yFour+\hh) node[midway] {$Y_2$};

    \def\rWidth{0.18}     
    \def\ynWidth{0.30}    
    \pgfmathsetmacro{\ynStart}{1 - \ynWidth - \rWidth}
    \pgfmathsetmacro{\ellipsisX}{(\uone+\utwo + \ynStart)/2}

    \node at (\ellipsisX, \yFour) {$\dots$};

    \draw[box, fillY] (\ynStart, \yFour-\hh) rectangle (\ynStart+\ynWidth, \yFour+\hh)
        node[midway, font=\scriptsize] {$Y_n=\left[\prod_{j<n}(1-\overline{U}_j)\right]\overline{U}_n$};

    \draw[box, fillRem] (\ynStart+\ynWidth, \yFour-\hh) rectangle (1, \yFour+\hh)
        node[midway, font=\scriptsize] {$\prod_{j\le n}(1-\overline{U}_j)$};

    \end{tikzpicture}
    \caption{Illustration of the iterative steps of a stick-breaking process, in which a stick of unit length is successively broken into infinitely many pieces from left to right. At step 1, a fraction $Y_1=\overline{U}_1$ is removed, leaving a segment of length $1 - \overline{U}_1$. At step~2, a fraction $\overline{U}_2$ of the remaining stick is broken off, producing a new piece of length $Y_2=\overline{U}_2(1 - \overline{U}_1)$ and leaving a residual segment of length $(1 - \overline{U}_2)(1 - \overline{U}_1)$. The procedure continues indefinitely, yielding pieces of lengths $Y_n$. Each $\overline{U}_n \in (0,1)$ is drawn independently from a distribution $\mathrm{Beta}(1,\alpha)$.}
    \label{SBP_fig}
\end{figure}
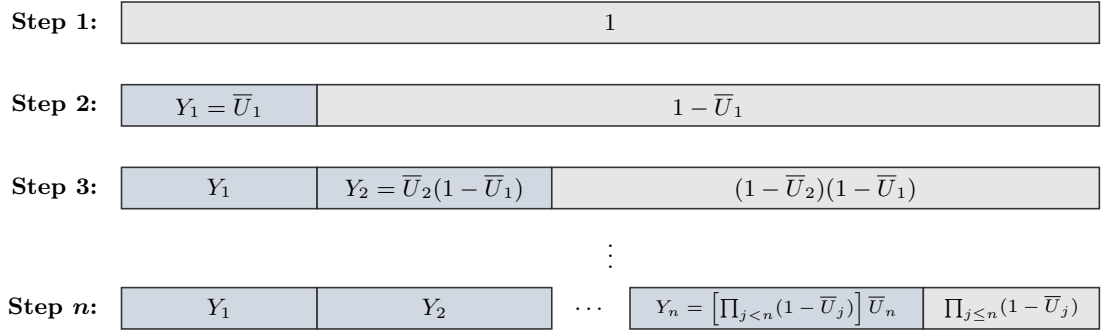

The recursion relation~(\ref{Kest_rel}) for $x_n$ can actually be solved explicitly. Starting from $x_0 = 0$, the first iteration gives
\begin{equation}
x_1 = \frac{{\sf v}_1}{\mu}\left(1 - U_1\right)\, .
\end{equation}
Using this result iteratively, we obtain for the next terms
\begin{eqnarray}
x_2 &=& U_2\,\frac{{\sf v}_1}{\mu}\left(1 - U_1\right) + \frac{{\sf v}_2}{\mu}\left(1 - U_2\right)\, , \nonumber\\
x_3 &=& U_3 U_2\,\frac{{\sf v}_1}{\mu}\left(1 - U_1\right)
     + U_3\,\frac{{\sf v}_2}{\mu}\left(1 - U_2\right)
     + \frac{{\sf v}_3}{\mu}\left(1 - U_3\right)\, .
\end{eqnarray}
By induction, the general expression after $n$ renewals reads
\begin{equation}\label{FiniteNsolKesten}
x_n = \sum_{m=1}^{n}
\left( \prod_{j=m+1}^{n} U_j \right)
\frac{{\sf v}_m}{\mu}\left(1 - U_m\right)\, .
\end{equation}
{Now, apply the index reversal $m' = n + 1 - m$, $j' = n + 1 - j$ and define the reversed sequences
$\tilde U_{j'} := U_{n+1-j'}$, $\tilde v_{m'} := v_{n+1-m'}$. Under this change of variables, $\prod_{j=m+1}^{n} U_j = \prod_{j' < m'} \tilde U_{j'}$. Since the sum is finite, we may reorder the terms. Relabeling dummy indices (and dropping tildes) finally gives
\begin{equation}\label{FiniteNsolKesten2}
x_n \overset{d}{=} \sum_{m=1}^{n}
\left( \prod_{j<m}^{} U_j \right)
\frac{{\sf v}_m}{\mu}\left(1 - U_m\right)\, ,
\end{equation}
where, by convention, $\prod_{j<1}^{} U_j=1$}. In the stationary limit $n \to \infty$, this yields
\begin{equation}
X \overset{d}{=}
\sum_{m=1}^{\infty}
\left( \prod_{j<m} U_j \right)
\frac{{\sf v}_m}{\mu}\left(1 - U_m\right)\, .
\label{Kesten_sol}
\end{equation}
Finding the steady-state distribution  $p_X(x)$ seems highly nontrivial, as it is expressed as an infinite sum involving a complicated function of the set of random variables  $\{U_n, {\sf v}_n\}$. Remarkably, we show below that by suitably rewriting this sum, one can identify a natural connection to a stick-breaking process~--~see Fig.~\ref{SBP_fig}. Let us define
\begin{equation}
\overline{U}_n = 1-U_n\, , \\ \qquad \qquad \overline{U}_n \sim \mathrm{Beta}\left(  1, \alpha\right) \in (0,1).
\end{equation}
This change of variable allows us to rewrite Eq.~(\ref{Kesten_sol}) as follows
\begin{eqnarray}
    X \overset{d}{=} \frac{1}{\mu} \sum_{n\geq1} {\sf{v}}_n Y_n  \, , \qquad \qquad  Y_n = \left[ \prod_{j<n} (1 - \overline{U}_j) \right] \overline{U}_n\, ,
    \label{x_yn_def}
\end{eqnarray}
where the random variables $Y_n$'s are exactly the weights of a stick-breaking process with parameter $\alpha$ (see again Fig.~\ref{SBP_fig}). {As is clear from Eq.~(\ref{x_yn_def}), the $Y_n$'s are not independent: they involve overlapping products of the same form $(1-\bar U_j)$, and are therefore strongly correlated}. By definition of the $Y_n$'s, one immediately finds that
\begin{eqnarray}
    \sum_{n=1}^{+\infty} Y_n = 1\, , \qquad \text{with}\,\, \text{probability}\,\, 1\, ,
    \label{SumYn}
\end{eqnarray}
since the series is telescoping. The solution~(\ref{x_yn_def}) admits an intuitive interpretation: the stationary position $X$ can be mapped as a weighted sum of ballistic displacements with velocity ${\sf v}_n$ over effective times $Y_n/\mu$. This representation is purely formal, however, since the actual dynamics is non-ballistic due to the harmonic force.\\

The variable $X$ defined in Eq.~(\ref{x_yn_def}) is a mathematical object known as a mean functional of a Dirichlet process~\cite{CifarelliRegazzini}. This connection is made explicit in Appendix~\ref{AppDirichlet}, where Dirichlet processes are introduced. The statistical properties of such objects have been studied in detail in the mathematical literature, in particular in Ref.~\cite{CifarelliRegazzini}. The MGF given in Eq.~(\ref{MGF_Main}) can be recovered using the so-called Cifarelli--Regazzini identity, as shown in Appendix~\ref{AppDirichlet}. Moreover, Ref.~\cite{CifarelliRegazzini} provides the cumulative distribution of $X$, which can be used to obtain the exact stationary state $p_X(x)$ (see again Appendix~\ref{AppDirichlet}).

\subsection{Stationary State and Moments for an Arbitrary Distribution $W(v)$}
Using the results in Ref.~\cite{CifarelliRegazzini} from Cifarelli and Regazzini, it is possible to invert the MGF in Eq.~(\ref{MGF_Main}) back in real space. One can show (see Appendix~\ref{AppDirichlet}) that for any $W(v)$, the stationary state is given by
\begin{equation}\label{SS_RTP_continuous}
p_X(x) = \dfrac{1}{\pi} 
\int_{v_{\min}/\mu}^x \! dt \, (x-t)^{\alpha-1}\, \frac{d\phi_\alpha(t)}{dt},
\qquad x \in \left[\tfrac{v_{\min}}{\mu}, \tfrac{v_{\max}}{\mu}\right], \qquad \forall \alpha > 0\, ,
\end{equation}
where
\begin{eqnarray}\label{phi_def}
\phi_{\alpha}(t) = \sin\!\left( \pi \alpha \int_{v_{\min}}^{\mu t}dv\, W(v) \right)
\exp\!\left( -\alpha \int_{v_{\min}}^{v_{\max}} dv\, 
\log\!\left| t - \tfrac{v}{\mu} \right|\, W(v)\right)\, . 
\end{eqnarray}
The moments of this distribution can be expressed explicitly in terms of Bell polynomials~\cite{BellPoly}. In Appendix~\ref{MomentApp}, we show that they are given by
\begin{eqnarray}
\langle X^{n} \rangle = \frac{1}{\mu^n} \frac{\Gamma(\alpha)}{\Gamma(\alpha+n)} \;
B_n\left(1!\, \alpha\, \frac{\langle v \rangle}{1}, \ldots,n!\, \alpha\, \frac{\langle v^{n} \rangle}{n} \right)\, .
\label{exactmoments}
\end{eqnarray}

We end this subsection with two remarks:
\begin{itemize}
    \item {\bf The special case $\alpha = 1$}. The stationary state takes a particularly simple form when $\alpha = 1$. This corresponds to the case where the switching rate of the velocity matches the mean relaxation time inside the harmonic well. {Using Eq.~(\ref{SS_RTP_continuous}), the kernel $(x-t)^{\alpha-1}$ then simplifies straightforwardly, and after integration, the stationary state is explicitly obtained as}
\begin{eqnarray}
    p_X(x) = \frac{\phi_1(x)}{\pi} = \frac{1}{\pi}\sin\!\left( \pi \int_{v_{\min}}^{\mu x}dv\, W(v) \right)
\exp\!\left( - \int_{v_{\min}}^{v_{\max}} dv\, 
\log\!\left| x - \tfrac{v}{\mu} \right|\, W(v)\right)\, , \qquad \alpha = 1 \, .
\label{generalBeta1}
\end{eqnarray}
These explicit and general formulae in Eqs. (\ref{SS_RTP_continuous}), (\ref{phi_def}), (\ref{exactmoments}) and (\ref{generalBeta1}) constitute the main results of the present paper. In the next section, we will use them to compute explicitly the stationary PDF for the RTP in a $d$-dimensional harmonic potential.
\item{ {\bf Joint statistics of the position and the velocity in the stationary state}. Besides the position $X$, 
another interesting observable is the (total) velocity $\dot{X} = - \mu X + v$ of the particle in the stationary state. In principle, using the approach presented here it is possible to compute the joint distribution of $X$ and $v$, from which the joint law of $X$ and $\dot{X}$ can be obtained. We refer the reader to Ref. \cite{Frydel2} for further discussions of the marginal distribution of $\dot{X}$ in the context of the RTP in a $d$-dimensional harmonic potential.}

\end{itemize}

\subsection{Large and Small $\alpha$ Behavior of the Stationary State}\label{asymptoticsSSSection}

{It is interesting to study how $p_X(x)$ behaves in the two extreme cases $\alpha=\gamma/\mu\to0$ (strongly active) and $\alpha\to+\infty$ (strongly passive).
To quantify the crossover, we consider the mean stick-breaking weights $Y_n$ in Eq.~(\ref{x_yn_def}):
\begin{eqnarray}
\langle Y_n\rangle = \frac{1}{1+\alpha}\left(\frac{\alpha}{1+\alpha}\right)^{n-1}\, .
\label{meanYn}
\end{eqnarray}
Together with $\sum_{n\ge1}Y_n=1$, Eq.~(\ref{meanYn}) implies that the sum is dominated by the first few terms for $\alpha\to0$,
while it is spread across many components for $\alpha\to+\infty$.\\
}

\noindent \textbf{Small $\alpha$ limit (strongly active) -- few dominant velocities.} For small $\alpha$, the first weight $Y_1$ is large on average, while the subsequent weights $Y_{i>1}$ decrease rapidly. As a result, the particle spends most of its stationary lifetime in a single velocity state. Therefore, the motion of the RTP effectively reduces to ballistic motion, which in the stationary state leads to a Dirac delta-function located at the fixed-point of the dynamics, i.e., $\delta\!\left(x-\frac{v}{\mu}\right)$. Averaging over realizations then gives
\begin{eqnarray}
p_X(x)\underset{\alpha \to 0}{\approx}\int_{v_\mathrm{min}}^{v_\mathrm{max}} dv\, W(v)\, \delta\!\left(x-\frac{v}{\mu}\right) = \mu \, W(\mu x)\, .
\label{samllbetapX}
\end{eqnarray}

\noindent \textbf{Large $\alpha$ limit (strongly passive) -- highly mixed environment.}
For large $\alpha$, all $Y_n$'s are small on average, meaning that the particle samples many different velocity states. The environment thus becomes a highly mixed combination of values drawn from $W(v)$, and in the infinite-$\alpha$ limit, the stationary distribution of the RTP position $x$ concentrates around $\langle v \rangle / \mu$.
To see this, using the expressions of the moments in Eq.~(\ref{exactmoments}), we have
\begin{equation}
\langle x \rangle = \frac{\langle v \rangle}{\mu}\, ,
\end{equation}
while the variance of $x$ reads
\begin{equation}
\langle x^2 \rangle_c = \frac{\langle v^2 \rangle_c}{\mu^2(1+\alpha)}
\ \underset{\alpha \to +\infty}{\longrightarrow}\ 0\, ,
\end{equation}
where the notation $\langle . \rangle_c$ denotes the cumulants of a random variable. Therefore, when $\alpha \to +\infty$, the stationary distribution becomes concentrated in the following Dirac delta function
\begin{eqnarray}
p_X(x)\underset{\alpha \to +\infty}{\approx}\delta\!\left(x-\frac{\langle v\rangle}{\mu}\right)\, .
\label{diracasympt}
\end{eqnarray}
It is also instructive to analyze how the stationary state converges to the Dirac delta function. 
In particular, one can show that the cumulants $\langle x^{2n}\rangle_c$ decrease as $1/\alpha^{\, n-1}$. This implies that the Dirac delta function in Eq.~(\ref{diracasympt}) 
is approached by the following Gaussian distribution, valid for {$(x - \langle v \rangle/\mu)\sim O(1/\sqrt{\alpha})$}, namely
\begin{eqnarray}
    p_X(x)\underset{\alpha \to +\infty}{\approx}\sqrt{\frac{\mu^2 \alpha }{2\pi\, \langle v^2 \rangle_c}}
\exp\!\left[
-\frac{\mu^2\alpha}{2\,\langle v^2 \rangle_c}
\left(x - \frac{\langle v \rangle}{\mu}\right)^{2}
\right]\, .
\label{GaussianAsympt}
\end{eqnarray}\\

Based on the asymptotic forms given in Eqs.~(\ref{samllbetapX}) and (\ref{GaussianAsympt}), the asymptotic behavior of the moments~(\ref{exactmoments}), as $\alpha$ is varied, can also be obtained and is given by
\begin{equation}
\label{moment_asympt}
\langle x^{n}\rangle \approx
\begin{cases}
\frac{\langle v^n \rangle}{\mu^n} + O(\alpha)\quad , \quad \alpha \ll 1 , \\
\\
\frac{\langle v \rangle^n}{\mu^n} 
+ \frac{n(n-1)}{2}
\left( \frac{\langle v \rangle}{\mu} \right)^{n-2}
\frac{\langle v^2 \rangle_c}{\mu^{2}\alpha}
+ O(\alpha^{-2})\quad , \quad \alpha \gg 1 \;.
\end{cases}
\end{equation}
{Hence, at leading order, $\langle x^n \rangle$ crosses over from $\langle v^n\rangle/\mu^n$ as $\alpha \to 0$ to $\langle v \rangle^n/\mu^n$ as $\alpha \to \infty$.}


\section{Stationary State of a Run-and-Tumble Particle in 1D, 2D, and 3D}
\label{2dSSSection}

In Section~\ref{ModelSection}, we introduced the $d$-dimensional run-and-tumble particle and showed that each of its components follows the dynamics of the generalized one-dimensional RTP discussed in Section~\ref{GeneralizedRTPSection}. In particular, we established that the stationary distribution of this generalized RTP is given by Eq.~(\ref{SS_RTP_continuous}). Hence, in Eq.~(\ref{SS_RTP_continuous}), by setting $v_{\min} = -v_0$, $v_{\max} = v_0$, and using the appropriate velocity distribution $W(v)$, {i.e., setting $W(v) = W_{\rm proj}(v)$ given in Eq. (\ref{WprojEq})}, we can directly obtain the stationary statistics of a component of a $d$-dimensional RTP. For example, in two and three dimensions, the projected velocity distributions read -- from Eq. (\ref{WprojEq}) -- 
\begin{equation}
W_{\text{proj}}(v) = \frac{1}{\pi \sqrt{v_0^2 - v^2}}\, , \qquad |v|\leq v_0\, ,\qquad d=2\, ,
\label{Arcsine2}
\end{equation}
and
\begin{equation}
W_{\text{proj}}(v) = \frac{1}{2 v_0}\, , \qquad |v|\leq v_0\, ,\qquad d=3\, .
\label{Unif3}
\end{equation}
Due to isotropy, all the statistical properties of the stationary state in $d$ dimensions can be directly obtained from those of a single component. Before studying the stationary state in two and three dimensions, it is useful to first recall the results obtained in one dimension.

\subsection{Stationary State of a One-Dimensional RTP}\label{SS1dRTPnoD}
\begin{figure}[t]
    \centering
    \includegraphics[width=0.32\linewidth]{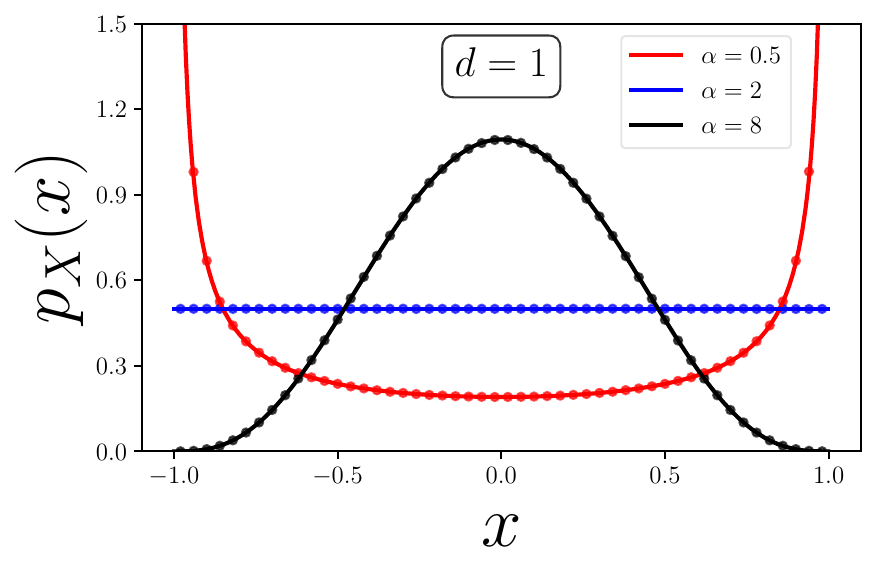}
    \includegraphics[width=0.32\linewidth]{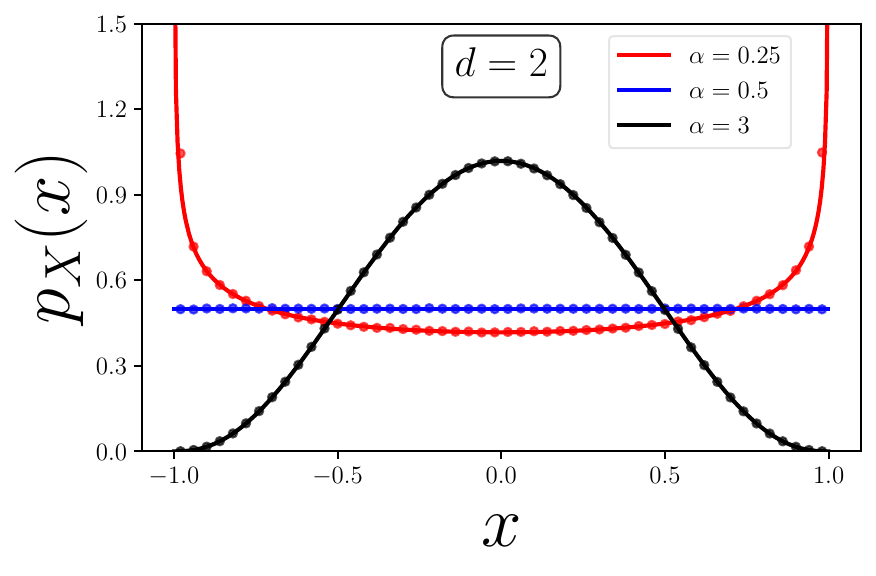}
    \includegraphics[width=0.32\linewidth]{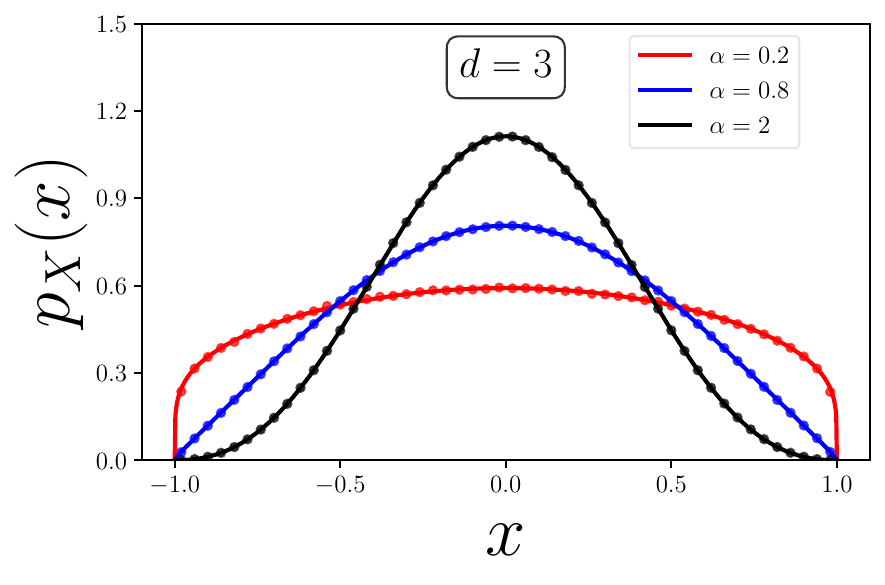}
    \caption{Stationary probability density $p_X(x)$ of a single  component of the RTP in dimensions $d=1$ (left), $d=2$ (center), and $d=3$ (right), for different values of the activity parameter $\alpha = \gamma/\mu$. In $d=1$ and $d=2$, the distribution undergoes a shape transition at $\alpha=2$ and $\alpha=1/2$, respectively, signaled by the emergence or disappearance of singularities at the turning points $\pm v_0/\mu$. In contrast, in $d=3$ the distribution remains finite and vanishes algebraically at the boundaries for all $\alpha$. Solid lines are exact analytical results, while symbols correspond to numerical simulations of the RTP dynamics. Parameters are $\mu = 1$ and $v_0 = 1$.}
    \label{FigpXsimu}
\end{figure}

We first recall the simplest realization of a run-and-tumble particle in one dimension, characterized by two possible velocities $v = \pm v_0$, chosen with equal probability,
\begin{equation}
    W(v) = \frac{1}{2}\,\delta(v - v_0) + \frac{1}{2}\,\delta(v + v_0)\, .
\end{equation}
This minimal two-state model has been studied in~\cite{DKMSS19,Klyatskin, LHKI, TC2009}\footnote{In Ref.~\cite{DKMSS19}, the mapping of parameters is given by $\gamma \to 2\gamma$, hence $\gamma/\mu \to \alpha/2$. This correspondence arises because, in the standard RTP model, each tumbling event necessarily induces a change of velocity. In contrast, in our formulation, at a tumbling event the particle remains in the same state with probability $1/2$.}. 
In that case, the stationary distribution of the particle position is known exactly and takes the form of a beta distribution supported on the finite interval $|x| < v_0/\mu$
\begin{equation}
    p_X(x) = \frac{\mu}{v_0}\,
    \frac{\Gamma\!\left(\frac{\alpha + 1}{2}\right)}
         {\sqrt{\pi}\,\Gamma\!\left(\frac{\alpha}{2}\right)}\,
    \left[1 - \left(\frac{\mu x}{v_0}\right)^2 \right]^{\alpha/2 - 1}\, .
    \label{eq:px_1d_rtp}
\end{equation}
This result can be recovered as a special case of the general $N$-state RTP analyzed below in Sec.~\ref{NstateSection}, upon setting $N=2$. The two edges of the support, at $\pm v_0/\mu$, correspond to the turning points of the dynamics, i.e., the positions where the velocity of the RTP vanishes. Interestingly, the stationary state exhibits a qualitative transition as the parameter $\alpha$ is varied. For $\alpha < 2$, corresponding to persistent motion with infrequent tumbling, the particle tends to accumulate near the turning points, leading to integrable divergences of $p_X(x)$ at the edges of the support -- an ``active-like'' regime dominated by persistence. At the critical value $\alpha = 2$, the stationary distribution becomes uniform over the interval $[-v_0/\mu,\, v_0/\mu]$. In contrast, for larger values $\alpha > 2$, frequent tumbling events effectively randomize the motion, and the stationary distribution gradually approaches a Gaussian form, characteristic of a passive-like diffusive regime. These limiting behaviors of the stationary distribution can be made explicit through the following asymptotic forms~(see Section~\ref{asymptoticsSSSection})
\begin{equation}
p_X(x) \approx
\begin{cases}
\displaystyle 
\frac{1}{2}\,\delta\!\left(x-\frac{v_0}{\mu}\right)
+ \frac{1}{2}\,\delta\!\left(x+\frac{v_0}{\mu}\right)\, , & \alpha \to 0 \, ,\\[8pt]
\displaystyle 
\frac{\mu}{2v_0}\, , & \alpha = 2 \, ,\\[8pt]
\displaystyle 
\frac{\mu}{v_0}\sqrt{\frac{\alpha}{2\pi}}\exp\!\left[-\,\frac{\alpha}{2}\left(\frac{\mu x}{v_0}\right)^2\right]\, , & \alpha \to +\infty  \, .
\end{cases}
\label{asymptotics_px}
\end{equation}
In the left panel of Fig.~\ref{FigpXsimu}, we illustrate this shape transition and compare the analytical prediction in Eq.~\eqref{eq:px_1d_rtp} with numerical simulations. In the following subsections, we investigate whether this shape transition of the stationary distribution, from an active-like to a passive-like form, persists in higher dimensions.\\

In higher dimensions, a key observable is the distribution of the radius $R$, i.e., the distance of the particle from the origin. For comparison, it is useful to first consider the one-dimensional case. When $d=1$, since $R = |X|$, we have
\begin{eqnarray}\label{1dRXp}
    p_R(r) = p_X(r)+p_X(-r) = 2\, p_X(r)\, ,
\qquad 0 \le r \le \frac{v_0}{\mu}\, .
\end{eqnarray}
The radial distribution exhibits the same qualitative transition as in $p_X(x)$ -- see the left panel of Fig.~\ref{FigpR}. 
For $\alpha \to 0$, the particle is confined at the boundary $r = v_0/\mu$. For $\alpha = 2$, the distribution is uniform, while for large $\alpha$, it becomes Gaussian around the origin 
\begin{equation}
p_R(r) \;\approx\;
\begin{cases}
\displaystyle 
\delta\!\left(r-\frac{v_0}{\mu}\right),
& \alpha \to 0 \, , \\[10pt]
\displaystyle 
\frac{\mu}{v_0},
& \alpha = 2 \, , \\[10pt]
\displaystyle 
\frac{\mu}{v_0}\,\sqrt{\frac{2\alpha}{\pi}}\,
\exp\!\left[-\,\frac{\alpha}{2}
\left(\frac{\mu r}{v_0}\right)^2\right],
& \alpha \to +\infty \, .
\end{cases}
\label{asymptotics_pr1d}
\end{equation}

\subsection{Stationary State of a Two-Dimensional RTP}

\begin{figure}[t]
    \centering
    \includegraphics[width=0.32\linewidth]{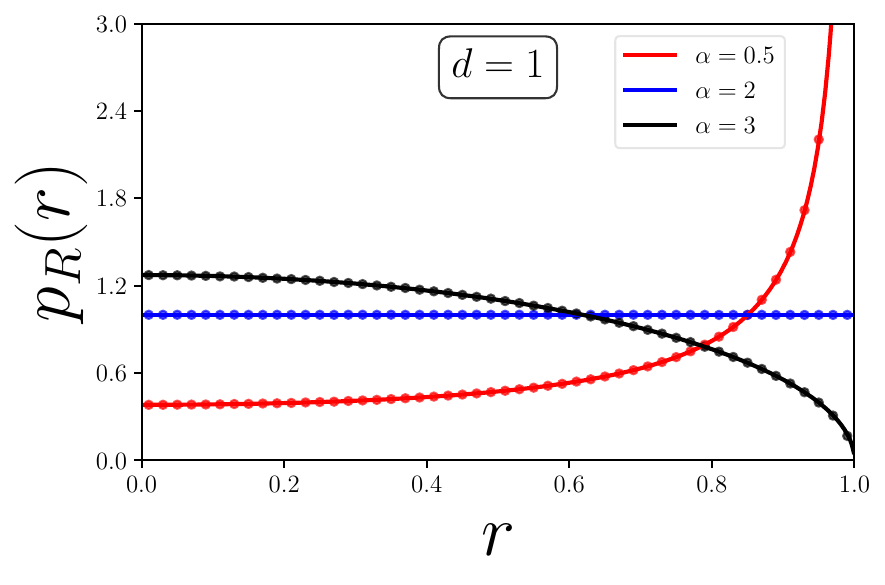}
    \includegraphics[width=0.32\linewidth]{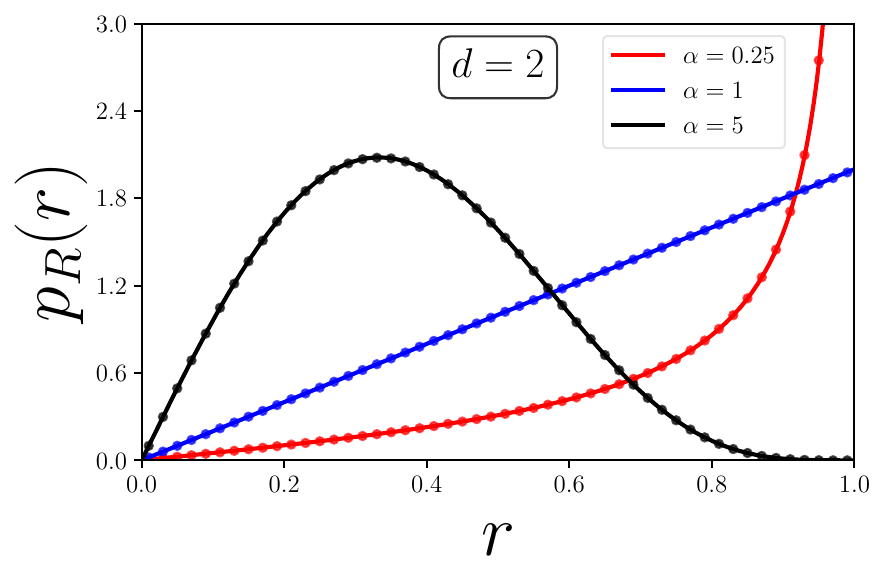}
    \includegraphics[width=0.32\linewidth]{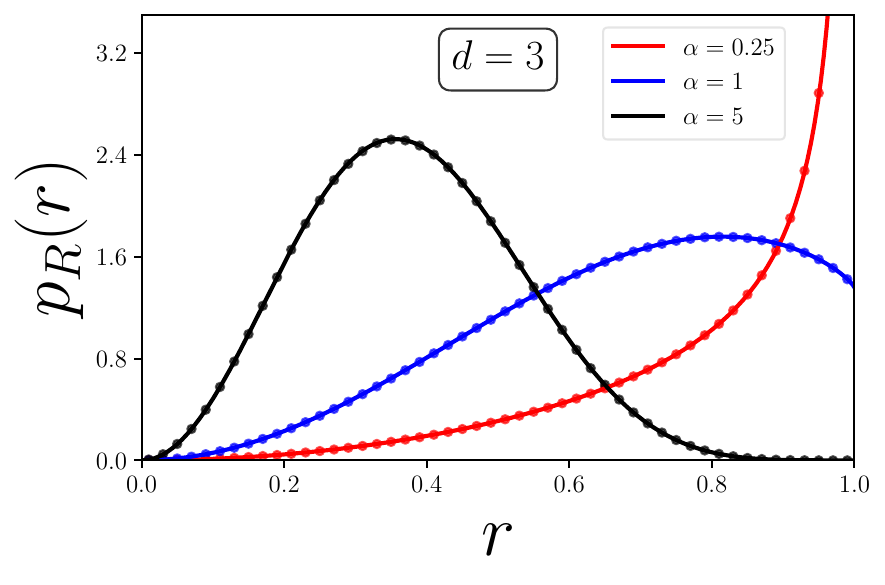}
    \caption{Stationary radial probability density $p_R(r)$ for an RTP confined in a harmonic trap, shown for dimensions $d = 1$ (left), $d = 2$ (center), and $d = 3$ (right), and for different values of the activity parameter $\alpha = \gamma/\mu$. In all panels, we set $\mu = 1$ and $v_0 = 1$. For small $\alpha$ (high persistence), the probability mass accumulates near the turning radius $r = v_0/\mu$. At the critical value of $\alpha$ (dimension-dependent), the distribution has finite mass there. For large $\alpha$, tumbling events become more frequent, and the distribution contracts toward the origin. Solid lines are exact analytical predictions, while markers correspond to numerical simulations.
    }
    \label{FigpR}
\end{figure}


When Eq.~(\ref{SS_RTP_continuous}) is specialized to the arcsine distribution~(\ref{Arcsine2}), the integrals involved can be evaluated explicitly and it is possible to show that
\begin{equation}
\phi_\alpha(t)
= \left(\frac{2\mu}{v_0}\right)^{\alpha}
  \sin\!\left(\frac{\pi\alpha}{2}+\alpha\,\arcsin\!\frac{\mu t}{v_0}\right)\, .
\end{equation}
Consequently, for $|x|\le v_0/\mu$, the stationary distribution reads -- from Eq. (\ref{SS_RTP_continuous}) -- 
\begin{equation}
p_X(x)
= \frac{\alpha\mu}{\pi}\!\left(\frac{2\mu}{v_0}\right)^{\alpha}
  \int_{-\frac{v_0}{\mu}}^{x}dt\, 
  \frac{(x-t)^{\alpha-1}}{\sqrt{v_0^2-\mu^2 t^2}}\,
  \cos\!\left(\frac{\pi\alpha}{2}+\alpha\,\arcsin\!\frac{\mu t}{v_0}\right)\, .
\end{equation}
Surprisingly, the integral can be evaluated exactly, and we show in Appendix~\ref{BetaApp} that the stationary state reduces to a beta distribution, just as in $d=1$, although with a different exponent (see also~\cite{Frydel3})
\begin{equation}
p_X(x) = \frac{\mu}{v_0}\frac{\Gamma(1+\alpha)}{\sqrt{\pi}\, \Gamma(1/2+\alpha)}\, \left[1-\left(\frac{\mu x}{v_0}\right)^2\right]^{\alpha - 1/2} \quad, \quad |x| < \frac{v_0}{\mu} \;.
\label{2dpX}
\end{equation}
Our result coincides with that of Ref.~\cite{Frydel1}, obtained independently using a different approach, namely a Fokker–Planck approach. The stationary distribution exhibits the same asymptotic behaviors as in Eq.~(\ref{asymptotics_px}), although the transition between the active-like and passive-like regimes now occurs at $\alpha = 1/2$ -- see the middle panel of Fig.~\ref{FigpXsimu} for a comparison with simulations. This shows that the shape transition identified in one dimension persists in the single-component dynamics of the two-dimensional RTP in the presence of a harmonic trap.\\

We now turn to the stationary distribution of the radius $R = \sqrt{x^2 + y^2}$. 
From Eq.~(\ref{pXtopRd2d3}), we see that it can be obtained directly from the single-component stationary state. 
Interestingly, it again takes the form of a beta distribution, but with a different exponent and an additional linear prefactor
\begin{equation}\label{Radial2d}
p_R(r) = -2r\int_r^{\frac{v_0}{\mu}} dx\, \frac{p_X'(x)}{\sqrt{x^2-r^2}} = 2\alpha \frac{\mu^2 r}{v_0^2}
\left[ 1 - \left( \frac{\mu r}{v_0} \right)^2 \right]^{\alpha - 1}\, ,
\qquad 0 \le r \le \frac{v_0}{\mu}\, .
\end{equation}
In contrast to the one-dimensional case, the transition of the radial distribution $p_{R}(r)$ in two dimensions occurs at a different value of $\alpha$ than the transition of the single-component distribution $p_{X}(x)$. 
While $p_{X}(x)$ changes behavior at $\alpha = 1/2$, the radial distribution $p_{R}(r)$ undergoes its transition at $\alpha = 1$: for $\alpha < 1$, $p_{R}(r)$ diverges at the edge $r = v_{0}/\mu$, whereas for $\alpha > 1$ it vanishes there (see the middle panel of Fig.~\ref{FigpR}). 
At the critical point $\alpha = 1$, the distribution becomes linear in $r$ near the origin. These features are made explicit by computing the behavior close to the boundary $r = v_{0}/\mu$. Writing $r = v_{0}/\mu - \epsilon$ with $\epsilon \to 0^{+}$, one finds
\begin{eqnarray}
p_R(r) = \alpha\left(\frac{2\mu}{v_{0}}\right)^{\!\alpha}\,
\varepsilon^{\,\alpha-1}\left(1 + O\!\left(\varepsilon\right)\right) \;.
\label{2dTPbehavior}
\end{eqnarray}
As $\alpha$ is varied, $p_R(r)$ interpolates between the following asymptotic forms
\begin{equation}
p_R(r) \;\approx\;
\begin{cases}
\displaystyle 
\delta\!\left(r-\frac{v_0}{\mu}\right)\, ,
& \alpha \to 0 \, , \\[10pt]
\displaystyle 
2\alpha\frac{\mu^2}{v_0^2}r\, ,
& \alpha = 1 \, , \\[10pt]
\displaystyle 
2\alpha\frac{\mu^2}{v_0^2}r
\exp\!\left[-\alpha
\left(\frac{\mu r}{v_0}\right)^2\right],
& \alpha \to +\infty \, .
\end{cases}
\label{asymptotics_pr1d}
\end{equation}\\

\begin{figure}[t]
    \centering
    \includegraphics[width=1.\linewidth]{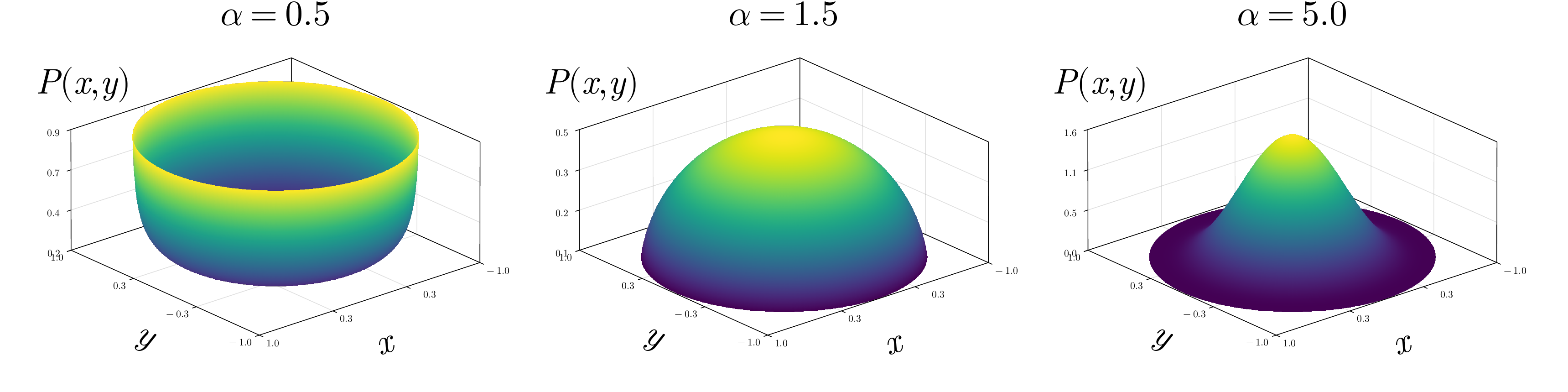}
    \caption{Stationary joint distribution $P(x,y)$ of the RTP in $d=2$, shown for increasing values of the activity
parameter $\alpha$. For $\alpha<1$, the probability is concentrated near the turning circle $r=v_0/\mu$. At the
critical point $\alpha=1$, the density becomes uniform along the circle. For $\alpha>1$, the distribution progressively
contracts toward the origin, approaching a Gaussian form in the passive limit. Parameters are $\mu = 1$ and $v_0 =1$.}
    \label{Figpxy}
\end{figure}

From Eq.~(\ref{generalJointPDF}), the corresponding two-dimensional stationary distribution then follows as (see also~\cite{Frydel1})
\begin{equation}\label{Joint2d}
P(x,y) = f_R\!\left(r = \sqrt{x^2 +y^2}\right)=\frac{p_R(r)}{2\pi r}  = \frac{\alpha}{\pi} \frac{\mu^2}{v_0^2}
\left[ 1 - \left( \frac{\mu r}{v_0} \right)^2 \right]^{\alpha - 1}\, ,
\qquad 0 \le r \le \frac{v_0}{\mu}\, .
\end{equation}
This shows that the density is maximal at the boundary for $\alpha < 1$, becomes flat at $\alpha = 1$, and concentrates near the origin for $\alpha > 1$. In particular, the stationary state interpolates continuously between an ``active'' regime 
where the particle spends most of its time at the turning circle $r = v_{0}/\mu$, and a ``passive'' regime dominated by frequent reorientations -- see Fig.~\ref{Figpxy}. It is thus a two-dimensional generalization of the shape transition discussed in Ref. \cite{DKMSS19}.  


\subsection{Stationary State of a Three-Dimensional RTP}
For a three-dimensional RTP, the projected velocity of a given component is piecewise constant and uniformly distributed over the interval $[-v_0, v_0]$. In this case, it is also possible to compute the stationary state of the individual component from Eq.~(\ref{SS_RTP_continuous}).  
For $x$ in the range $-v_0/\mu < x < v_0/\mu$, the distribution $p_X(x)$ is given by
\begin{eqnarray}
p_X(x)
= \frac{\alpha\,e^{\alpha}}{2\pi}\,\frac{\mu}{v_0}
\int_{0}^{\frac{1}{2}\left(1+\frac{\mu x}{v_0} \right)} \!\!du\,
\left[\frac{1}{2}\left(1+\frac{\mu x}{v_0}\right)-u\right]^{\alpha-1}\, f(u)\, ,
\label{3dcomponentdistrib}
\end{eqnarray}
where
\begin{eqnarray}
f(u)= \frac{1}{\alpha}\frac{d}{du} \left[ \sin(\pi \alpha u) e^{-\alpha\,[\,u\ln u + (1-u)\ln(1-u)\,]} \right] = 
e^{-\alpha\,[\,u\ln u + (1-u)\ln(1-u)\,]}\,
\left[\pi\cos(\pi\alpha u)
-\ln\!\left(\frac{u}{1-u}\right)\,\sin(\pi\alpha u)\right]\, .\quad
\label{3dfu}
\end{eqnarray}
In general, it is difficult to perform explicitly the integral over $u$ in Eq. (\ref{3dcomponentdistrib}). However its moments can be explicitly computed from the formula~(\ref{exactmoments}) leading to  
\begin{eqnarray}
\langle X^{n} \rangle = \frac{1}{\mu^n} \frac{\Gamma(\alpha)}{\Gamma(\alpha+n)} \;
B_n\left(1!\, \alpha\, \frac{\langle v \rangle}{1}, \ldots,n!\, \alpha\, \frac{\langle v^{n} \rangle}{n} \right)\, , \qquad \langle v^{n}\rangle = \frac{v_0^n}{2(n+1)}\left[1+(-1)^{n}\right]\, . 
\label{exactmoments3d}
\end{eqnarray}
We have explicitly checked that this formula yields back the results up to $\langle X^{10} \rangle$ given in Table I of Ref. \cite{Frydel2}.\\

Interestingly, in the special case $\alpha = 1$ [see also Eq. (\ref{generalBeta1})] it is possible to compute explicitly the stationary distribution in (\ref{3dcomponentdistrib}), yielding
\begin{eqnarray}\label{pX_a1}
p_X(x) = \frac{e}{\pi} \frac{\mu}{v_0} \cos{\left(\frac{\mu \pi x}{2 v_0}\right) \left(1+\frac{\mu x}{v_0}\right)^{-(1+\frac{\mu x}{v_0})/2} \left(1-\frac{\mu x}{v_0}\right)^{-(1-\frac{\mu x}{v_0})/2} } \quad, \quad -\frac{v_0}{\mu} \leq x \leq \frac{v_0}{\mu} \;.
\end{eqnarray}
In Ref. \cite{Frydel3}, it was shown that $p_X(x)$ satisfies an integro-differential equation, namely  
\begin{eqnarray} \label{int_eq}
0 = - p_X(x) \frac{v_0}{2} \ln \left( \frac{v_0/\mu+x}{v_0/\mu-x}\right) - \frac{v_0}{2} {\rm PV} \int_{-v_0/\mu}^{v_0/\mu} \frac{p_X(x')}{x-x'} \, dx' - \frac{v_0^2}{\mu} p_X'(x) \;,
\end{eqnarray} 
which was however not solved there. In fact, as also noticed in~\cite{Frydel3}, this equation is similar to, though different from, the type of integral equations encountered in random matrix theory and related Coulomb gas systems \cite{Forrester}. However, the standard techniques to solve such equations, like Tricomi's formula, can not be used straightforwardly here, and such equations are thus difficult to solve. Here, we have checked numerically that $p_X(x)$ given in Eq.~(\ref{pX_a1}) indeed satisfies the integro-differential equation (\ref{int_eq}). \\

For general $\alpha > 0$, it is instructive to examine the behavior of $p_X(x)$ in (\ref{3dcomponentdistrib}) near the turning points $\pm v_0/\mu$. This allows us to address two questions: (i) whether a shape transition still occurs in three dimensions, as in $d=1$ and $d=2$, and (ii) whether the rather involved expression~(\ref{3dcomponentdistrib}) eventually reduces to a simple beta distribution, as it did in lower dimensions. Since the distribution is symmetric, it is sufficient to analyze the behavior near the left boundary at $x = -v_0/\mu$.  
Let $\epsilon > 0$ and write $x = -v_0/\mu +\epsilon$. With the change of variable $v = u \frac{2v_0}{\mu \epsilon}$, Eq.~(\ref{3dcomponentdistrib}) becomes
\begin{equation}
    p_X(x)= \frac{\alpha\,e^{\alpha}}{2\pi}\,\frac{\mu}{v_0}\left(\frac{\mu \epsilon}{2v_0}\right)^{\alpha}\int_{0}^{1} \!\!dv\,
\left(1-v\right)^{\alpha-1}\, f\!\left(\frac{\mu \epsilon}{2v_0}v\right)\, , 
\end{equation}
For small argument, the function $f(u)$ admits the expansion
\begin{equation}
    f(u\to 0) \approx \pi + u \bigl[ -2\pi\alpha \ln u + \pi\alpha \bigr]\, .
\end{equation}
Using this expansion, one finds that for $x$ close to either turning point, $|x| = | v_0/\mu- \epsilon|$, the stationary distribution behaves as
\begin{equation}
    p_X(x)
\approx \frac{e^{\alpha}}{2}\,\frac{\mu}{v_0}
\left( \frac{\mu |\epsilon|}{2 v_0} \right)^{\alpha}
\;-\;
\frac{\alpha e^{\alpha}}{\alpha+1}\,\frac{\mu}{v_0}
\left( \frac{\mu |\epsilon|}{2 v_0} \right)^{\alpha+1}
\left[
\ln\!\left(\frac{\mu |\epsilon|}{2 v_0}\right)  + \tfrac{1}{2}
- \gamma_E - \psi(\alpha+2)
\right]\, ,
\label{3dTPbehavior}
\end{equation}
where $\gamma_E$ is the Euler constant and $\psi(x)$ is the digamma function. We first observe that, here, no shape transition occurs for the PDF of the projected component: as $\epsilon \to 0$, the distribution $p_X(x)$ always vanishes at the turning points, in contrast with the one- and two-dimensional cases. Moreover, the logarithmic correction appearing at the next order clearly signals a more intricate behavior than a simple beta law. Hence, already at the level of a single component, the three-dimensional case departs qualitatively from lower dimensions. 

\vspace*{0.5cm}

The absence of a shape transition for the projected component is further confirmed by the asymptotic forms of the distribution that one can compute using the result of Section~\ref{asymptoticsSSSection}. One finds
\begin{equation}
p_X(x) \approx
\begin{cases}
\displaystyle 
\frac{\mu}{2v_0}\, , & \alpha \to 0 \, ,\\[15pt]
\displaystyle 
\sqrt{\frac{3\,\mu^2 \alpha}{2\pi v_0^2}}\,
    \exp\!\left[-\,\frac{3\,\mu^2 \alpha}{2\,v_0^2}\,x^2\right]\, , & \alpha \to +\infty  \, .
\end{cases}
\label{asymptotics_pxd3}
\end{equation}
Thus, neither the uniform limit ($\alpha \to 0$) nor the Gaussian limit ($\alpha \to \infty$) exhibits the edge divergence characteristic of $d=1$ and $d=2$. These analytical predictions are fully consistent with numerical simulations, as shown in the right panel of Fig.~\ref{FigpXsimu}.
\\
\begin{figure}[t]
    \centering
    \includegraphics[width=0.9\linewidth]{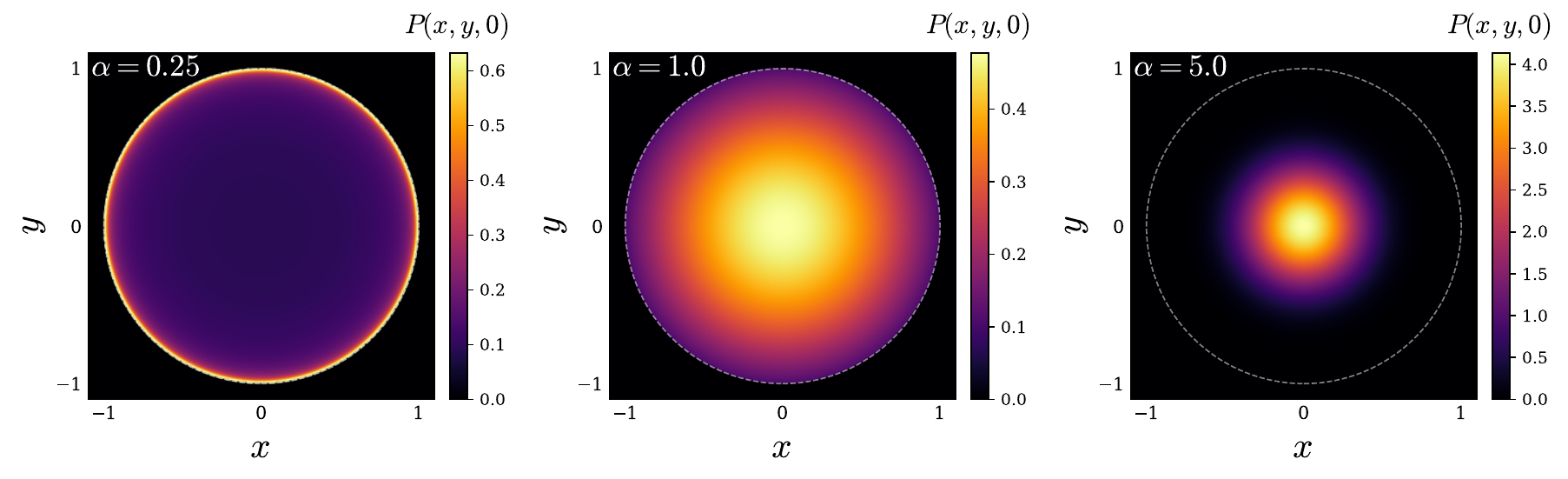}
    \caption{Cross sections of the stationary density $P(x,y,0)$ of a three-dimensional run-and-tumble particle for selected values of the parameter $\alpha$. The distribution is supported on the sphere of radius $v_{0}/\mu$ and is obtained from the exact result given in Eq.~(\ref{Joint3d}). For small $\alpha$, the density accumulates near the boundary because of the high persistence. For $\alpha\sim 1$, it becomes broad and nearly flat, and for large $\alpha$, it develops a peak at the origin.
    }
    \label{FigpR3d}
\end{figure}

Using isotropy, one can deduce the distribution of the radius $R = \sqrt{x^2 + y^2 + z^2}$ from $p_X(x)$ using the formula given in Eq.~(\ref{pXtopRd2d3}), i.e.,
\begin{equation}
    p_R(r) \;=\; -\,2r\,p'_X(r)\, .
\end{equation}
It is convenient to introduce the notation $A(x) = \tfrac{1}{2}\!\left(\tfrac{\mu x}{v_0} + 1\right)$ and, with the change of variable $t = A(x) - u$, we can rewrite
\begin{eqnarray}
    p_X(x) \;=\; \frac{\alpha\,e^{\alpha}}{2\pi}\,\frac{\mu}{v_0} \int_{0}^{A(x)} dt\, t^{\alpha-1}\,f(A(x)-t)\, .
\end{eqnarray}
Applying Leibniz's rule gives
\begin{eqnarray}
    p_X'(x) \;=\; \frac{\alpha\,e^{\alpha}}{2\pi}\,\frac{\mu}{v_0} A'(x)\left[A(x)^{\alpha-1}f(0) + \int_{0}^{A(x)} dt\, t^{\alpha-1}\,f'(A(x)-t)\right]\, .
\end{eqnarray}
Returning to the $u$-variable and noting that $f(0) = \pi$, we obtain
\begin{eqnarray}
    p_X'(x) \;=\; \frac{\alpha\,e^{\alpha}}{2\pi}\,\frac{\mu}{v_0} A'(x)\left[\pi A(x)^{\alpha-1} + \int_{0}^{A(x)} du\,  (A(x)-u)^{\alpha-1}f'(u)\right]\, .
\end{eqnarray}
In the end, the radial distribution reads
\begin{equation}
    p_R(r) = - \frac{\alpha\,e^{\alpha}}{2\pi}\,\frac{\mu^2}{v_0^2}\, r\left[\pi A(r)^{\alpha-1} + \int_{0}^{A(r)} du\,  (A(r)-u)^{\alpha-1}f'(u)\right]\, .
\end{equation}
As $\alpha$ is varied, $p_R(r)$ interpolates between the following behaviors
\begin{equation}
p_R(r) \approx
\begin{cases}
\displaystyle 
\delta\!\left(r-\frac{v_0}{\mu}\right)\, , & \alpha \to 0 \, ,\\[15pt]
\displaystyle 
\frac{\,e} {2\pi}\,\frac{\mu^2}{v_0^2}\, r\, |f\!\left(A(r)\right)|\, , & \alpha =1 \, ,\\[15pt]
\displaystyle 
\sqrt{\frac{2}{\pi}}
\left( \frac{3\mu^{2}\alpha}{v_{0}^{2}} \right)^{3/2}
r^{2} \exp\!\left( -\,\frac{3\mu^{2}\alpha}{2v_{0}^{2}}\, r^{2} \right)\, , & \alpha \to +\infty  \, ,
\end{cases}
\label{asymptotics_pxd3}
\end{equation}
where we recall that $A(r) = \tfrac{1}{2}\!\left(\tfrac{\mu r}{v_0} + 1\right)$, while $f(u)$ is defined in Eq.~(\ref{3dfu}) and is negative for $u\in [1/2,1]$. Hence, in contrast to the component distribution, {\it the radial distribution does exhibit a shape transition}. For small $\alpha$ (high persistence), the probability mass accumulates near the spherical shell of radius $v_0/\mu$, while for large $\alpha$ the density is concentrated close to the origin. The transition occurs at $\alpha = 1$, where the distribution remains finite at the turning point. For $\alpha > 1$, the radial distribution instead vanishes at the boundary. Thus, although each Cartesian component behaves differently from the one- and two-dimensional cases, the radial distribution in three dimensions is qualitatively similar to its two-dimensional counterpart.\\

To clarify why the Cartesian component does not exhibit a shape transition in $d=3$, we derive in Appendix~\ref{generalpotSec} the general edge behavior of $p_X(x)$ and $p_R(r)$ in arbitrary dimension ($d>1$) and for a general spherically symmetric confining potential. In the case of the harmonic potential, for $\delta >0$ and $\delta \to 0$, we obtain
\begin{eqnarray}
    p_X(r_0 -\delta) \sim \delta^{\alpha +(d-3)/2}\, , \qquad {\rm and}\qquad p_R(r_0-\delta) \sim \delta^{\alpha -1}\, ,
\end{eqnarray}
where $r_0 = v_0/\mu$ is the location of the turning surface. These asymptotic forms show that the radial distribution exhibits a shape transition in any dimension, since its edge exponent changes sign at $\alpha = \gamma/\mu=1$. By contrast, the Cartesian marginal contains the additional exponent $(d-3)/2$, which comes from the projection law~(\ref{WprojEq}) and is therefore purely geometrical in origin. As a consequence, for $d=2$ the Cartesian marginal can still display an edge singularity, while for $d\geq 3$ this singularity is suppressed by projection.\\

Finally, the joint distribution of the three components $(x,y,z)$ is obtained using the result given in Eq.~(\ref{generalJointPDF}). It yields
\begin{equation}\label{Joint3d}
P(x,y,z) = f_R\!\left(r = \sqrt{x^2 +y^2 + z^2}\right)=\frac{p_R(r)}{4\pi r^2}  =  - \frac{\alpha\,e^{\alpha}}{8\pi^2 r}\,\frac{\mu^2}{v_0^2}\left[\pi A(r)^{\alpha-1} + \int_{0}^{A(r)} du\,  (A(r)-u)^{\alpha-1}f'(u)\right]\, ,
\end{equation}
which does not reduce to a beta distribution. In Fig.~\ref{FigpR3d}, we plot cross sections $P(x,y,z=0)$ for different values of $\alpha$ as a function of $(x,y)$: this shows qualitatively different behaviors as $\alpha$ crosses the ``critical'' value $\alpha = 1$ where the shape transition occurs.

\section{Stationary State of an RTP with $D > 0$ in 1D, 2D, and 3D}\label{DonSection}

We now consider a generalized run–and–tumble particle evolving in a harmonic potential and driven in addition by a thermal white noise with a diffusion coefficient $D$. We denote by $z(t)$ its position, whose dynamics thus reads
\begin{equation}
\dot{z}(t)= -\mu z(t) + v(t) + \sqrt{2D}\,  \eta(t)\, ,
\qquad z(0)=0,
\label{eq:RTPWN_sde}
\end{equation}
where $\mu>0$ is the trap stiffness, and $\eta(t)$ is a standard white noise with $\langle\eta(t)\rangle = 0$ and $\langle \eta(t)\eta(t')\rangle=\delta(t-t')$. The velocity $v(t)$ is again piecewise constant and takes values ${\sf v}_i$'s drawn from a distribution $W(v)$ with support $[v_{\rm{\min}}, v_{\rm{\max}}]$ between tumbling events. If one considers the pure generalized RTP $x(t)$ evolving via Eq. (\ref{generalizedRTP}) with the {\it same} $v(t)$, i.e., 
\begin{equation} \label{gen_rtp}
\dot{x}(t)= -\mu x(t) + v(t) \, ,
\qquad x(0)=0\, ,
\end{equation}
it is easy to see that the variable $y(t) = z(t) - x(t)$ is a simple Ornstein-Uhlenbeck process, namely
\bea
\dot{y}(t) = -\mu y(t) + \sqrt{2D}\,  \eta(t)\, ,
\qquad y(0)=0 \;.
\eea
Since the noises $v(t)$ and $\eta(t)$ are independent, it is clear that the two random variables $x(t)$ and $y(t)$ are statistically independent, at all time $t$, including in the stationary state. Therefore, in the limit $t \to \infty$, one has $z(t) = x(t) + y(t)$ where the stationary distribution of $x(t)$ has been studied in Section~\ref{GeneralizedRTPSection}, and its distribution is given by $p_X(x)$ given in Eq.~(\ref{SS_RTP_continuous}), while the distribution of $y(t)$, at large time $t$ is simply a Gaussian of zero mean and variance $D/\mu$. It follows from this argument that the stationary distribution of $z(t)$ is given by the convolution
\begin{equation} \label{pZ2}
    p_Z(z) =\sqrt{\frac{\mu}{2\pi D}} \int_{\frac{v_{\rm min}}{\mu}}^{\frac{v_{\rm max}}{\mu}}\, e^{-\frac{\mu (z-x)^2}{2 D}}\, p_X(x)\,dx \;.
\end{equation}
The same convolution argument holds for the stationary distribution of the $d$-dimensional position vector $\mathbf{r}(t \to +\infty) = (x_1,\cdots,x_d)$ in Eq.~(\ref{ddimRTPLangeThermal}), but not for its norm.
In the following, we will apply this formula (\ref{pZ2}) to study the standard RTP in the presence of thermal noise in $d=1$ and higher dimensions $d>1$ separately.

\begin{figure}[t]
    \centering
    \includegraphics[width=0.32\linewidth]{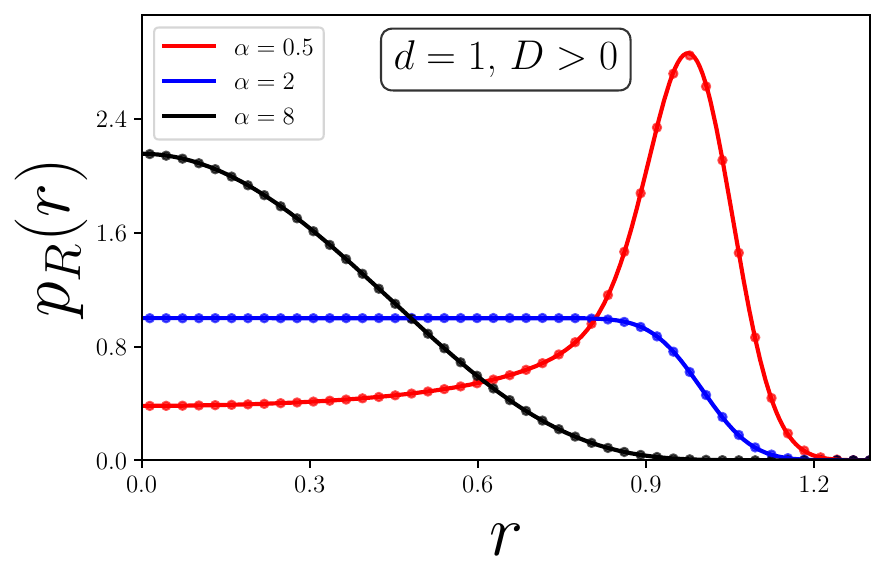}
    \includegraphics[width=0.32\linewidth]{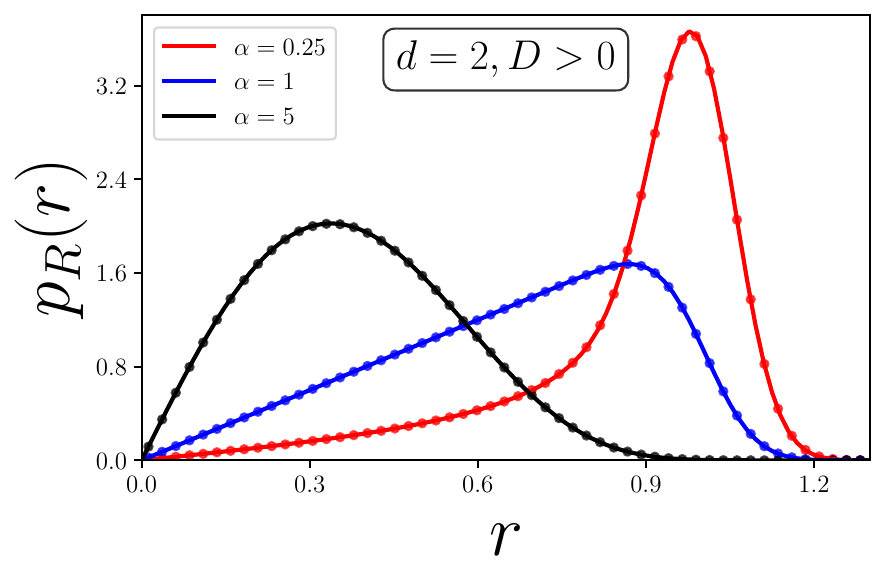}
    \includegraphics[width=0.32\linewidth]{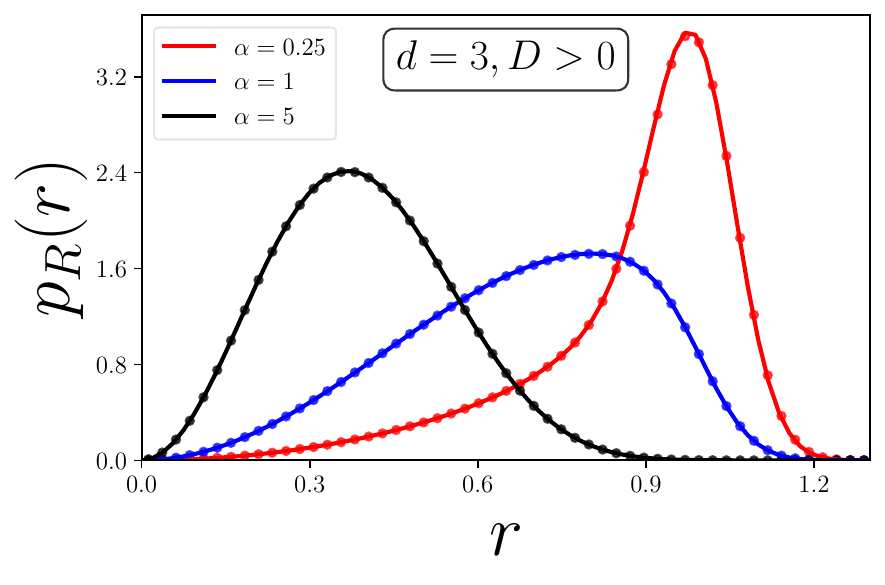}
    \caption{Stationary radial probability density $p_R(r)$ for an RTP confined in a harmonic trap with thermal noise ($D>0$), shown for dimensions $d = 1$ (left), $d = 2$ (center), and $d = 3$ (right). Parameters are $\mu = 1$, $v_0 = 1$, and $D=0.005$, for different values of the  parameter $\alpha = \gamma/\mu$. The presence of white noise smooths the singularities characteristic of the noiseless limit and extends the support of the distribution beyond the turning point $r_0 = v_0/\mu$. Solid lines correspond to the theoretical predictions, and markers represent Langevin simulations.}
    \label{FigpR_noise}
\end{figure}

\subsection{Stationary State of a One-Dimensional RTP with $D>0$}

In the one-dimensional case ($d=1$), the stationary position distribution $p_Z(z)$ is the convolution of the $D=0$ RTP density $p_X(x)$ -- given in Eq.~(\ref{eq:px_1d_rtp}) -- with the Gaussian distribution of zero mean and variance $D/\mu$. By substituting the expression for $p_X(x)$ into Eq.~(\ref{pZ2}), one obtains the exact stationary state for non-zero temperature (see also~\cite{sebastianActiveNoises})
\begin{equation} \label{pZ1d}
    p_Z(z) =\frac{\mu}{v_0}\,
    \frac{\Gamma\!\left(\frac{\alpha + 1}{2}\right)}
         {\sqrt{\pi}\,\Gamma\!\left(\frac{\alpha}{2}\right)}\,\sqrt{\frac{\mu}{2\pi D}} \int_{-\frac{v_0}{\mu}}^{\frac{v_0}{\mu}}\, e^{-\frac{\mu (z-x)^2}{2 D}}\, \left[1 - \left(\frac{\mu x}{v_0}\right)^2 \right]^{\alpha/2 - 1}\,dx \;.
\end{equation}
We note that in $d=1$, the radial distribution $p_R(r)$ is simply related to the position distribution via $p_R(r) = 2 p_Z(r)$ (since $R=|Z|$ -- see Eq.~(\ref{1dRXp})). In Fig.~\ref{FigpR_noise}, we show a comparison of our theoretical prediction (\ref{pZ1d}) with numerical simulations, showing a very good agreement. It is instructive to introduce the dimensionless scaling variables $\tilde z = \mu z /v_0$ and $\tilde x = \mu x /v_0$. In terms of these variables, Eq.~(\ref{pZ1d}) can be recast in the following scaling form
\begin{equation}
    p_Z(z) = \frac{\mu}{v_0} f_\theta\left(\tilde z = \frac{\mu z}{v_0}\right) \, , \qquad  f_\theta\left(\tilde z\right) =\frac{\Gamma\!\left(\frac{\alpha + 1}{2}\right)}
         {\pi\,\Gamma\!\left(\frac{\alpha}{2}\right)} \frac{1}{\sqrt{\theta}}\int_{-1}^{1} d\tilde x\, (1-\tilde x^2)^{\frac{\alpha}{2}-1} e^{-\frac{(\tilde x - \tilde z)^2}{\theta}}\, ,
         \label{PZscalingForm}
\end{equation}
Here, we have introduced the dimensionless parameter $\theta$, which quantifies the relative strength of thermal fluctuations to active noise (see Eq.~(\ref{theta_MR}))
\begin{equation}\label{thetaTXT}
    \theta= \frac{2 \mu D}{v_0^2} = \frac{2}{\alpha}\frac{D}{D_{\text{eff}}} \;,
\end{equation}
where $D_{\text{eff}}= v_0^2/\gamma$ denotes the effective diffusion coefficient of a free RTP with tumbling rate $\gamma$. For sufficiently small values of $\theta$, the shape of the distribution $p_Z(z)$ depends strongly on $\alpha$. For $\alpha > 2$, the system exhibits a passive-like regime where the distribution is a unimodal, bell-shaped function centered at $z=0$. In contrast, for $\alpha < 2$, the distribution undergoes a shape transition, exhibiting bimodality with two maxima near the boundaries of the active domain, $z = \pm v_0/\mu$ (see Fig.~\ref{fig_theta_c_d1}). {Interestingly, a very similar shape transition of the stationary distribution $p_Z(z)$ given in Eq.~(\ref{pZ1d}), was also found in Ref.~\cite{SabhaMajum}, albeit in a different context. There, the authors studied a (passive) Brownian particle in a harmonic trap whose center undergoes a telegraphic (dichotomous) process. The corresponding Langevin dynamics of that model turns out to be equivalent to Eq.~(\ref{eq:RTPWN_sde}) when $d=1$.}\\

One can easily derive the asymptotic behaviors of the scaling function $f_{\theta}(\tilde z)$ in Eq. (\ref{PZscalingForm}). At small $\tilde z$, the function approaches a constant, whereas for large $ \tilde z$ it has a Gaussian tail with a polynomial prefactor that is determined by noticing that when $\tilde z \to + \infty$, the integral in Eq.~(\ref{PZscalingForm}) is dominated by $\tilde x$ close to $1$. More precisely, one obtains
\begin{equation}
f_{\theta}(\tilde z) \sim
\begin{cases}
\dfrac{1}{\sqrt{\pi\,\theta}}\,
{}_{1}F_{1}\!\left(\dfrac{1}{2}\,;\,\dfrac{\alpha+1}{2}\,;\,-\dfrac{1}{\theta}\right)\, ,
& \text{when } \tilde z \to 0, \\[1.2em]
\frac{\Gamma \left( \frac{\alpha+1}{2} \right)}{2\pi} \theta^{\frac{\alpha-1}{2}} \tilde{z}^{-\frac{\alpha}{2}} e^{-\frac{(\tilde{z}-1)^2}{\theta}},
& \text{when } \tilde z \to +\infty\, .
\end{cases}
\label{asymptftheta}
\end{equation}
A plot of $p_R(r) = \frac{2\mu}{v_0} f_\theta\left(\frac{\mu r}{v_0}\right)$ is shown in the left panel of Fig. \ref{FigpR_noise} for $r\geq0$.

\subsubsection{Finite-$D$ Shape Transition}

\begin{figure}[t]
    \centering
    \includegraphics[width=0.32\linewidth]{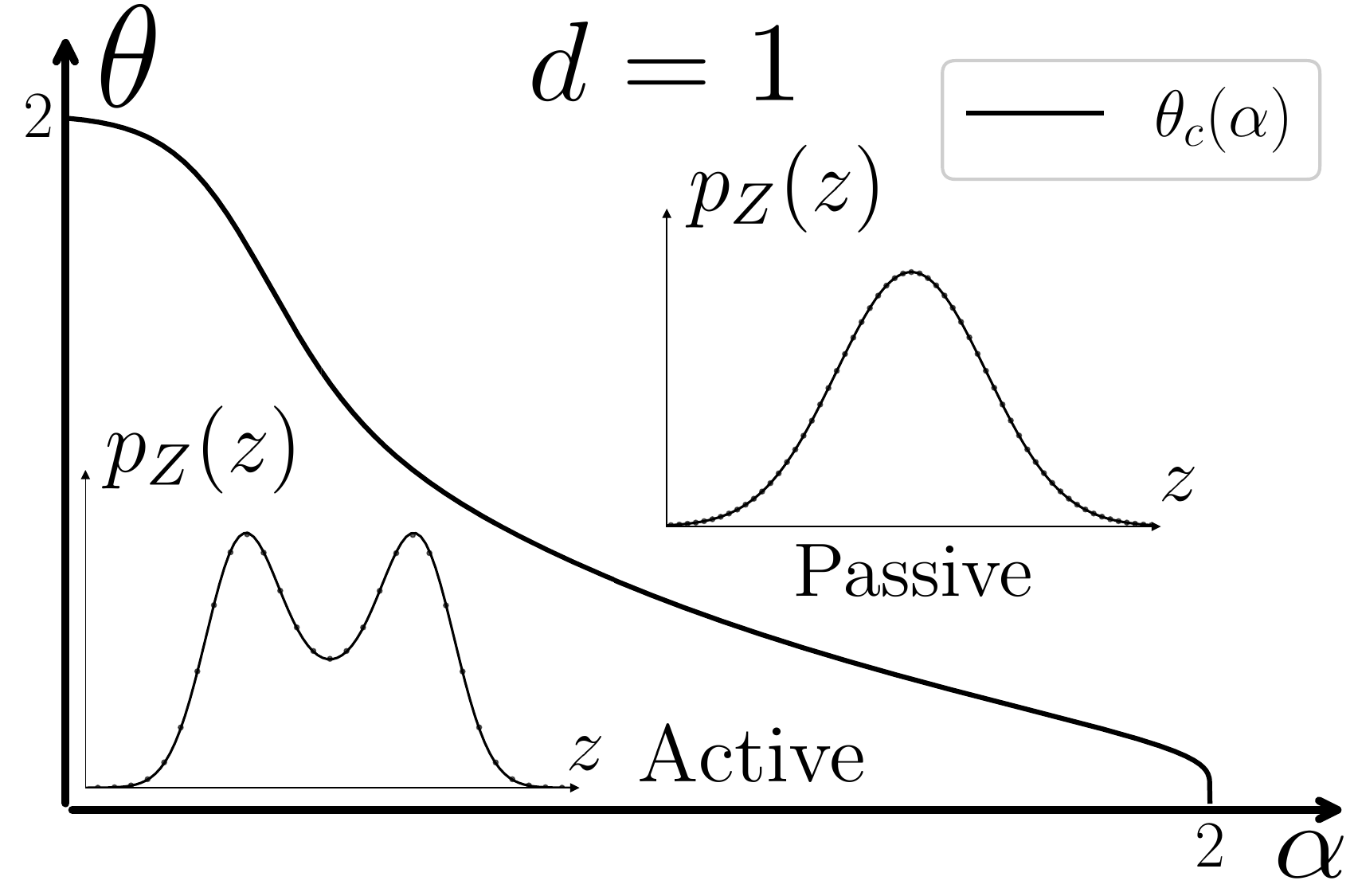}
    \includegraphics[width=0.32\linewidth]{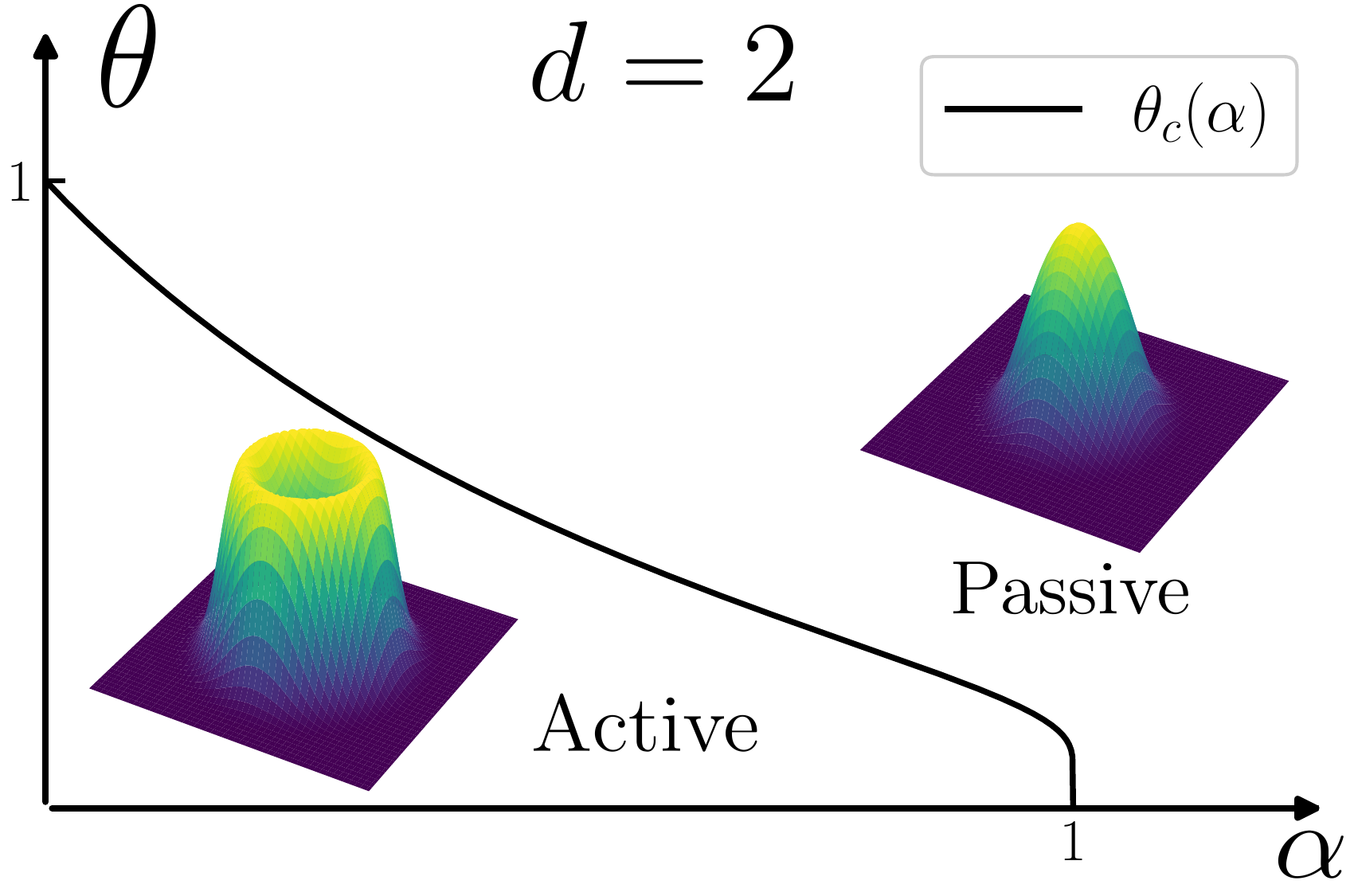}
    \caption{Diagrams in the $(\alpha, \theta)$ plane distinguishing between active and passive regimes for $d=1$ (left panel) and $d=2$ (right panel). The critical boundary $\theta_c(\alpha)$ is computed numerically using Eq.~(\ref{eq:critical_condition}) for $d=1$ and using Eq.~(\ref{thetacd2}) for $d=2$. 
\textbf{Left:} Insets display the stationary density $p_Z(z)$ for representative parameters in the active regime ($\theta =0.25, \alpha =0.5$, bottom left) and the passive regime ($\theta =4, \alpha =2$, top right). Solid lines correspond to the exact theoretical prediction from Eq.~(\ref{pZ1d}), while dots represent numerical simulation results. 
\textbf{Right:} Surface plots of the 2D distributions generated using the analytical prediction in Eq.~(\ref{2dJointD}). Here the parameters used are $\theta =0.1, \alpha =0.5$ in the active regime (bottom left) and $\theta =0.1, \alpha =2$ in the passive regime (top right).}
    \label{fig_theta_c_d1}
\end{figure}

As $D$ (or equivalently $\theta$) is decreased, the scaling function exhibits qualitatively different behaviors depending on whether $\alpha > 2$ or $\alpha < 2$ (see Fig.~\ref{fig_theta_c_d1}). For $\alpha > 2$, the point $\tilde{z} = 0$ remains the unique maximum of $f_{\theta}(\tilde{z})$, and the function retains a standard bell-shaped profile for all values of $\theta$. In contrast, for $\alpha < 2$, the origin $\tilde{z} = 0$ is a local maximum when $\theta$ is large, but becomes a local minimum when $\theta < \theta_c$, where $\theta_c$ denotes a critical value. In this regime, a peak develops away from the origin. As in Ref.~\cite{SabhaMajum}, to determine the critical temperature $\theta_c(\alpha)$ where this transition occurs, we analyze the curvature of the scaling function at the origin, i.e., the sign of the second derivative of $f_{\theta}(\tilde{z})$ at $\tilde{z} = 0$. We start by expanding the exponential term in the integrand of Eq.~(\ref{PZscalingForm}) for small $\tilde{z}$. Using the identity $e^{-(\tilde{x}-\tilde{z})^2/\theta} = e^{-\tilde{x}^2/\theta} e^{(2\tilde{x}\tilde{z} - \tilde{z}^2)/\theta}$, we expand up to second order in $\tilde{z}$ to get
\begin{align}
    e^{-\frac{(\tilde{x} - \tilde{z})^2}{\theta}} \approx e^{-\frac{\tilde{x}^2}{\theta}} \left[ 1 + \frac{1}{\theta}(2\tilde{x}\tilde{z} - \tilde{z}^2) + \frac{1}{2\theta^2}(2\tilde{x}\tilde{z})^2 + \mathcal{O}(\tilde{z}^3) \right] \;.
    \label{eq:expansion_exp}
\end{align}
Substituting this expansion into the definition of $f_\theta(\tilde{z})$ and collecting the terms of order $\mathcal{O}(\tilde{z}^2)$, we obtain the second derivative of the scaling function at the origin
\begin{align}
    f''_\theta(0) = \frac{2}{\pi} \frac{\Gamma\left(\frac{\alpha+1}{2}\right)}{ \Gamma\left(\frac{\alpha}{2}\right)} \frac{1}{\theta^{3/2}} \int_{-1}^{1} \mathrm{d}\tilde{x} \, (1-\tilde{x}^2)^{\frac{\alpha}{2}-1} \, e^{-\frac{\tilde{x}^2}{\theta}} \left( \frac{2\tilde{x}^2}{\theta} - 1 \right) \;.
    \label{eq:second_derivative}
\end{align}
The shape transition corresponds to the point where the convexity at the origin changes sign, i.e., $f''_{\theta_c}(0) = 0$~\cite{SabhaMajum}. This yields the following implicit equation for the critical temperature $\theta_c(\alpha)$ with $\alpha <2$
\begin{align}
    \int_{-1}^{1} \mathrm{d}x \, (1-x^2)^{\frac{\alpha}{2}-1} \, e^{-\frac{x^2}{\theta_c(\alpha)}} \left( \frac{2x^2}{\theta_c(\alpha)} - 1 \right) = 0 \;.
    \label{eq:critical_condition}
\end{align}
From this equation, one can obtain the asymptotic behaviors of $\theta_c(\alpha)$ in the two limits $\alpha \to 0$ and $\alpha \to 2$. In the limit $\alpha \to 0$, because of the function $(1-x^2)^{\alpha/2-1}$ the integral over $x$ is dominated by 
the boundaries $x \to \pm 1$. Evaluating the integrand at these boundaries leads to the condition $2/\theta_c(0) - 1 = 0$, yielding $\theta_c(0) = 2$. In the other limit $\alpha \to 2$, a careful asymptotic analysis of the integral in Eq.~(\ref{eq:critical_condition}) shows that the critical temperature vanishes logarithmically
\begin{align}
    \theta_c(\alpha) \underset{\alpha \to 2^-}{\sim} \frac{1}{|\ln(2-\alpha)|} \;.
\end{align}
The full phase diagram in the $(\alpha, \theta)$ plane is shown in Fig.~\ref{fig_theta_c_d1}, where the critical line $\theta_c(\alpha)$ separates the active and passive regimes of the RTP stationary state. 

\subsubsection{Universal Low-$D$ Scaling Form Around $z = \pm \frac{v_0}{\mu}$}\label{UniversalTP}

\begin{figure}
    \centering
    \includegraphics[width=0.4\linewidth]{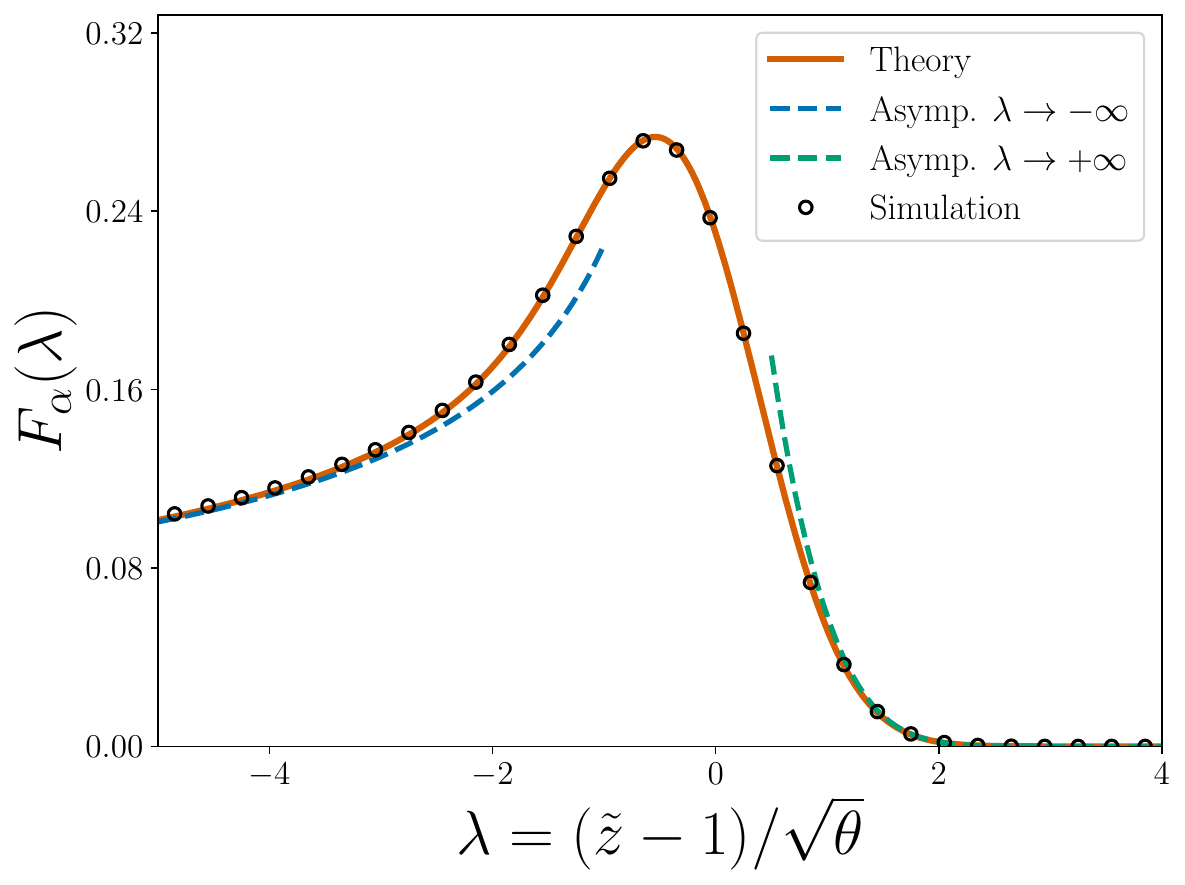}
    \caption{Universal low-$D$ turning-point scaling of the stationary position distribution for a one-dimensional run-and-tumble particle in a harmonic trap in the presence of thermal noise. Symbols show simulations for $\mu=1.0$, $v_{0}=1.0$, $\alpha=\gamma/\mu=1.0$, and $\theta=2\mu D/v_{0}^{2}=10^{-4}$, plotted versus the scaled coordinate $\lambda=(\tilde z-1)/\sqrt{\theta}$ with $\tilde z=\mu z/v_{0}$. The solid curve is the theoretical scaling function $F_{\alpha}(\lambda)$ from Eq.~(\ref{fthetaTP}), while the dashed curves show the asymptotic forms in Eq.~(\ref{Fbeta_asymp}) for $\lambda\to-\infty$ (interior matching) and $\lambda\to+\infty$ (exterior tail).}
    \label{fig:turning_point_scaling}
\end{figure}

As discussed in the previous section, in the vanishing noise limit $D \to 0$, the stationary density $p_Z(z)$ exhibits algebraic singularities at the turning points $z = \pm v_0/\mu$ for $\alpha < 2$. For finite $D > 0$, the thermal fluctuations regularize these divergences and allow the particle to perform excursions beyond the $D=0$ finite support $[-v_0/\mu, v_0/\mu]$. This smoothening mechanism manifests itself in the emergence of a boundary layer of width $\mathcal{O}(\sqrt{D})$ around the turning points. To characterize the behavior of the distribution in this regime, we focus on the right turning point $\tilde z = 1$ (corresponding to $z = v_0/\mu$). The analysis for the left turning point follows by symmetry. We introduce the scaling variable $\lambda = (\tilde z - 1)/\sqrt{\theta}$, where $\theta$ is given in Eq.~(\ref{thetaTXT}), to describe the vicinity of the turning point. By analyzing the integral representation in Eq.~(\ref{PZscalingForm}) in the scaling limit $\theta \to 0$, $\tilde z \to 1$ with $\lambda$ fixed, we obtain the following scaling form
\begin{align}
    f_\theta(\tilde z) \sim \theta^{\frac{\alpha}{4}-\frac{1}{2}} F_\alpha\left(\lambda = \frac{\tilde z -1}{\sqrt{\theta}}\right)\, , \qquad 
    \text{where } F_\alpha(\lambda) = \frac{\Gamma\left(\frac{\alpha+1}{2}\right)}{\pi \Gamma\left(\frac{\alpha}{2}\right)} 2^{\frac{\alpha}{2}-1} \int_0^{+\infty} \!\! dv\, e^{-(v+\lambda)^2}v^{\frac{\alpha}{2}-1} \, .
    \label{fthetaTP}
\end{align}
It is instructive to examine the asymptotic behaviors of this scaling function in the limits $\lambda \to \mp \infty$, which describe the matching with the bulk and the tail of the distribution respectively. A straightforward asymptotic analysis of the integral representation of $F_\alpha(\lambda)$ in Eq. (\ref{fthetaTP}) yields
\begin{equation}
    F_\alpha(\lambda) \sim 
    \begin{cases} 
       \displaystyle \frac{\Gamma\left(\frac{\alpha+1}{2}\right)}{\sqrt{\pi}\,\Gamma\left(\frac{\alpha}{2}\right)} \, 2^{\frac{\alpha}{2}-1} \, |\lambda|^{\frac{\alpha}{2}-1}, & \text{as } \lambda \to -\infty, \\[15pt]
       \displaystyle \frac{\Gamma\left(\frac{\alpha+1}{2}\right)}{2\pi} \, \lambda^{-\frac{\alpha}{2}} \, e^{-\lambda^2}, & \text{as } \lambda \to +\infty.
    \end{cases}
    \label{Fbeta_asymp}
\end{equation}
These limits ensure a smooth crossover between the different regimes of the density. For $\lambda \to -\infty$ (corresponding to the interior of the active domain), substituting the asymptotic form back into Eq.~(\ref{fthetaTP}) recovers the algebraic divergence of the $D=0$ RTP distribution $p_X(x)$ near the edge, as given in Eq.~(\ref{eq:px_1d_rtp}). Conversely, for $\lambda \to +\infty$ (the exterior region), the decay of $F_\alpha(\lambda)$ matches the form derived in Eq.~(\ref{asymptftheta}). In Fig.~\ref{fig:turning_point_scaling}, we test the scaling prediction against simulations by plotting the universal function $F_{\alpha}(\lambda)$ from Eq.~(\ref{fthetaTP}) together with its asymptotic forms in Eq.~(\ref{Fbeta_asymp}), finding a perfect agreement.

\subsection{Stationary State of a Two-Dimensional RTP with $D>0$}\label{HighDimDon2}

In two dimensions, the single component stationary state $p_Z(z)$ of the RTP with $D>0$ is given by Eq.~(\ref{pZ2}) where the function $p_X(x)$ is the beta function from Eq.~(\ref{2dpX}). In the presence of the thermal noise, the support of $p_Z(z)$ is now extended to the whole real line $(-\infty,+\infty)$. Hence, using isotropy, the distribution of the radius can then be computed using the first line of Eq.~(\ref{pXtopRd2d3}), where the upper bound of the integral is instead $+\infty$. Straightforward calculations then lead to 
\begin{eqnarray}
p_R\!\left(\tilde r = \frac{\mu r}{v_0}\right) =  \sqrt{\frac{2}{D\mu}} \frac{v_0^2}{ \pi D} \frac{\Gamma\!\left(1+\alpha\right)}{ \Gamma\!\left(\frac{1}{2}+\alpha\right)} \, \tilde r \int_{\tilde r}^{+\infty}\frac{d\tilde z}{\sqrt{\tilde z^2-\tilde r^2}}\int_{-1}^{1}d\tilde x\,  (\tilde z - \tilde x)(1-\tilde x^2)^{\alpha-\frac{1}{2}}\, e^{- \frac{v_0^2}{2 D \mu}(\tilde x- \tilde z)^2}\, .
\end{eqnarray}
For a comparison with simulation, see the middle panel of Fig.~\ref{FigpR_noise}. From the radial distribution, we find that the joint distribution in $d=2$ is given by
\begin{eqnarray}
    P(x,y) &=& \frac{p_R\!\left(r=\sqrt{x^2+y^2}\right)}{2\pi r}=\left(\frac{\mu}{v_0}\right)^2 g_{\theta}\!\left(\tilde r = \frac{\mu r}{v_0}\right)\, ,\\
    g_\theta(\tilde r) &=& \frac{2}{\pi^2} \frac{1}{\theta^{\frac{3}{2}}}\frac{\Gamma\!\left(1+\alpha\right)}{ \Gamma\!\left(\frac{1}{2}+\alpha\right)} \int_{\tilde r}^{+\infty}d\tilde z\int_{-1}^{1}d\tilde x \frac{(\tilde z - \tilde x)}{\sqrt{\tilde z^2-\tilde r^2}}(1-\tilde x^2)^{\alpha-\frac{1}{2}}\, e^{- \frac{(\tilde x- \tilde z)^2}{\theta}}\, .\label{2dJointD}
\end{eqnarray}
In the right panel of Fig.~\ref{fig_theta_c_d1} we show a 3D plot of $P(x,y)$ as a function of $(x,y)$ for two different sets of parameters. In the top-right (for high-$D$ or equivalently high-$\theta$), $P(x,y)$ has a bell shape with a unique maximum at the origin $(x=0,y=0)$ while in the bottom left (for low-$D$ or equivalently low-$\theta$) the origin is a local minimum with a local maximum on a ring of radius $r_0\approx v_0/\mu$. This is a manifestation of the shape transition, similar to the one found in $d=1$, which we now describe in more detail.

As in $d=1$, the passive to active shape transition is characterized by the value of $\theta$ at which the maximum of $P(x,y)$ at the origin becomes a local minimum. Equivalently, the critical line $\theta_c(\alpha)$ is determined by the condition $g''_{\theta}({\tilde r = 0})=0$. Using the change of variables $u=\sqrt{\tilde z^{\,2}-\tilde r^{\,2}}$, one finds that the resulting equation that determines $\theta_c(\alpha)$ can be written, for $\alpha < 1$, as
\begin{eqnarray}
\int_{0}^{+\infty}\!\! \!du\! \int_{0}^{1} \!\!d{x}\, (1 - x^{2})^{\alpha - \tfrac{1}{2}}
\left[
\frac{x}{2u^{3}}
\left(
e^{-\frac{(u-x)^{2}}{\theta_c(\alpha)}}
-
e^{-\frac{(u+x)^{2}}{\theta_c(\alpha)}}
\right)
-
\frac{1}{u^{2}\theta_c(\alpha)}
\left(
(u-x)^{2} e^{-\frac{(u-x)^{2}}{\theta_c(\alpha)}}
+
(u+x)^{2} e^{-\frac{(u+x)^{2}}{\theta_c(\alpha)}}
\right)
\right]= 0\, .\qquad
\label{thetacd2}
\end{eqnarray}
By solving numerically this equation for $\theta_c(\alpha)$ one finds the black solid line shown in Fig.~\ref{fig_theta_c_d1}, which separates the passive and active in the $(\alpha, \theta)$ plane. One can numerically check that $\theta_c(\alpha) \to 1$ when $\alpha \to 0$ and $\theta_c(\alpha) \to 0$ when $\alpha \to 1$ -- see Fig.~\ref{fig_theta_c_d1}.

\subsection{Stationary State of a Three-Dimensional RTP with $D>0$}\label{HighDimDon3}
\begin{figure}
    \centering
    \includegraphics[width=0.4\linewidth]{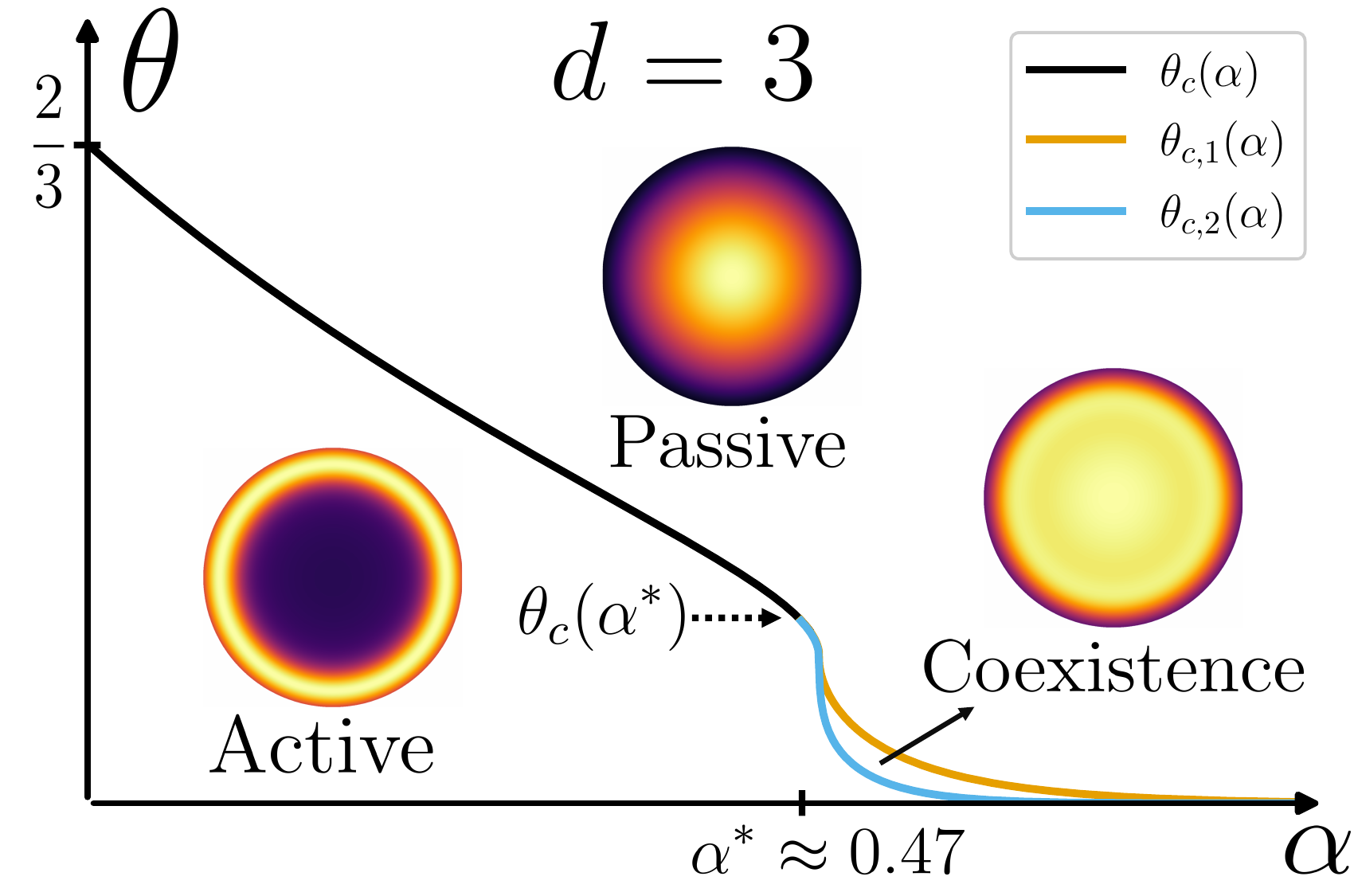}
    \includegraphics[width=0.35\linewidth]{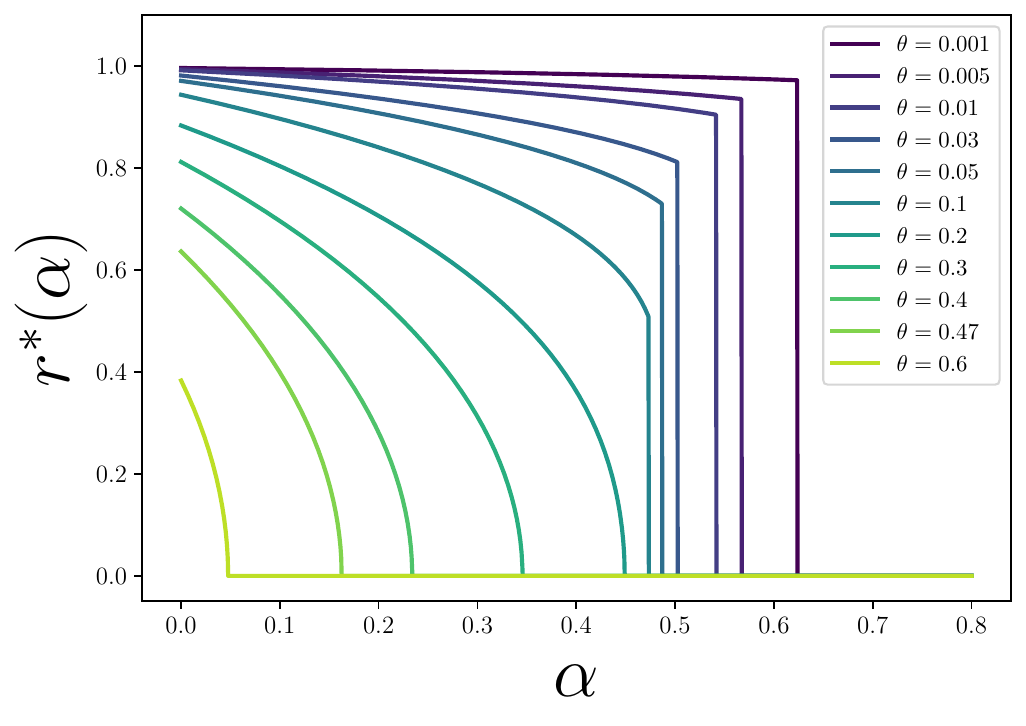}
    \caption{\textbf{Left:} The three plots are cross sections $P(x,y,0)$ as a function of $(x,y)$ calculated using Eq.~(\ref{jointd3D}) illustrating the finite-$D$ shape transition in $d=3$. The black curve shows the analytic critical line obtained from Eq.~(\ref{d3critical}): it exists only up to the value $\alpha^*\simeq 0.47$ (i.e., there is no solution to this equation for $\alpha>\alpha^*$). For $\alpha\le \alpha^*$, this line separates a phase with a single maximum at the origin from a phase where the origin becomes a local minimum and a shell-like maximum develops near the turning sphere $r\simeq v_0/\mu$. For $\alpha>\alpha^*$, the transition occurs instead in two steps and across two critical lines computed numerically. The upper line $\theta_{c,1}(\alpha)$ (orange) marks the emergence of an outer local maximum while the origin remains the global maximum. The lower line $\theta_{c,2}(\alpha)$ (blue) corresponds to an abrupt change of the location of the global maximum from the origin to the outer shell. In the intermediate region $\theta_{c,2}(\alpha)<\theta<\theta_{c,1}(\alpha)$, $P(x,y,z)$ is nearly flat for $r\in[0,v_0/\mu]$ and exhibits two maxima (a global one at the origin and a local one at $r\simeq v_0/\mu$).
\textbf{Right:} For several fixed values of $\theta$, we show the numerically evaluated position of the global maximum $r^*(\alpha)$ of $h(r)$ given in Eq.~\eqref{jointd3D}. For $\theta > \theta_c(\alpha^* \approx 0.47)\simeq 0.15$, $r^*(\alpha)$ moves continuously from the origin to the turning sphere of radius $r_0=v_0/\mu$ as $\alpha$ is varied. In contrast, for $\theta < \theta_c(\alpha^*)$, $r^*(\alpha)$ exhibits an abrupt jump between $r=0$ and $r\simeq v_0/\mu$ upon crossing $\theta_{c,2}(\alpha)$.}
    \label{figcriticald3}
\end{figure}

In the three dimensional case, the expressions are more involved. One has to inject the formula at zero temperature given in Eq.~(\ref{3dcomponentdistrib}) inside Eq.~(\ref{pZ2}) to obtain the single component distribution with $D>0$. That is,
\begin{equation} \label{pZ3d}
    p_Z(z) = \frac{\alpha\,e^{\alpha}}{2\pi}\,\frac{\mu}{v_0}\sqrt{\frac{\mu}{2\pi D}} \int_{-\frac{v_{0}}{\mu}}^{\frac{v_0}{\mu}}dx\, \, e^{-\frac{\mu (z-x)^2}{2 D}}\,
\int_{0}^{\frac{1}{2}\left(1+\frac{\mu x}{v_0} \right)} \!\!du\, \left[\frac{1}{2}\left(1+\frac{\mu x}{v_0}\right)-u\right]^{\alpha-1}\, f(u)\, ,
\end{equation}
where $f(u)$ is the function given in Eq.~(\ref{3dfu}).
Then, when $d=3$, one finally has to compute $p_R(r) = -2rp_Z'(r)$ to obtain the radial distribution. We check that the result agrees with simulations in the right panel of Fig.~\ref{FigpR_noise}. The joint law is then given by
\begin{eqnarray}\label{jointd3D}
    P(x,y,z) = h\!\left(r=\sqrt{x^2+y^2+z^2}\right) = \frac{p_R(r)}{4\pi r^2} = -\frac{p_Z'(r)}{2 \pi r}\, .
\end{eqnarray}
A plot of $P(x,y,z=0)$ as a function of $(x,y)$ is shown in the left panel of Fig. \ref{figcriticald3} for different values of the parameters. For small values of $\alpha < \alpha^*$ (to be discussed below), we find a shape transition that is very similar to the one found in $d=1$ and $d=2$. As above, the transition occurs when the maximum at the origin becomes a local minimum. It is therefore given by the condition $h''(r)=0$ which here is equivalent to $p_Z^{(4)}(r) = 0$ (since $p_Z'(z=0) =0$ by symmetry).  

Using the change of variables $\tilde x = \mu x/v_0$ and $\tilde z = \mu z/v_0$ as well as the identity
\begin{equation}
\left.\frac{\partial^{4}}{\partial \tilde z^{4}} e^{-\frac{(\tilde z-\tilde x)^{2}}{\theta}}\right|_{z=0}
= \frac{4}{\theta^2}\left( 3\theta^2-12\theta \tilde x^2+4\tilde x^4 \right)
e^{-\frac{\tilde x^{2}}{\theta}}\, ,
\end{equation}
one finds that the condition in $d=3$ reads
\begin{equation}\label{d3critical}
     \int_{-1}^{1}d\tilde x\,\left( 3\theta_c(\alpha)^2-12\theta_c(\alpha) \tilde x^2+4\tilde x^4 \right)e^{-\frac{\tilde x^2}{\theta_c(\alpha)}}\,
\int_{0}^{\frac{1}{2}\left(1+\tilde x \right)} \!\!du\, \left[\frac{1}{2}\left(1+\tilde x\right)-u\right]^{\alpha-1}f(u) = 0\, .
\end{equation}
Numerically, we find that Eq.~(\ref{d3critical}) admits a unique solution only up to the value $\alpha^*\simeq 0.47$ -- see the left panel of Fig.~\ref{figcriticald3}. Hence, for $\alpha < \alpha^*$, the behavior remains analogous to $d=1,2$, and Eq.~(\ref{d3critical}) yields a single critical line separating a phase with a unique maximum at $r=0$ from a phase where the origin becomes a local minimum when a maximum
of $P(x,y,z)$ develops on a shell near the ``turning sphere'' $r\simeq v_0/\mu$.

Interestingly, Eq.~(\ref{d3critical}) can be studied analytically in the limit $\alpha\to 0$ as follows. Using the $\alpha\to 0$ asymptotics of $p_X$ in Eq.~(\ref{samllbetapX}), we find in $d=3$
\begin{equation}
p_X(x)\xrightarrow[\alpha\to 0]{}\mu\,W(\mu x)=\frac{\mu}{2v_0}\, ,
\end{equation}
since $W(v)=1/(2v_0)$ on $[-v_0,v_0]$. Substituting this limiting form into the convolution formula Eq.~(\ref{pZ2}) yields an explicit expression for $p_Z$. Imposing the transition criterion $p_Z^{(4)}(0)=0$ then gives
\begin{equation}
\theta_c(\alpha\to 0)=\frac{2}{3}\, .
\end{equation}

{

For $\alpha>\alpha^*$, Eq.~(\ref{d3critical}) has no solution, and, for fixed $\alpha > \alpha^*$ the shape transition occurs instead in two steps as $\theta$ crosses the two distinct lines $\theta_{c,1}(\alpha)$ and $\theta_{c,2}(\alpha)<\theta_{c,1}(\alpha)$ in the $(\alpha, \theta)$ plane -- see Fig. \ref{figcriticald3}. These two lines are found by numerically identifying the location of the maxima of $P(x,y,z)$ given in Eq.~(\ref{jointd3D}). For $\theta>\theta_{c,1}(\alpha)$ the distribution has a single maximum at the origin. For $\theta_{c,2}(\alpha)<\theta<\theta_{c,1}(\alpha)$, an additional outer local maximum appears near $r\simeq v_0/\mu$ while the origin remains the global maximum, leading to a broad, nearly flat profile on $r\in[0,v_0/\mu]$ with two competing maxima (see again Fig.~\ref{figcriticald3}, left panel). Below the lower critical line $\theta< \theta_{c,2}(\alpha)$ 
the global maximum of $P(x,y,z)$ is located at the outer shell near $r\simeq v_0/\mu$. Interestingly, as $\theta$ crosses the lower critical line $\theta= \theta_{c,2}(\alpha)$ the location of the global maximum jumps (discontinuously) from the origin to the outer shell, similar to a first-order transition. This two-step transition can also be probed by varying $\alpha$, for a fixed value of $\theta$. In this case, for sufficiently small value of $\theta$, the location of the global maximum $r^*(\alpha)$ jumps discontinuously from a finite value $r^*(\alpha) > 0$ to  $r^*(\alpha) = 0$ as $\alpha$ increases and crosses some critical value, as shown in the right panel of Fig.~\ref{figcriticald3}. 
}

\section{Stationary Distribution of a $N$-state RTP}\label{NstateSection}

Up to now, we have only considered the case where $W(v)$ is a continuous distribution. However, in many cases, 
it is relevant to consider a case where the velocities can only take discrete values. This amounts to consider a $N$-state model where $W(v)$ is given by
\begin{equation}
    W(v) \;=\; \sum_{i=1}^N p_i \, \delta(v - v_i)\,,
    \qquad 0<p_i<1, \qquad \sum_{i=1}^N p_i = 1 \, .
    \label{NWstate}
\end{equation}
Without loss of generality, we assume the velocities to be ordered: $v_1< \cdots < v_N$. As mentioned above, the one dimensional RTP corresponds to a two-state process with velocities $\pm v_0$ and probabilities $p_1=  p_2 = 1/2$. Different extensions, including RTP models with multiple internal velocity states have also been considered. For instance, three-state \cite{RTP3state, thesis} and four-state \cite{Frydel4} cases have been investigated. In both cases, the stationary distributions were obtained by solving coupled Fokker-Planck equations, which quickly becomes a hard task as $N$ increases.
Here, using the Dirichlet-process (stick-breaking) representation and the general formula~(\ref{MRSSX}) obtained in this paper,
we obtain the stationary state for an arbitrary $N$-state RTP in a direct and systematic way.\\

\noindent {\bf Cifarelli-Regazzini Representation.}
The stationary state can be obtained from Eq.~(\ref{SS_RTP_continuous}). For the $N$-state distribution~(\ref{NWstate}), one first shows that
\begin{equation}\label{phiNstate}
\phi_\alpha(t)=\sin\!\left(\pi \alpha \sum_{i=1}^kp_i \Theta\left(t-\frac{v_i}{\mu}\right)\right)\,
\prod_{i=1}^{N} \lvert t-v_i/\mu\rvert^{-\alpha p_i}\, .
\end{equation}
To avoid the appearance of delta-function terms in Eq.~(\ref{MRSSX}) when deriving the Heaviside functions in~(\ref{phiNstate}), one can consider the calculation of the derivative of $\phi_\alpha(t)$ for values $t \neq v_i/\mu$. One finds
\begin{equation}\label{dphiNstate}
\frac{d\phi_\alpha(t)}{dt}=-\alpha \, \phi_\alpha(t)\, \sum_{i=1}^{N}\frac{p_i}{t-v_i/\mu}\, .
\end{equation}
Then, using~(\ref{dphiNstate}), we find that the stationary density~(\ref{MRSSX}) is piecewise continuous on the intervals $x \in (v_k/\mu,v_{k+1}/\mu)$, and its explicit form is
\begin{eqnarray}\label{CRRepNState}
p_X(x) = -\,\frac{\alpha}{\pi}\Bigg[
&&\!\!\!\!\!\sum_{m=1}^{k-1} \sin\left(\pi\alpha \sum_{i=1}^{m}p_i\right)
\int_{v_m/\mu}^{v_{m+1}/\mu}dt\,  (x-t)^{\alpha-1}\,\left(\sum_{i=1}^{N} \frac{p_i}{t-\frac{v_i}{\mu}}\right)\prod_{i=1}^{N} \lvert t-v_i/\mu\rvert^{-\alpha p_i}\!
\;\nonumber\\
&&+\;
\sin\left(\pi\alpha \sum_{i=1}^{k}p_i\right)
\int_{v_k/\mu}^{x}dt\,  (x-t)^{\alpha-1}\,\left(\sum_{i=1}^{N} \frac{p_i}{t-\frac{v_i}{\mu}}\right)\prod_{i=1}^{N} \lvert t-v_i/\mu\rvert^{-\alpha p_i}\!
\Bigg].
\end{eqnarray}\\

\noindent {\bf The case $\alpha = 1$.}
In that case, we can directly use the general expression~(\ref{generalBeta1}). It yields
\begin{eqnarray}
    p_X(x) = \frac{\phi_1(x)}{\pi}= \frac{1}{\pi}\sin\left(\pi \sum_{i=1}^{k}p_i\right)\prod_{i=1}^{N} \lvert x-v_i/\mu\rvert^{- p_i}\, , \qquad \qquad x \in (v_k/\mu,v_{k+1}/\mu)\, .
\end{eqnarray}\\

\noindent {\bf Dirichlet Representation}. Another explicit formula for the stationary state can be found from the result~(\ref{x_yn_def}) which gives the stationary position of the generalized RTP in terms of stick-breaking weights. In the case of this $N$-state model, for a given trajectory and a given sequence of velocities $\{{\sf v}_1\cdots,{\sf v}_n \}$, distinct samples ${\sf v}_i$ drawn from $W(v)$ may take identical values $v_j$. Hence, one must aggregate the corresponding weights in the equation~(\ref{x_yn_def}), leading to
\begin{eqnarray}
    X = \frac{1}{\mu} \sum_{n\geq1} {\sf{v}}_n Y_n  =\frac{1}{\mu} \sum_{n=1}^N v_n Z_n\,  ,
   \qquad 
   Z_n \;=\; \sum_{i\geq 1}  Y_i \, \mathbf{1}_{{\sf v}_i = v_n}\, .
\end{eqnarray}
It is known that these aggregated weights $Z_n$ follow a Dirichlet distribution~\cite{Ferguson, CifarelliRegazzini}
\begin{equation}
   (Z_1,\ldots,Z_N) \;\overset{d}{=}\; 
   \mathrm{Dir}\!\left(\alpha\,  p_1,\,\ldots,\,\alpha\, p_N\right),
\end{equation}
where $\mathrm{Dir}(\cdot)$ denotes the Dirichlet distribution.
The corresponding joint probability density reads
\begin{eqnarray}
    f_{\text{Dir}}(z_1,\cdots,z_N) \;=\; 
\frac{\Gamma(\alpha)}{\prod_{i=1}^N \Gamma\left(\alpha\, p_i\right)}\,
\left( \prod_{i=1}^N z_i^{\alpha p_i - 1} \right)\,
\delta\!\left( 1 - \sum_{i=1}^N z_i \right)\,
\mathbf{1}_{\{z_i \geq 0\}} \, ,
\end{eqnarray}
and since the position of the RTP is $X = \frac{1}{\mu} \sum_{n=1}^N  v_n Z_n$, its probability density can be expressed as
\begin{eqnarray}
    p_X(x)  &=& \Big\langle \delta\left(x-\frac{1}{\mu}\sum_{n = 1}^N  v_n z_n\right)\Big\rangle_z \\
    &=& \frac{\Gamma(\alpha)}{\prod_{i=1}^N \Gamma\left(\alpha\, p_i\right)}\, \int_0^1dz_1\cdots\int_0^1dz_N  \, \delta\left(x-\frac{1}{\mu}\sum_{n = 1}^N  v_n z_n\right)\, 
\left( \prod_{i=1}^N z_i^{\alpha p_i - 1} \right)\,
\delta\!\left( 1 - \sum_{i=1}^N z_i \right)\, .
\label{NstateExact}
\end{eqnarray}
Note that this result can also be derived directly from the expression of the MGF given in Eq.~(\ref{MGF_Main}). In Appendix~\ref{3state}, we compute the stationary state of a general 3-state RTP and compare our theoretical predictions to simulation results. We finally mention that these results can be straightforwardly generalized in the presence of an additional thermal noise using the convolution structure in Eq.~(\ref{pZ2}).

\section{Conclusion}\label{ConclusionSection}

In this work, we have obtained the exact nonequilibrium stationary state distribution of a run-and-tumble particle confined by an isotropic harmonic potential in arbitrary spatial dimension $d$, including a closed-form characterization in $d=3$. Exploiting rotational invariance, we showed that the full stationary distribution is determined by a single Cartesian marginal $p_X(x)$: the radial distribution $p_R(r)$ and the joint density $P(x_1,\dots,x_d)$ follow from $p_X(x)$ through explicit projection/inversion transforms. Our central analytical result is an exact solution of a generalized one-dimensional RTP in a harmonic trap where post-tumble velocities are sampled from an arbitrary law $W(v)$. Obtaining the stationary solution via the standard Fokker–Planck approach leads to a nonlocal integro-differential equation that does not appear analytically tractable. Although this route proves unproductive, the problem can be solved by a completely different approach. In particular, the position in the stationary state yields a Kesten recursion relation. Solving it explicitly shows that the position of this generalized RTP in the stationary state identifies with a mean functional of a Dirichlet process via a stick-breaking construction. This connection to Dirichlet processes allowed us to obtain closed-form expressions for $p_X(x)$ and for its stationary moments, for an arbitrary law $W(v)$. Specializing $W(v)$ to the projected velocity distribution of an isotropic RTP then provides the complete stationary statistics of an RTP in a harmonic well in any dimension $d$.\\

A key physical scale of the dynamics is the turning radius $r_0=v_0/\mu$, where self-propulsion velocity $v_0$ and harmonic restoring force of strength $\mu$ balance. Varying the dimensionless activity parameter $\alpha=\gamma/\mu$, where $\gamma$ is the tumbling rate, controls the    
shape of the stationary distribution around this turning surface. In $d=1$ and $d=2$, the radial law reduces to a beta distribution and develops integrable edge singularities in the persistent regime, with accumulation near $r_0$. In $d=3$, we have shown that 
the stationary distribution remains explicitly computable but no longer collapses to a beta distribution. The stationary distribution $p_R(r)$ of the radius of the position, both in $d=2$ and $d=3$, exhibits a shape transition at $\alpha=1$: we find that $p_R(r)$ diverges as $r \to r_0$ for $\alpha < 1$ while $p_R(r) \to 0$ as $r \to r_0$ for $\alpha > 1$.\\

We have also studied the effects of thermal fluctuations by adding a Gaussian white noise of diffusivity $D>0$. For an RTP in a harmonic well, it turns out that the stationary state for $D>0$ is a Gaussian convolution of the $D=0$ stationary state. This provides an exact description of how temperature rounds off turning-point singularities and extends the support of the stationary distribution of the position beyond $r_0$. This representation reveals a crossover between persistence-dominated and diffusion-dominated regimes and yields universal low-$D$ scaling forms near the turning surface. In $d=1$ and $d=2$, lowering $D$ at fixed activity produces a clear unimodal-to-bimodal crossover of the stationary density, with two maxima that approach the turning points as $D\to0$. In $d=3$, the same trend holds up to a critical value of $\alpha$, beyond which two peaks coexist and the global maximum jumps discontinuously from the origin to the outer shell at $r=r_0$.\\

Throughout the paper, we have assumed a constant run speed $v_0$ in $d>1$ (see Eq.~(\ref{ddimRTPLange})). It would be interesting to extend this framework if one allows $v_0$ to be drawn from a prescribed distribution, which modifies the projected component of the velocity. Importantly, Eq.~(\ref{MRSSX}) remains unchanged. This extension is particularly relevant for experimental systems, such as bacteria exhibiting run-and-tumble motion, in which run speeds are indeed heterogeneous~\cite{BergVelocity, Taute}.\\

Starting from the exact stationary state obtained via the Dirichlet-process (stick-breaking) representation, several extensions are natural. One is to relax rotational invariance and study anisotropic confinement, for instance, quadratic traps of the form $V(\bm r)=\frac12\sum_{i=1}^d \mu_i x_i^2$ with different $\mu_i$'s. In this setting, the stationary distribution (and moments) of each Cartesian component can be obtained exactly using Eq.~(\ref{MRSSX}). Although reconstructing the full radial distribution from these marginals is nontrivial, the isotropic stationary distribution $p_R(r)$ derived in this paper still provides a convenient baseline for perturbative treatments of the corresponding Fokker-Planck equation around the isotropic limit.\\

Finally, we expect the analytical techniques developed here to be relevant to study other types of switching dynamics, which have generated some recent interest. A notable example is switching diffusion, i.e., Brownian motion with a diffusivity that changes randomly in time according to a prescribed law~\cite{SwitchingDiff, thesis}. More broadly, the same approach should apply to linear Langevin dynamics in switching environments, including many-particle systems undergoing Brownian motion in a harmonic potential whose stiffness switches in time~\cite{SabhaMajum, switchingtrap, switchingtrapExp}. In such cases, the switching dynamics of the trap induces dynamical emergent correlations between the particles and it will be interesting to extend the techniques used here to such many-body correlated systems.

\vspace*{0.5cm}
\noindent{\bf Acknowledgments.} We acknowledge support from ANR Grant No. ANR-23-CE30-0020-01 EDIPS. MG is grateful to Christina Kurzthaler for many interesting discussions on related problems.

\appendix
\section*{Appendices}

\section{Identities and Isotropy-Based Derivations}

\subsection{Derivation of $W_{\rm{proj}}(v)$}\label{WprojApp}

{In this appendix, for the sake of completeness, we provide a derivation of the projected velocity law of an isotropic RTP $W_{\rm{proj}}(v)$ given in Eq.~(\ref{WprojEq}). This derivation can also be found in Ref.~\cite{generalRTP}}. Consider a random unit vector $\mathbf{n} = (n_1,n_2, \cdots, n_d)\in\mathbb{R}^d$ drawn uniformly from the unit sphere $S^{d-1}$, and we assume $d\geq 2$. During a run, the particle governed by the Langevin dynamics in Eq.~(\ref{ddimRTPLange}) experiences the active force $\mathbf{v} = v_0\,\mathbf{n}$. Due to isotropy, all components are statistically equivalent. Without loss of generality, we therefore focus on the first component, i.e., the scalar projection $v_1=v_0 n_1$. We want to compute its PDF which we denote as $W_{\rm{proj}}(v)$. The PDF of $\mathbf{v}$ with length $v_0$ is simply given by
\begin{equation}
    P(\mathbf{v}) = \frac{1}{\Omega_{d-1} v_0^{d-1}}\delta(|\mathbf{v}| - v_0)\, , \qquad \Omega_{d-1} = \frac{2\pi^{d/2}}{\Gamma(d/2)}\, ,
\end{equation}
where $\Omega_{d-1}$ is the surface area of the unit sphere $S^{d-1}$. Using the standard identity for the Dirac delta under a change of variables, $\delta(|\mathbf{v}| - v_0) = 2v_0 \delta(|\mathbf{v}|^2 - v_0^2)$, we can equivalently rewrite the density in terms of $|\mathbf{v}|^2$, which will be convenient when integrating out components:
\begin{equation}
P(\mathbf{v})=\frac{2}{\Omega_{d-1}v_0^{\,d-2}}\delta\!\left(|\mathbf{v}|^2-v_0^2\right)\, ,
\qquad 
|\mathbf{v}|^2=v_1^2+v_2^2+\cdots+v_d^2\, ,
\end{equation}
where $v_i = v_0n_i$. We now compute the marginal distribution of $v_1$ by integrating over the remaining components. Introducing a delta constraint $\delta(v_1-v)$ to fix the first component, we obtain
\begin{equation}
    W_{\rm{proj}}(v) = \int P(\mathbf{v})\delta(v_1-v)dv_1dv_2\cdots dv_d = \frac{2}{\Omega_{d-1} v_0^{d-2}}\int\delta(|\mathbf{v}|^2 - v_0^2)dv_2dv_3\cdots dv_d\, .
\end{equation}
To evaluate the remaining integral, we switch to hyperspherical coordinates in the $(d-1)$-dimensional subspace spanned by $(v_2,\ldots,v_d)$. Let $\rho^2=v_2^2+\cdots+v_d^2$ so that the $(d-1)$-dimensional volume element becomes $dv_2\cdots dv_d = \Omega_{d-2}\,\rho^{d-2}\,d\rho$. With this substitution the marginal reduces to
\begin{equation}
    W_{\rm{proj}}(v) = \frac{2\Omega_{d-2}}{\Omega_{d-1} v_0^{d-2}}\int_0^{+\infty}\delta\!\left(\rho^2- (v_0^2-v^2)\right)\rho^{d-2}d\rho\, .
\end{equation}
Next, we set $u=\rho^2$, so that $du=2\rho\,d\rho$ and hence
\begin{equation}
    W_{\rm{proj}}(v) = \frac{\Omega_{d-2}}{\Omega_{d-1} v_0^{d-2}}\int_0^{+\infty}\delta\!\left(u- (v_0^2-v^2)\right)u^{\frac{d-3}{2}}du\, .
\end{equation}
Performing the integration and simplifying the prefactor, we obtain the result given in the main text
\begin{equation}
    W_{\rm proj}(v) \;=\; \frac{1}{v_0}\,\frac{\Gamma\!\bigl(\tfrac{d}{2}\bigr)}{\sqrt{\pi}\,\Gamma\!\bigl(\tfrac{d-1}{2}\bigr)}\,
    \left(1-\frac{v^2}{v_0^2}\right)^{\frac{d-3}{2}}, \qquad -v_0\le v\le v_0 \, .
    \label{AppWproj}
\end{equation}

\subsection{Identities Between $p_X(x)$ and $p_R(r)$}\label{radial_single_App}

As the motion is isotropic, all components of the $d$-dimensional position $x_i$'s  are identically distributed. 
Focusing on the first component $x \equiv x_1$, we have 
\begin{equation}
    X = R\,n_1\, ,
    \label{xProjApp}
\end{equation}where $n_1$ is the first component of a random unit vector 
$\mathbf{n}=(n_1,n_2,\ldots,n_d)\in\mathbb{R}^d$ drawn uniformly from the unit sphere $S^{d-1}$ ($d\ge 2$). 
From the above derivation of $W_{\rm proj}(v)$ in Eq.~(\ref{AppWproj}), it follows that the conditional density of $X=x$ at fixed radius $R=r$ 
is given by the same projection law with the replacement $v_0\to r$, namely
\begin{equation}
p(x|r)
=\frac{1}{r}\,\frac{\Gamma\!\left(\frac d2\right)}{\sqrt{\pi}\,\Gamma\!\left(\frac{d-1}{2}\right)}
\left(1-\frac{x^2}{r^2}\right)^{\frac{d-3}{2}},
\qquad -r\le x\le r\, .
\end{equation}
Averaging over $R$, whose PDF $p_R(r)$ is supported on $0\le r\le v_0/\mu$, we obtain the marginal PDF of $x$ {(see also~\cite{MoriCondensation})}
\begin{equation}
p_X(x) \;=\; \frac{\Gamma\!\bigl(\tfrac{d}{2}\bigr)}{\sqrt{\pi}\,\Gamma\!\bigl(\tfrac{d-1}{2}\bigr)}
\int_{|x|}^{v_0/\mu} dr \, \frac{p_R(r)}{r}\left(1-\frac{x^2}{r^2}\right)^{\frac{d-3}{2}} \, .
\label{distributionprojection}
\end{equation}
In particular, $p(x)=0$ for $|x|>v_0/\mu$. We are now interested in inverting Eq.~(\ref{distributionprojection}) in $d=2$, and $d=3$, to obtain a relation for $p_R(r)$ given $p_X(x)$.\\

\noindent\textbf{Inversion in $d=2$.} 
Let us focus on the case $x>0$ (this is sufficient since $p_X(x)$ is symmetric). We have
\begin{equation}
    p_X(x) = \frac{1}{\pi} \int_{x}^{v_0/\mu} dr \, \frac{p_R(r)}{\sqrt{r^2-x^2}} \, .
\end{equation}
Now introduce $u=x^2$, $v=r^2$, $u^* = (v_0/\mu)^2$, and the functions
\begin{equation}
    g(u) = \pi p_X(\sqrt{u})\, , \qquad f(v) = \frac{p_R(\sqrt{v})}{2\sqrt{v}}\, ,
    \label{changeFct}
\end{equation}
so that
\begin{equation}
    g(u) = \int_{u}^{u^*} dv \, \frac{f(v)}{\sqrt{v-u}}\, , \qquad 0\leq u \leq u^* \, .
\end{equation}
Define
\begin{equation}
I(s) := \int_{s}^{u^*} du \frac{g(u)}{\sqrt{u-s}}= \int_{s}^{u^*}du \frac{1}{\sqrt{u-s}}
\left(\int_{u}^{u^*} \frac{f(v)}{\sqrt{v-u}}\,dv\right)\, ,
\qquad 0 \le s \leq u^* .
\end{equation}
On the integration domain we have $s\leq u\leq v \leq u^*$. Swapping the order of integration yields
\begin{equation}
I(s)
= \int_{s}^{u^*} f(v)\,
\underbrace{\left(\int_{s}^{v} \frac{du}{\sqrt{(u-s)(v-u)}}\right)}_{=\ \pi}
\,dv .
\label{eq:I-underbrace}
\end{equation}
Hence, we obtain
\begin{equation}
    I'(s) = -\pi f(s) \implies f(s) = -\frac{1}{\pi} \frac{d}{ds}\left(\int_{s}^{u^*} du \frac{g(u)}{\sqrt{u-s}}\right)\, .
\end{equation}
Performing the change of variable $u=s+t^2$, and applying Leibniz’s rule for differentiation under the integral sign yields
\begin{equation}
f(s) = -\frac{1}{\pi}\int_{s}^{u^*}du \frac{g'(u)}{\sqrt{u-s}}\, ,
\qquad 0 \leq s \leq u^*\, .
\label{inversion-f}
\end{equation}
Reverting to the original variables, we thus arrive at the result~\eqref{pXtopRd2d3} reported in the main text
\begin{equation}
    p_R(r) = -2r\int_r^{\frac{v_0}{\mu}} dx\, \frac{p_X'(x)}{\sqrt{x^2-r^2}} \, ,\qquad 0 \leq r \leq v_0/\mu\, .
\end{equation}

\noindent\textbf{Inversion in $d=3$.} In this case, the calculation is straightforward. Setting $d=3$ in Eq.~(\ref{distributionprojection}) gives
\begin{equation}
    p_X(x) \;=\; \frac{1}{2}
\int_{|x|}^{v_0/\mu} dr \, \frac{p_R(r)}{r}\, .
\end{equation}
Differentiating with respect to $x$ then yields the inversion formula given in~\eqref{pXtopRd2d3}, i.e.,
\begin{equation}
    p_R(r)=-2r\, p'_X(r)\, .
\end{equation}

\subsection{Moments}\label{momentsXRapp}
From Eq.~(\ref{xProjApp}), $X=R\, n_1$, and using that the radial variable $R$ and the uniform
orientation $\mathbf n$ are independent, the moments factorize as
\begin{equation}
\langle |X|^{n}\rangle=\langle R^{n}\rangle\,\langle |n_1|^{n}\rangle\, ,
\qquad
\langle |n_1|^{n}\rangle=
\frac{\Gamma\!\left(\frac d2\right)\Gamma\!\left(\frac{n+1}{2}\right)}
{\Gamma\!\left(\frac12\right)\Gamma\!\left(\frac{n+d}{2}\right)}\, .
\label{eq:A20}
\end{equation}
Here, $\langle |n_1|^{n}\rangle$ is the $n$-th moment of the projection law $W_{\rm proj}$ in Eq.~(\ref{AppWproj}) evaluated at $v_0=1$. Eliminating $\langle |n_1|^{n}\rangle$
in Eq.~\eqref{eq:A20} yields
\begin{equation}
\langle R^{n}\rangle=
\frac{\Gamma\!\left(\frac12\right)\Gamma\!\left(\frac{n+d}{2}\right)}
{\Gamma\!\left(\frac d2\right)\Gamma\!\left(\frac{n+1}{2}\right)}\, \langle |X|^{n}\rangle .
\label{MomentsApp}
\end{equation}
The even moments of $R$ follow directly from the moments of $X$ given in Eq.~(\ref{exactmoments}):
\begin{equation}
\langle R^{2n}\rangle=
\frac{\Gamma\!\left(\frac12\right)\Gamma\!\left(n+\frac{d}{2}\right)}
{\Gamma\!\left(\frac d2\right)\Gamma\!\left(n+\frac{1}{2}\right)}\, \langle X^{2n}\rangle .
\label{MomentsEven}
\end{equation}

\subsection{Relation Between the Joint Law and the Radial Law}\label{jointToRadial}
Using the invariance of the system under rotation, the joint density of an RTP in $d$ dimensions depends only on the
radius $r=\sqrt{x_1^2+\cdots+x_d^2}$, so there exists a function $f_R$ such that
\begin{equation}
    P(x_1,\ldots,x_d)=f_R\!\left(r=\sqrt{x_1^2+\cdots+x_d^2}\right).
\end{equation}
Let $R=\|X\|$ denote the radial component and let $p_R(r)$ be its density.
Using spherical coordinates, we have
\begin{equation}
    \mathbb{P}(R<r)= \int_{\|x\|<r} P(x)\,d^dx \nonumber= \int_{0}^{r}\int_{S^{d-1}} f_R(\rho)\,\rho^{d-1}\, d\rho\, d\Omega_{d-1} = \Omega_{d-1}\int_{0}^{r} f_R(\rho)\,\rho^{d-1}\,d\rho\, ,
\end{equation}
where $\Omega_{d-1}$ is the surface area of the unit
sphere in $\mathbb{R}^d$. Differentiating with respect to $r$ yields
\begin{equation}
    f_R(r)=\frac{p_R(r)}{\Omega_{d-1}\,r^{d-1}},
\end{equation}
which is exactly relation~(\ref{generalJointPDF}) in the main text.

\section{Limit $D\to 0$ of the Radial Distribution in Two Dimensions}\label{2dDto0App}
Here, we show that the two dimensional radial distribution in the presence of a diffusive thermal noise with amplitude $\sqrt{2D}$ in Eq. (\ref{pR2dDneq0}) reduces to the one without diffusion in the limit $D\to 0$ in Eq. (\ref{MR2}). We have
\begin{eqnarray}\label{apppr2dproof}
    p_R\!\left(\tilde r = \frac{\mu r}{v_0}\right) =  \sqrt{\frac{2}{D\mu}} \frac{v_0^2}{ \pi D} \frac{\Gamma\!\left(1+\alpha\right)}{ \Gamma\!\left(\frac{1}{2}+\alpha\right)} \, \tilde r \int_{\tilde r}^{+\infty}\frac{d\tilde z}{\sqrt{\tilde z^2-\tilde r^2}}\int_{-1}^{1}d\tilde x\,  (\tilde z - \tilde x)(1-\tilde x^2)^{\alpha-\frac{1}{2}}\, e^{- \frac{v_0^2}{2 D \mu}(\tilde x- \tilde z)^2}\, .
\end{eqnarray}
We perform the change of variable
\begin{eqnarray}
    u = \frac{v_0}{\sqrt{2D\mu}}(\tilde x - \tilde z) \Longleftrightarrow \tilde x =\tilde z +\frac{\sqrt{2D\mu}}{v_0} u\, .
\end{eqnarray}
The inner integral becomes
\begin{eqnarray}\label{proofappinterm2dpR}
    \int_{-1}^{1}d\tilde x\,  (\tilde z - \tilde x)(1-\tilde x^2)^{\alpha-\frac{1}{2}}\, e^{- \frac{v_0^2}{2 D \mu}(\tilde x- \tilde z)^2} = -\frac{{2D\mu}}{v_0^2}\int_{-\frac{v_0}{\sqrt{2D\mu}}(1+\tilde z)}^{\frac{v_0}{\sqrt{2D\mu}}(1-\tilde z)}du\, u\left[1-\left(\tilde z +\frac{\sqrt{2D\mu}}{v_0} u\right)^2\right]^{\alpha-\frac{1}{2}}\, e^{-u^2}\, .
\end{eqnarray}
For small $D$, making the expansion
\begin{eqnarray}
  \left[1-\left(\tilde z +\frac{\sqrt{2D\mu}}{v_0} u\right)^2\right]^{\alpha-\frac{1}{2}} &\approx& (1-\tilde z^2)^{\alpha -\frac{1}{2}}\left[1-\left(\frac{2\tilde z}{1-\tilde z^2}\frac{\sqrt{2D\mu}}{v_0} u\right)\right]^{\alpha-\frac{1}{2}}\\
  &\approx& (1-\tilde z^2)^{\alpha -\frac{1}{2}} -\left(\alpha-\frac{1}{2}\right)\frac{2\sqrt{2D\mu}}{v_0}\tilde z u(1-\tilde z^2)^{\alpha -\frac{3}{2}}\, ,
\end{eqnarray}
we find, after performing Gaussian integrals,
\begin{eqnarray}
   - \frac{{2D\mu}}{v_0^2}\int_{-\frac{v_0}{\sqrt{2D\mu}}(1+\tilde z)}^{\frac{v_0}{\sqrt{2D\mu}}(1-\tilde z)}du\, u\left[1-\left(\tilde z +\frac{\sqrt{2D\mu}}{v_0} u\right)^2\right]^{\alpha-\frac{1}{2}}\, e^{-u^2} \approx \frac{(2D\mu)^{\frac{3}{2}}\sqrt{\pi}}{v_0^3}\left(\alpha-\frac{1}{2}\right)\tilde z(1-\tilde z^2)^{\alpha -\frac{3}{2}}\, .
\end{eqnarray}
Note that, in the limit $D\to 0$, the lower bound of the $u$-integral in Eq.~(\ref{proofappinterm2dpR}) tends naturally to $-\infty$, since $\tilde z>0$. The behavior of the upper bound depends on $\tilde z$: it tends to $+\infty$ for $\tilde z<1$, while it tends to $-\infty$ for $\tilde z>1$. In the latter case, the Gaussian contribution vanishes. Therefore, the limit $D\to0$ restricts the relevant range to $\tilde z\in[\tilde r,1]$, so that the upper bound of the $\tilde z$-integral in Eq.~(\ref{apppr2dproof}) becomes $1$.
Hence,
\begin{eqnarray}
    p_R\!\left(\tilde r = \frac{\mu r}{v_0}\right) = \frac{4\mu}{ v_0\sqrt{\pi} } \frac{\Gamma\!\left(1+\alpha\right)}{ \Gamma\!\left(\frac{1}{2}+\alpha\right)} \left(\alpha-\frac{1}{2}\right)\, \tilde r \int_{\tilde r}^{1}\frac{d\tilde z}{\sqrt{\tilde z^2-\tilde r^2}} \tilde z(1-\tilde z^2)^{\alpha -\frac{3}{2}}\, .
\end{eqnarray}
We then use the identity 
\begin{eqnarray}\label{betaidentity}
   \left(\alpha-\frac{1}{2}\right)\, \tilde r \int_{\tilde r}^{1}\frac{d\tilde z}{\sqrt{\tilde z^2-\tilde r^2}} \tilde z(1-\tilde z^2)^{\alpha -\frac{3}{2}}
=
\frac{\sqrt{\pi}}{2}
\frac{\Gamma\left(\alpha+\frac{1}{2}\right)}{\Gamma(\alpha)}
\left(1-\tilde r^{2}\right)^{\alpha-1}\, .
\end{eqnarray}
Note that for $0<\alpha\le 1/2$, the integral in Eq.~(\ref{betaidentity}) is not convergent as an ordinary improper integral because of the endpoint singularity at $\tilde z=1$. In this range, Eq.~(\ref{betaidentity}) is understood through the analytic continuation of its right-hand side, which is well defined for all $\alpha>0$. Finally, we obtain
\begin{equation}
p_R(r)  = 2\alpha \frac{\mu^2 r}{v_0^2}
\left[ 1 - \left( \frac{\mu r}{v_0} \right)^2 \right]^{\alpha - 1}\, ,
\qquad 0 \le r \le \frac{v_0}{\mu}\, ,
\end{equation}
which is exactly the expected result given in Eq.~(\ref{MR2}).
\section{Moment Generating Function of the Generalized RTP}\label{MGFapp}

\subsection{Kesten Approach}\label{KestenApp}

In Section~\ref{KestenSubsect}, we derived an integral equation satisfied by the stationary state distribution via the Kesten relation~(\ref{Kest_rel}). We recall it here (see Eq. (\ref{kesten.RTPG1})):
\begin{equation}
p_X(x)= \int dU \int dV \int  dx'\, \alpha\, U^{\alpha-1}\, \frac{\mu}{1-U}\, W\left(\frac{\mu V}{1-U}\right)\, 
p_X(x')\,  \delta(x-U\, x'-V)\, .
\label{kesten.RTPG1_App}
\end{equation}
We also recall that $U\in [0,1]$ and $V \in \big[(1-U)v_{\min}/\mu,\,(1-U)v_{\max}/\mu\big]$. We introduce the moment generating function (MGF) as the bilateral Laplace transform (BLT) of the stationary state of the RTP, i.e.,
\begin{equation}
    \tilde{p}_X(q) = \langle e^{qx} \rangle = \int_{-\infty}^{+\infty}dx\,  e^{qx} p_X(x)\, .
\label{MGF_BLT}
\end{equation}
Taking the BLT of Eq.~(\ref{kesten.RTPG1_App}) yields
\begin{equation}
    \tilde{p}_X(q)  = \int dU \int dV \int  dx'\, \alpha\, U^{\alpha-1}\, \frac{\mu}{1-U}\, W\left(\frac{\mu V}{1-U}\right)\, 
p_X(x')\, e^{q(Ux'+V)}\, .
\end{equation}
Performing the integral over the variable $x'$ leads to
\begin{equation}
    \tilde{p}_X(q)  = \int dU \int dV\, \alpha\, U^{\alpha-1}\, \frac{\mu}{1-U}\, W\left(\frac{\mu V}{1-U}\right)\, 
\tilde{p}_X(qU) \, e^{qV}\, .
\end{equation}
Next, integrating over $V$ gives
\begin{equation}
    \tilde{p}_X(q)  =\int_{0}^{1} dU \alpha\, U^{\alpha-1}\, \tilde W\left(\frac{1-U}{\mu}q\right)\, 
\tilde{p}_X(qU) \, .
\end{equation}
The change of variable $\bar q = qU$ allows for further analytical progress. We have
\begin{equation}
    q\left[q^{\alpha -1}\tilde{p}_X(q)\right]  =\alpha \int_{0}^{q} d\bar q\left[\bar q^{\alpha -1}\tilde{p}_X(\bar q)\right]\, \tilde W\left(\frac{q-\bar q}{\mu}\right)\, .
    \label{MGFintermediate}
\end{equation}
Introducing the function $f_\alpha(q) = q^{\alpha -1}\tilde{p}_X(q)$ and the Laplace transform
\begin{equation}
    \hat f_\alpha(s)= \mathcal{L}_{q\to s}\left[f_\alpha(q)\right]=\int_0^{+\infty}dq\, e^{-sq}f(q)\, ,
\end{equation}
and taking the Laplace transform of Eq.~(\ref{MGFintermediate}) -- using the property $\mathcal{L}_{q\to s}\left[qf_\alpha(q)\right] = - \hat f_\alpha'(s)$ -- we find
\begin{equation}
    \hat f_\alpha'(s) = -\alpha \mu \hat f_\alpha(s) \hat W(\mu s)\, .
\end{equation}
Solving this differential equation yields
\begin{equation}
    \tilde{p}_X(q) \propto q^{1-\alpha}\mathcal{L}^{-1}_{s\to q}\left[\exp\left(-\alpha \mu \int_0^s du\, \hat W(\mu u)\right)\right]\, ,
\label{solODEMGF}
\end{equation}
where $\mathcal{L}^{-1}$ denotes the inverse Laplace transform. The multiplicative factor is finally determined by the normalization condition $\tilde p_X(0)=1$ and is given by $\Gamma(\alpha)$.
The solution~(\ref{solODEMGF}) was first derived in Ref.~\cite{thesis}. Finally, the argument of the exponential can be rewritten as
\begin{equation}
    \int_0^s du\, \hat W(\mu u) = \int_0^s du \int_0^{+\infty}dq\,  e^{-\mu u q}\int_{v_{\text{min}}}^{v_{\text{max}}}dv \, e^{qv\, }W(v) =\frac{1}{\mu}\int_{v_{\text{min}}}^{v_{\text{max}}}dv \,  \log\!\left( s - \frac{v}{\mu} \right) W(v)  \, .
\end{equation}
Hence,
\begin{equation}
    \tilde{p}_X(q) =\Gamma(\alpha)\,  q^{1-\alpha}\mathcal{L}^{-1}_{s\to q}\left[\exp\left(-\alpha\int_{v_{\text{min}}}^{v_{\text{max}}}dv \,  \log\!\left( s - \frac{v}{\mu} \right) W(v)\right)\right]\, .
\end{equation}
It can be straightforwardly shown that this is equivalent to
\begin{eqnarray}
\tilde{p}_X(q)
= \Gamma\!\left( \alpha\right)
\int_{\Gamma_B} \frac{ds}{2\pi i}\, 
\exp\!\left(s -\alpha \int_{v_{\min}}^{v_{\max}} dv\, \log\!\left( s - \frac{v}{\mu}q \right) W(v) \right)\, ,
\label{MGF}
\end{eqnarray}
where $\Gamma_B$ is a Bromwich contour. From Eq.~(\ref{MGF}), one can readily verify that $\tilde p_X(0)=1$.

\subsection{Moments in The Stationary State}\label{MomentApp} To extract the expression for the moments from the integral representation of the MGF~(\ref{MGF}), we first expand the logarithm inside the exponential. This yields
\begin{eqnarray}
\tilde{p}_X(q)
= \Gamma\!\left( \alpha\right)
\int_{\Gamma_B} \frac{ds}{2\pi i}\, 
e^s s^{-\alpha}\exp\!\left[\alpha \sum_{n=1}^{+\infty} \frac{\langle v^n \rangle}{n} \left(\frac{q}{\mu s}\right)^n \right]\, ,
\end{eqnarray}
where $\langle v^n \rangle$ are the moments of $W(v)$. The exponential inside the integral is precisely the generating function of the complete Bell polynomials $B_n$~\cite{BellPoly}. It can therefore be rewritten as
\begin{equation}
   \exp\!\left[\alpha \sum_{n=1}^{+\infty} \frac{\langle v^n \rangle}{n} \left(\frac{q}{\mu s}\right)^n \right] = \sum_{n=1}^{+\infty} \frac{q^n}{n!}\left(\frac{1}{\mu s}\right)^n B_n\left(1!\, \alpha\, \frac{\langle v \rangle}{1}, \ldots,n!\, \alpha\, \frac{\langle v^{n} \rangle}{n} \right) \;.
\end{equation}
The contour integral over $s$ can then be evaluated using the standard identity
\begin{equation}
    \int_{\Gamma_B} \frac{ds}{2\pi i}\, 
e^s s^{-\alpha-n} = \frac{1}{\Gamma(\alpha+n)}
\end{equation}
Finally, using the definition of the moment generating function,
$\tilde{p}_X(q)
=\sum_{n=1}^{+\infty} (q^n/n!)\,  \langle X^n\rangle$,
we identify the coefficients of $q^n$ and obtain the exact expression for the moments:
\begin{eqnarray}
\langle X^{n} \rangle = \frac{1}{\mu^n} \frac{\Gamma(\alpha)}{\Gamma(\alpha+n)} \;
B_n\left(1!\, \alpha\, \frac{\langle v \rangle}{1}, \ldots,n!\, \alpha\, \frac{\langle v^{n} \rangle}{n} \right)\,,
\label{exactmoments_app}
\end{eqnarray}
which is the formula given in Eq. (\ref{exactmomentsMR}).

\section{Dirichlet Process}\label{AppDirichlet}

We have shown in Section~\ref{KestenSectionExplicit} that the position in the stationary state is a random variable that can be written as follows:
\begin{eqnarray}
    X = \frac{1}{\mu} \sum_{n\geq1} {\sf{v}}_n Y_n  \, , \qquad \qquad  Y_n = \left[ \prod_{j<n} (1 - \overline{U}_j) \right] \overline{U}_n\, .
    \label{x_yn_def_App}
\end{eqnarray}
For a given trajectory of the RTP, one obtains a specific realization of the velocities $\{{\sf v}_1, {\sf v}_2, \ldots, {\sf v}_n\}$ (where ${\sf v}_i$'s are drawn from $W({\sf v}_i)$), which can be associated with the corresponding weights $\{Y_1, Y_2, \ldots, Y_n\}$ of a stick-breaking process (see Fig.~\ref{SBP_fig}). It is then possible to introduce the random empirical measure
\begin{equation}
G(v) = \sum_{n\ge1} Y_n\, \delta(v - {\sf v}_n)
= Y_1\, \delta(v - {\sf v}_1) + Y_2\, \delta(v - {\sf v}_2) + \cdots\, .
\label{DirichletProcess_App}
\end{equation}
With this definition, the position $x$ can be expressed as the mean of $v/\mu$ with respect to $G(v)$, namely,
\begin{equation}
X = \int_{v_\mathrm{min}}^{v_\mathrm{max}} dv\, \frac{v}{\mu}\, G(v)\, .
\label{xDirichlet_App}
\end{equation}
The random measure $G(v)$ is precisely a realization of a Dirichlet process with concentration parameter $\alpha$ and base distribution $W(v)$~\cite{Ferguson, Sethuraman}: we denote this as $G(v) \sim\text{DP}(\alpha, W)$.
Correspondingly, the position $x$ can be interpreted as the expectation of the function $v/\mu$ with respect to this random empirical measure -- that is, a linear mean functional of a Dirichlet process in the terminology of the mathematical literature~\cite{CifarelliRegazzini}.\\

In Ref.~\cite{CifarelliRegazzini}, Cifarelli and Regazzini have derived the cumulative distribution of a linear mean functional of a Dirichlet process (see Theorem 1 of~\cite{CifarelliRegazzini}), i.e., the cumulative distribution $M(x)$ of $x$ given in Eq.~(\ref{xDirichlet_App}). It is given by
\begin{eqnarray}\label{MTcumu_App}
    M(x) = \frac{1}{\pi}\,\int_{v_{\min}/\mu}^{x}dt\,  (x - t)^{\alpha - 1}\phi_{\alpha}(t)\, ,\qquad 
\phi_{\alpha}(t) = \sin\!\left( \pi \alpha \int_{v_{\min}}^{\mu t}dv\, W(v) \right)
\exp\!\left( -\alpha \int_{v_{\min}}^{v_{\max}} dv\, 
\log\!\left| t - \tfrac{v}{\mu} \right|\, W(v)\right)\, . \quad\end{eqnarray}
Using that $(x-t)^{\alpha-1} = -\frac{1}{\alpha}\frac{d}{dt}(x-t)^{\alpha}$ and integrating by parts, one can then compute $p_X(x) = M'(x)$ to obtain the stationary state distribution given in the main text in Eq.~(\ref{SS_RTP_continuous}).

\vspace*{0.5cm}
\noindent{\bf Cifarelli–Regazzini Identity for the MGF.} An important property of the mean functional of a Dirichlet process $\mathrm{DP}(\alpha, W)$ is given by the Cifarelli–Regazzini identity.
For any $\alpha > 0$, $s > -v_{\min}/\mu$, and $W(v)$ (discrete or continuous), this identity yields the expression for the following expectation value, depending on $X$ as defined in Eq.~(\ref{xDirichlet_App}):
\begin{eqnarray}
    \langle(s + X)^{-\alpha}\rangle 
=  \exp\!\left(
    -\alpha \int_{v_\mathrm{min}}^{v_\mathrm{max}} dv\, \log\!\left( s + \frac{v}{\mu}\right) W(v) 
\right)\, .
\label{CRidentity}
\end{eqnarray}
From this identity, we now derive an expression for the MGF of the stationary state of the RTP defined in Eq.~(\ref{MGF_BLT}). Using the following identity,
\begin{eqnarray}
(s+X)^{-\alpha} 
= \frac{1}{\Gamma(\alpha)} \int_{0}^{\infty}dq\, e^{-q s} q^{\alpha-1}  e^{-q X}\, ,
\end{eqnarray}
and taking the expectation, we obtain
\begin{eqnarray}
\Gamma(\alpha) \,  \langle(s + X)^{-\alpha}\rangle
= \int_{0}^{\infty} e^{-q s} q^{\alpha-1} 
\tilde{p}_X(-q) \, dq\, ,
\end{eqnarray}
where we recognize the right-hand side as the Laplace transform in $s$ of $q^{\alpha-1}\, \tilde{p}_X(-q)$.
Therefore, taking the inverse Laplace transform and using the Cifarelli–Regazzini identity~(\ref{CRidentity}) yields
\begin{eqnarray}
\tilde{p}_X(q)
&=& \Gamma\!\left( \alpha \right)\left(-q\right)^{1-\alpha} 
\mathcal{L}^{-1}_{s \to -q} \!\left[
\exp\!\left(
    -\alpha \int_{v_\mathrm{min}}^{v_\mathrm{max}} dv\, \log\!\left( s + \frac{v}{\mu} \right) W(v) 
\right)
\right]\, ,
\end{eqnarray}
where $\mathcal{L}^{-1}$ is the inverse Laplace transform. It is possible to show that it can be rewritten exactly as~Eq.~(\ref{MGF}) which is exactly the solution that we have found via the Kesten approach.

\section{Derivation of the Beta Distribution in $d = 2$}\label{BetaApp}

In Section~\ref{2dSSSection}, we showed that the stationary distribution of a component of an RTP in two dimensions can be expressed as
\begin{equation}
p_X(x)
= \frac{\alpha\mu}{\pi}\!\left(\frac{2\mu}{v_0}\right)^{\alpha}
  \int_{-\frac{v_0}{\mu}}^{x}dt\, 
  (x-t)^{\alpha-1}\psi(t)\, , \qquad \qquad \psi(t) = \frac{\cos\!\left(\tfrac{\pi\alpha}{2}+\alpha\,\arcsin\!\tfrac{\mu t}{v_0}\right)}{\sqrt{v_0^2-\mu^2 t^2}}
\,.
\label{appPxd2}
\end{equation}
It turns out that this integral over $t$ in Eq.~(\ref{appPxd2}) can be performed explicitly. To see that, we first notice that $\psi(t)$ in Eq. (\ref{appPxd2}) can be written as a hypergeometric function, namely
\bea \label{psi_2F1}
\psi(t) = \frac{\cos\!\left(\tfrac{\pi\alpha}{2}+\alpha\,\arcsin\!\tfrac{\mu t}{v_0}\right)}{\sqrt{v_0^2-\mu^2 t^2}} = \frac{1}{v_0\sqrt{2\left(1+\frac{\mu t}{v_0}\right)}}\, \,_2F_1 \left(\frac{1}{2}+ \alpha, \frac{1}{2}-\alpha, 1/2; \frac{1+\frac{\mu t}{v_0}}{2} \right) \;, 
\eea
where we recall that the hypergeometric function $\,_2F_1(a,b,c;z)$ admits the series expansion
\bea \label{def_2F1}
\,_2F_1(a,b,c;z) = \sum_{k=0}^\infty \frac{(a)_k\, (b)_k}{(c)_k}\, z^k \;,
\eea
where $(x)_n = \Gamma(x+1)/\Gamma(x-n+1)$ is the Pochhammer symbol. This can be shown by observing that $\psi(t)$ satisfies a second-order differential equation of the hypergeometric type, namely
\bea \label{edif}
(\mu^2 t^2-v_0^2) \psi''(t) + 3\mu^2 t \psi'(t) + \mu^2(1-\alpha^2) \psi(t) = 0 \;,
\eea
and eventually rewriting this equation in terms of the variable $u=(1+\mu t/v_0)/2$.\\

Substituting this expression (\ref{psi_2F1}) in (\ref{appPxd2}) to compute $p(x =v_0(-1+v)/\mu)$ and performing the change of variable $t= v_0(-1 + v y)/\mu$ one finds
\bea \label{pofv}
p_X(x =v_0(-1+v)/\mu) = \frac{\alpha}{\pi} 2^{\alpha - 1/2} \frac{\mu}{v_0} v^{\alpha -1/2} \int_0^1 dy (1-y)^{\alpha - 1} \frac{1}{\sqrt{y}} \,_2 F_1 \left(\frac{1}{2}+\alpha, \frac{1}{2}-\alpha, \frac{1}{2}, \frac{v\, y}{2} \right) \;.
\eea
We now insert the series expansion (\ref{def_2F1}) in (\ref{pofv}) and integrate term by term, using the identity
\begin{eqnarray}
\int_0^1 dy (1-y)^{\alpha -1} y^{k-1/2} = \frac{\Gamma(\alpha) \Gamma(1/2+k)}{\Gamma(1/2+\alpha+k)}    \label{identity} \;.
\end{eqnarray}
One gets, after simplification of the Pochhammer symbols and gamma functions
\bea \label{pofv2}
p_X(x =v_0(-1+v)/\mu) = \frac{\alpha}{\pi} 2^{\alpha-1/2}\frac{\mu}{v_0}  v^{\alpha-1/2} \sum_{k=0}^\infty \left( \frac{v}{2}\right)^k \frac{\cos{(\alpha \pi)} \Gamma(\alpha) \Gamma(1/2-\alpha+k)}{\sqrt{\pi}\, \Gamma(k+1)} \;.
\eea
The sum over $k$ can then be performed straightforwardly and we obtain
\bea
\label{pofv3}
p_X(x =v_0(-1+v)/\mu) = \frac{\alpha}{\pi^{3/2}} \frac{\mu}{v_0}\cos{(\alpha \pi)} \Gamma(1/2-\alpha) \Gamma(\alpha) v^{\alpha-1/2}(2-v)^{\alpha - 1/2} \;.
\eea
Returning back to the variable $v = 1 + \frac{\mu x}{v_0}$ and using the reflection formula $\Gamma(1/2 + \alpha) \Gamma(1/2-\alpha) = \pi/\cos(\pi \alpha)$, one finally finds
\bea \label{pofx_final}
p_X(x) = \frac{\mu}{v_0}\frac{\Gamma(1+\alpha)}{\sqrt{\pi}\, \Gamma(1/2+\alpha)}\, \left[1-\left(\frac{\mu x}{v_0}\right)^2\right]^{\alpha - 1/2} \quad, \quad |x| < \frac{v_0}{\mu} \;,
\eea
which is the formula given in Eq. (\ref{2dpX}) in the main text.

\section{Behaviors of $p_X(x)$ and $p_R(r)$ Close to the Turning Point in a General Spherically Symmetric Potential in $d>1$}\label{generalpotSec}

In this section, we study the case of an RTP moving in a $d$-dimensional spherically symmetric potential of the form
\bea \label{spherical_V}
V(x_1, \cdots, x_d) = \tilde V(r) =\tilde V\left(\sqrt{x_1^2+ \cdots + x_d^2}\right)
\eea
where $\tilde V(r)$ is a smooth function that grows as $r \to \infty$, typically $\tilde V (r) \simeq a\, r^p$ with $a >0$ and $p>0$. The equations of motion read
\bea \label{gen_Langevin}
\frac{d x_i}{dt} = f_i(x_1, \cdots, x_d)+ v_i(t) + \sqrt{2D}\, \eta_i(t) \;,
\eea
where $v_i(t)$ is a piecewise constant velocity that changes value at the tumbling times, as studied before -- see Section \ref{GeneralizedRTPSection} -- and $\eta_i(t)$ are independent Gaussian white noises satisfying
\begin{equation} \label{GWN}
    \langle \eta_i(t)\rangle=0\, ,\qquad \langle \eta_i(t)\eta_j(t')\rangle=\delta_{ij}\,\delta(t-t')\, .
\end{equation} 
In Eq. (\ref{gen_Langevin}), $f_i(x_1, \cdots, x_d)$ denotes the force field, namely
\bea \label{def_force}
f_i(x_1, \cdots, x_d) = - \frac{\partial V(x_1, \cdots, x_d)}{\partial x_i} = - \frac{x_i}{\sqrt{x_1^2 + \cdots + x_d^2}} \tilde V'\left(\sqrt{x_1^2+ \cdots + x_d^2}\right) = - \frac{x_i}{r} \tilde V'(r) \;.
\eea

\vspace*{0.5cm}
\noindent{\bf The case $D=0$.} In this case, the stationary density (if it exists) has support inside the ball of radius $r_0$ such that $\tilde V'(r_0) = v_0$. Let us first focus on the stationary PDF of the component $p(x_1)$ -- of course, from the spherical symmetry the PDFs of all the components are identical. To obtain the singularity of $p(x_1)$ near $x_1 = r_0$, we argue that this singularity is due to trajectories where the velocity of the particles remain close to $v_0$, i.e., $v_i(t) \simeq v_0$ in Eq.~(\ref{gen_Langevin}) (in $d=1$ this is a strict equality) for a long time. In this case, the position of the particle will be close to the hyper-sphere of radius $r_0$. With no loss of generality, we assume that the position of the particle is close to the vector $(r_0,0,\cdots, 0)$. By parameterising the position of the particle by ${\bf r}(t) = (r_0 + \epsilon_1(t), \epsilon_2(t), \cdots, \epsilon_d(t))$ with $|\epsilon_i(t)| \ll 1$, linearizing the equation of motion (\ref{gen_Langevin}) and assuming that $v_1(t) \equiv v_1 \simeq v_0$ independent of $t$, one finds (for $D=0$) 
\bea \label{gen_Langevin_linear}
\frac{d \epsilon_1}{dt} = f_1(r_0,0,\cdots,0) + \sum_{j=1}^d \epsilon_j \frac{\partial f_1}{\partial x_j} \Big \vert_{(r_0,0,\cdots,0)} + v_1 \;.
\eea
Because of the spherical symmetry of the potential, it is easy to see from Eq. (\ref{def_force}) that 
\bea \label{deriv_force}
\frac{\partial f_1}{\partial x_j} \Big \vert_{(r_0,0,\cdots,0)} = 0 \quad {\rm for} \quad j=2,\cdots,d \;,
\eea
while 
\bea \label{fprime0}
f'_0 =  \frac{\partial f_1}{\partial x_1} \Big \vert_{(r_0,0,\cdots,0)} \neq 0 \;.
\eea
Here we consider the case where $f_0' = - |f'_0|<0$, which corresponds to a stable fixed point \cite{MFPT_1D_RTP, MFPTlong, leo_active_dbm}. Hence Eq.~(\ref{gen_Langevin_linear}) simply becomes
\bea \label{gen_Langevin_linear2}
\frac{d \epsilon_1}{dt} = f'_0 \epsilon_1 + v_1 - v_0 \;.
\eea
Integrating over a time interval $\tau$ one gets
\bea \label{gen_Langevin_linear2}
r_0-x_1(\tau)  = -\epsilon_1(\tau) = \frac{v_0 - v_1}{|f_0'|} + A\, e^{-|f_0'| \tau} \;,
\eea
where $A$ is some constant, independent of $v_1$, which is irrelevant for the present argument. If $\tau$ is distributed according to an exponential $p(\tau) = \gamma e^{-\gamma \tau}$ the random variable $Y = A\, e^{f_0' \tau} = A\, e^{-|f_0'| \tau}$ has a power law behavior for small $y$, namely $P_Y(y) \sim y^{-1-\gamma/(f'_0)} = y^{-1+\gamma/(|f'_0|)}$ \footnote{Note that in $d=1$, the above argument remains correct with the change $p(\tau) = (\gamma/2) \,e^{-\gamma\tau/2}$ since here $\tau$ denotes the time spent in a given velocity state and not the duration between two tumbling events. In this case, one finds $p_X(r_0 - \delta)  \sim \delta^{\gamma/(2 |f'_0|)-1}$, in agreement with the exact result for the harmonic case for which $|f'_0| = \mu$ -- see Eq.~(\ref{eq:px_1d_rtp}).}. \\

Besides, for $d>1$, the distribution $P_Z(z)$ of the random variable $Z = (v_0 - v_1)/|f_0'|$ in Eq. (\ref{gen_Langevin_linear2}) has an algebraic behavior near $z=0$, namely $P_Z(z) \sim z^{(d-3)/2}$ -- see Eq. (\ref{WprojEq}). From Eq. (\ref{gen_Langevin_linear2}) we see that, in the stationary state, the distance from the turning point $\Delta = r_0-x_1$ is the convolution of the two distributions $p_Y(y)$ and $p_Z(z)$, since the two random variables $Y$ and $Z$ are statistically independent. Hence, for small $\delta$,  the distribution $p_X(x)$ behaves as
\bea \label{convol_delta}
p_X(r_0 - \delta) \sim \int_0^{\delta} P_Y(\delta - z) P_Z(z) dz \sim \int_0^{\delta} (\delta-z)^{-1+\gamma/|f_0'|}\, z^{(d-3)/2} dz \sim \delta^{\gamma/|f_0'|
+(d-3)/2} \;.
\eea
In the harmonic case $|f_0'|=\mu$, this simple analysis yields back the correct exponent~(\ref{convol_delta}), which can be computed from an exact solution of the stationary profile in the harmonic case -- see Eq.~(\ref{2dpX}) in $d=2$ and Eq.~(\ref{3dTPbehavior}) in $d=3$. \\

It is also interesting to compute the behavior of the radial distribution $p_R(r)$ for $r$ close to $r_0$. The two distributions are related through (for $d>1$) Eq.~(\ref{pXtopR}). Hence, for $x = r_0 - \delta$, writing $r = r_0 - z$, one has
\begin{equation}
    p_X(r_0 - \delta) \;=\; \frac{\Gamma\!\bigl(\tfrac{d}{2}\bigr)}{\sqrt{\pi}\,\Gamma\!\bigl(\tfrac{d-1}{2}\bigr)}
    \int_{0}^{\delta} dz \, \frac{p_R(r_0 - z)}{r_0}\left(\frac{2(\delta - z)}{r_0}\right)^{\frac{d-3}{2}} \, .
\end{equation}
Now, if one assumes that $p_R(r_0 - z) \sim z^s$ (here $z\to 0$), and makes the change of variable $u = z/\delta$, we have 
\begin{eqnarray}
      p_X(r_0 - \delta)  \sim  \delta^{s + \frac{d-1}{2}}\, .
\end{eqnarray}
Using the result~(\ref{convol_delta}), we find the following condition for $s$
\begin{equation}
    \frac{\gamma}{|f_0'|} + \frac{d-3}{2}
=
    s + \frac{d-1}{2} \, .
\end{equation}
Finally, we find for $r \to r_0$
\begin{equation}
    p_R(r_0- \delta) \sim \delta^{\gamma/|f_0'| -1}\, ,
\end{equation}
which is independent of $d$. Therefore, for $d>1$, there is always a shape transition (at the level of the radial PDF) when $ \gamma/|f_0'|=1$.

\vspace*{0.5cm}
\noindent{\bf The case $D>0$.} It is now simple to extend the analysis presented above in the case $D>0$. Indeed in this case the dynamics of the RTP linearized close to a turning point leads to (see Eq. (\ref{gen_Langevin_linear2}))
\bea \label{gen_Langevin_linear_Dfinite}
\frac{d \epsilon_1}{dt} = -|f'_0| \epsilon_1 + v_1 - v_0 + \sqrt{2 D}\, \eta_1(t) \;.
\eea
Since one simply needs to take the additional convolution of this result (\ref{convol_delta}) with a Gaussian random variable ${\cal N}(0, D/|f'_0|)$. All the results obtained for the scaling form around $r_0$ in the harmonic case (see Section~\ref{UniversalTP}) hold when replacing $\mu \to |f'_0|$.

\section{Stationary State Distribution for a Three-State RTP}\label{3state}

\begin{figure}[t]
    \centering
    \includegraphics[width=0.45\linewidth]{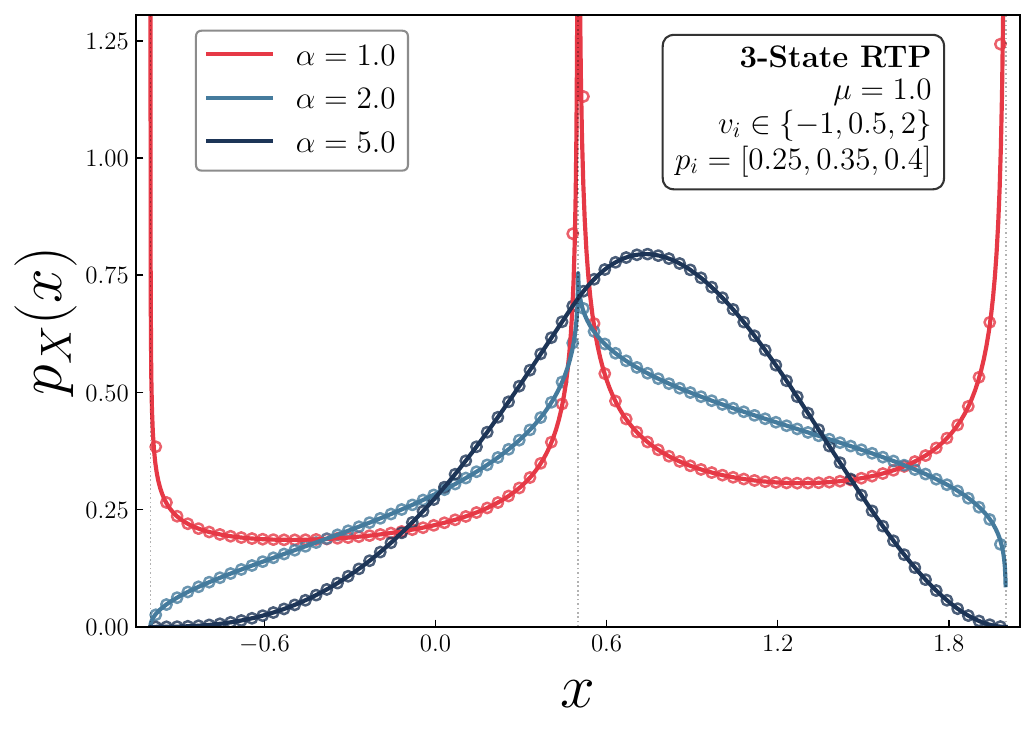}
    \caption{Comparison of the analytical prediction Eq.~\eqref{3stateapp} for the stationary distribution of an RTP with three velocity states in a harmonic trap compared to simulation results. The solid lines represent the exact theory, while the symbols denote numerical data obtained from simulations. The distributions are shown for different values of the dimensionless switching rate $\alpha = \gamma/\mu$. Here the parameters are $\mu=1$, $\{v_1, v_2, v_3\} = \{-1, 0.5, 2\}$, and $\{p_1, p_2, p_3\} = \{0.25, 0.35, 0.4\}$. The vertical dotted line separates the region I, for $x \in [v_1/\mu, v_2/\mu]$, from the region II for $x \in [v_2/\mu, v_3/\mu]$.}
    \label{3statefig}
\end{figure}

In this appendix, we provide an illustrative and nontrivial example of a discrete state RTP model, as presented in Section \ref{NstateSection}. Let us indeed consider a model with $N=3$ states 
\begin{eqnarray}
    W(v) = p_1\delta(v-v_1) + p_2\delta(v-v_2) + p_3\delta(v-v_3) \, ,
\end{eqnarray}
where $v_3 > v_2 > v_1$ and $p_1 + p_2 + p_3 = 1$. In this case, we start with the exact stationary state PDF given in Eq. (\ref{NstateExact}) where we integrate explicitly over the variables $z_1$ and $z_2$. For $v_1 / \mu < x < v_3 / \mu$, this simplifies to
\begin{eqnarray}
  p_X(x) =  \frac{\Gamma(\alpha)}{\Gamma(\alpha p_1)\Gamma(\alpha p_2)\Gamma(\alpha p_3)} \frac{\mu}{v_2-v_1} \int_{ z_3^{-}}^{ z_3^{+}} &dz_3&
\left(1 - \frac{\mu x - v_1 - z_3 (v_3 - v_1)}{v_2 - v_1}
- z_3 \right)^{\alpha p_{1} - 1}  \nonumber \\
&& \times \left( \frac{\mu x - v_1 - z_3 (v_3 - v_1)}{v_2 - v_1} \right)^{\alpha p_{2} - 1} \!\!z_3^{\alpha p_{3} - 1}\, ,\quad
\label{3stateapp}
\end{eqnarray}
where we have introduced
\begin{eqnarray} \label{z3m}
    z_3^{-} = \max\!\left( 0,\; \frac{\mu x - v_2}{v_3 - v_2} \right)\, ,
\qquad
z_3^{+} = \frac{\mu x - v_1}{v_3 - v_1} \, .
\label{bounds}
\end{eqnarray}
To proceed, we note that the integral over $z_3$ can be rewritten as follows
\begin{eqnarray}
 p_X(x) &=& \frac{\Gamma(\alpha)}{\Gamma(\alpha p_1)\Gamma(\alpha p_2)\Gamma(\alpha p_3)} \frac{\mu}{v_2-v_1} \left(\frac{v_2-\mu x}{v_2-v_1}\right)^{\alpha p_{1} - 1}\left(\frac{\mu x-v_1}{v_2-v_1}\right)^{\alpha p_{2} - 1}I(z_3^{-},z_3^{+})\, ,
\end{eqnarray}
with
\begin{eqnarray}
I(z_3^{-},z_3^{+}) &=& \int_{z_3^{-}}^{z_3^{+}} dz_3
\left[1 +\left(\frac{v_3-v_2}{v_2-\mu x}\right) z_3\right]^{\alpha p_{1} - 1} \left[1 +\left(\frac{v_3-v_1}{v_1-\mu x}\right) z_3\right]^{\alpha p_{2} - 1} \!\!z_3^{\alpha p_{3} - 1}\\
&=&\left.
\frac{z^{\alpha p_3}}{\alpha p_3}
F_1\!\left(
\alpha p_3;\,
1 - \alpha p_1,\,
1 - \alpha p_2;\,
\alpha p_3 + 1;\,
\frac{v_3 - v_2}{ \mu x-v_2}z,\,
\frac{v_3 - v_1}{ \mu x - v_1}z
\right)
\right|_{z = z_3^-}^{z = z_3^+}\, ,
\end{eqnarray}
where $F_1$ is the Appell hypergeometric function \cite{Appell}. Interestingly, because of the singular dependence of $z_3^-$ on $x$ in Eq. (\ref{z3m}), the function $p_X(x)$ takes different functional forms on the intervals $x \in [v_1/\mu, v_2/\mu]$ (where $z_3^-=0$) and $x \in [v_2/\mu, v_3/\mu]$ (where $z_3^-={(\mu x - v_2)}/{(v_3 - v_2)}$). \\

\noindent\textbf{Region I: $x \in [v_1/\mu, v_2/\mu]$.}
In this region, the boundaries of the integral are
\begin{eqnarray}
    z_3^{-} = 0\, ,
\qquad
z_3^{+} = \frac{\mu x - v_1}{v_3 - v_1} \, .
\end{eqnarray}
This leads to
\begin{align}
p_X(x) 
&= 
\frac{\Gamma(\alpha)}{
    \Gamma(\alpha p_1)\Gamma(\alpha p_2)\Gamma(\alpha p_3)
}
\,
\frac{\mu}{v_2 - v_1}
\left(\frac{v_2 - \mu x}{v_2 - v_1}\right)^{\alpha p_1 - 1}
\left(\frac{\mu x - v_1}{v_2 - v_1}\right)^{\alpha p_2 - 1}
\left(\frac{\mu x - v_1}{v_3 - v_1}\right)^{\alpha p_3}
\frac{1}{\alpha p_3}
\nonumber\\[6pt]
&\quad \times 
F_1\!\left(
    \alpha p_3;\,
    1 - \alpha p_1,\,
    1 - \alpha p_2;\,
    \alpha p_3 + 1;\,
    \frac{v_3 - v_2}{\mu x - v_2}
    \frac{\mu x - v_1}{v_3 - v_1},\,
    1
\right).
\end{align}

\noindent\textbf{Region II: $x \in [v_2/\mu, v_3/\mu]$.}
Here, the boundaries~(\ref{bounds}) read
\begin{eqnarray}
    z_3^{-} = \frac{\mu x - v_2}{v_3 - v_2}\, ,
\qquad
z_3^{+} = \frac{\mu x - v_1}{v_3 - v_1} \, .
\end{eqnarray}
We obtain
\begin{align}
p_X(x)
&=
\frac{\Gamma(\alpha)}{
    \Gamma(\alpha p_1)\Gamma(\alpha p_2)\Gamma(\alpha p_3)
}
\,
\frac{\mu}{v_2 - v_1}
\left(\frac{v_2 - \mu x}{v_2 - v_1}\right)^{\alpha p_1 - 1}
\left(\frac{\mu x - v_1}{v_2 - v_1}\right)^{\alpha p_2 - 1}
\frac{1}{\alpha p_3}
\nonumber\\[6pt]
&\quad \times
\Bigg[
\left(\frac{\mu x - v_1}{v_3 - v_1}\right)^{\alpha p_3}
F_1\!\left(
    \alpha p_3;\,
    1 - \alpha p_1,\,
    1 - \alpha p_2;\,
    \alpha p_3 + 1;\,
    \frac{v_3 - v_2}{\mu x - v_2}
    \frac{\mu x - v_1}{v_3 - v_1},\,
    1
\right)
\nonumber\\[6pt]
&\qquad -
\left(\frac{\mu x - v_2}{v_3 - v_2}\right)^{\alpha p_3}
F_1\!\left(
    \alpha p_3;\,
    1 - \alpha p_1,\,
    1 - \alpha p_2;\,
    \alpha p_3 + 1;\,
    1,\,
    \frac{v_3 - v_1}{\mu x - v_1}
    \frac{\mu x - v_2}{v_3 - v_2}
\right)
\Bigg].
\end{align}
In Fig.~\ref{3statefig}, we compare our theoretical prediction with simulation results.

\end{document}